%% file: WZrap.tex
\documentclass[a4paper,11pt]{article}
\pdfoutput=1 
\usepackage{jheppub}  
\usepackage{graphicx,color}
\usepackage{amsmath}
\usepackage{autobreak}
\allowdisplaybreaks
\newcommand{\ben}{\begin{enumerate}}
\newcommand{\een}{\end{enumerate}}
\newcommand{\beq}{\begin{equation}}
\newcommand{\eeq}{\end{equation}}
\def\g0#1DY{{g^V_{_{d,0#1}} }}
\def\gSVN#1{\widetilde{\Delta}^{\rm f.o.(#1)}_{d,ab}}
\usepackage[dvipsnames]{xcolor}
\newcommand{\LNoNtb}{{\overbar{\boldsymbol{L}}}}
\newcommand{\as}{a_S}
\newcommand{\als}{\alpha_S}

\def\z#1{\zeta_{#1}}
\def\bt#1{\beta_{#1}}
\newcommand{\CA}{C_A}
\newcommand{\CF}{C_F} 
\newcommand{\NF}{n_f}
\newcommand{\Cf}{C_F}
\newcommand{\Ca}{C_A}
\newcommand{\nf}{n_f}

\newcommand{\dFA}{\frac{d_{FA}^{(4)}}{N_F}}
\newcommand{\dFF}{\frac{d_{FF}^{(4)}}{N_F}}

\newcommand{\Lqr}{L_{qr}}
\newcommand{\Lfr}{L_{fr}}
\newcommand{\Lqrfr}{L_{qr}L_{fr}}
\newcommand{\omg}{\omega}
\newcommand{\nfv}{n_{fv}}

\newcommand{\eq}[1]{eq.\ (\ref{#1})}
\newcommand{\fig}[1]{fig.\ (\ref{#1})}

\newcommand{\sect}[1]{sec.\ (\ref{#1})}
\newcommand{\app}[1]{app.\ (\ref{#1})}

\newcommand{\nn}{\nonumber\\}
\newcommand{\df}{{\rm{d}}}
\newcommand{\mur}{\mu_r}
\newcommand{\muf}{\mu_f}
\newcommand{\overbar}[1]{\,\overline{\!{#1}}}
\newcommand{\Nbar}{\overbar{N}}
\newcommand{\Nbari}{\overbar{N}_i}
\def\Nbar#1{\overbar{N}_{#1}}

\def\GNDY#1{{\boldsymbol{g}^{'q}_{d,#1}}}
\def\Ap#1{\boldsymbol{A}^{'}_{#1}}
\def\Dp#1{\boldsymbol{D}^{'}_{#1}}
\def\btp#1{{\boldsymbol{\beta}^{'}_{#1}}}

\newcommand{\zlmmte}{\ln^3(\widetilde{\omg})}
\newcommand{\zlmmto}{\ln^2(\widetilde{\omg})}
\newcommand{\zlmmo}{\ln(\widetilde{\omg})}

\newcommand{\wminv}{{\widetilde{\omg}_i}}
\newcommand{\w}{\overbar{\omg}}

\definecolor{amber}{rgb}{1.0, 0.49, 0.0}
\def\ca{C_A}
\def\tf{T_F}
\def\cf{C_F}
\def\cas{C^{2}_{A}}
\def\cat{C^{3}_{A}}
\def\caf{C^{4}_{A}}
\def\tfs{T^{2}_{F}}
\def\tft{T^{3}_{F}}

\def\cfs{C^{2}_F}
\def\cft{C^{3}_F}
\def\nfs{n^{2}_{f}}
\def\nft{n^{3}_{f}}
\title{\bf $Z$, $W^{\pm}$ rapidity distributions 
at NNLL and beyond}
\author[]{Goutam Das}
\affiliation[]{
        Institut f{\"u}r Theoretische 
        Teilchenphysik und Kosmologie,\\
        RWTH Aachen University,
        D-52056 Aachen, Germany
}
\emailAdd{goutam@physik.rwth-aachen.de}
\abstract{
In this article, we have studied threshold effects 
on rapidity distributions of massive gauge bosons 
($Z,W^{\pm}$) in the Standard Model at the Large 
Hadron Collider. By exploiting the universal behavior
of soft gluon emissions in the threshold region, 
we resum the large threshold logarithms arising in the 
rapidity distribution at 
next-to-next-to leading logarithmic accuracy 
and match them to next-to-next-to leading order 
in QCD. We adapt the double Mellin approach to resum 
both threshold variables corresponding to 
partonic threshold and rapidity consistently within 
the standard QCD framework. Furthermore, we have 
studied 
in detail the numerical impact of these threshold 
effects on the rapidity distribution of massive gauge 
bosons and found a better perturbative 
convergence in the resummed rapidity spectrum.
As a by-product, we also provide all the 
perturbative ingredients to extend the analysis to 
next-to-next-to-next-to leading logarithmic accuracy. 
As a first application of these third order 
ingredients, we have estimated their effects by
matching them with the third order soft-virtual 
results. Our results will be useful to understand 
and possibly constrain parton distributions
using the rapidity spectrum. 
}

\begin{document} 
%%%Preprint
\preprint{
        TTK-23-05
        \\
        \vspace{-2cm}
        \hspace{13cm}P3H-23-018}
\maketitle
\section{Introduction} \label{intro}
The Standard Model (SM) electroweak vector boson 
($Z,W^\pm$) production processes are important to 
understand  
perturbative Quantum Chromodynamics (QCD) at
hadron colliders such as the Large Hadron Collider (LHC). 
They have been subjected to intensive studies
\cite{Altarelli:1978id,Altarelli:1979ub} 
from early days of QCD which was possible 
due to their simple 
theoretical structure. A precise understanding of 
these massive boson production processes is also critical for the search 
of new physics beyond the SM (BSM). Often they 
provide important backgrounds to different signals of 
BSM models. 
The $Z$ boson production and its subsequent decay to 
leptons is one of the cleanest processes to study
at the LHC. The clean signature with a large event 
rate makes it possible to have very small experimental 
errors over a large energy range. 
As a result, the process is considered 
to be a standard candle at the LHC for 
luminosity monitoring. 
Similar leptonic decay modes for $W^\pm$ bosons 
involve a charged lepton and a neutrino. 
The latter escapes detection, 
leaving behind missing energy signatures
which pose additional experimental challenges.
Despite experimental issues, the study 
of $W^\pm$ bosons is equally important to 
understand the gauge structure of the SM, and also 
they provide important backgrounds to different BSM 
models involving $Z', W'$ or a spin-2 graviton.\footnote{See 
e.g.\ \cite{Fiaschi:2022wgl,Das:2016pbk}
and the references therein.}
Thus, a precise understanding of these processes 
could be the key to unravel physics beyond the SM.

The rapidity distributions of electroweak gauge 
bosons provide important information 
on parton distribution functions (PDFs) at 
different values of the invariant mass ($Q$) of the 
final state  and momentum fraction ($x$) of partons.
Thus, a precise
knowledge of rapidity distributions is necessary to 
constrain PDFs inside a proton. In this context,
an important observable is the charge-asymmetry
\cite{Berger:1988tu,Martin:1988aj} of $W^\pm$ boson, 
which directly probes the $u$ and 
$d$ quark distributions inside the proton. 
The rapidity distributions of $W^\pm$ bosons therefore 
can provide quantitative information on the size 
of valence quark densities. 

There has been a significant effort to improve 
theoretical predictions for 
these processes to match the experimental data. 
The inclusive cross section is known  
\cite{Hamberg:1990np,Harlander:2002wh} 
to next-to-next-to leading order (NNLO) accuracy 
for a long time.
The dominant singular contributions have been obtained 
beyond NNLO using the soft-virtual (SV) approximation 
at next-to-next-to-next-to leading order (N3LO)
in \cite{Ahmed:2014cla,Li:2014bfa,Catani:2014uta}. 
Within the threshold resummation framework 
\cite{Catani:1996yz,Contopanagos:1996nh,Magnea:2000ss,Catani:2003zt,Manohar:2003vb,Eynck:2003fn,Moch:2005ba,Moch:2005ky,Laenen:2005uz,Ravindran:2005vv,Ravindran:2006cg,Idilbi:2006dg,Becher:2006mr,deFlorian:2012za}
they have been further improved 
by resumming large threshold logarithms at N3LL
\cite{Catani:2014uta,Bonvini:2015ira,Ajjath:2020rci,Das:2022zie}
and beyond 
\cite{Das:2019btv,Das:2019uvh,Das:2020adl,Ajjath:2021lvg}.
Additionally, the electroweak (EW) corrections are 
also known up to NLO 
\cite{Dittmaier:2001ay,Baur:2001ze,Baur:2004ig,Arbuzov:2005dd,CarloniCalame:2006zq,Zykunov:2005tc,CarloniCalame:2007cd,Arbuzov:2007db,Dittmaier:2009cr}
and recently to NNLO in mixed QCD-EW 
\cite{Bonciani:2020tvf,Bonciani:2021zzf,Bonciani:2021iis,Armadillo:2022bgm}.
Further, the N3LO QCD result has been obtained 
recently for the inclusive cross sections 
\cite{Duhr:2020seh,Duhr:2020sdp,Duhr:2021vwj}
of these processes.

The complete NNLO rapidity distributions for gauge 
bosons are also known
\cite{Anastasiou:2003yy,Anastasiou:2003ds,Catani:2009sm,Melnikov:2006kv,Gavin:2012sy}
for quite some time.
The NNLO K-factor defined with respect to 
next-to-leading order (NLO) 
amounts to corrections below $1\%$ for 
a wide range of rapidity.
The NLO rapidity distribution describes the 
kinematics very well. The NNLO 
corrections do not change the shape 
of the distribution dramatically, rather its 
effect is quite flat over the rapidity range.
The scale uncertainties at NNLO amount to below $1\%$
at the LHC. The NNLO corrections already stabilize the 
perturbative convergence for these processes.
Beyond NNLO, the SV corrections have also been obtained 
at the third order 
\cite{Ravindran:2006bu,Ravindran:2007sv,Ahmed:2014uya} 
in QCD exploiting the universality of soft 
radiation \cite{Anastasiou:2014vaa} and available 
mass factorization kernels \cite{Moch:2004pa,Vogt:2004mw},
and process-dependent third order form factors \cite{Gehrmann:2010ue}.
The SV corrections in $x$ space at the 
third order amount to around 
$4\%$ of the NNLO contribution at higher rapidities.
The mixed QCD-EW corrections are also 
studied in \cite{Dittmaier:2009cr,Li:2012wna}, adding 
a correction of about $-1\%$ near the $Z$ resonance.
Using the $q_T$ subtraction formalism 
\cite{Catani:2007vq,Catani:2009sm,Catani:2010en,Catani:2011qz}, 
the rapidity distributions have been obtained recently 
for photon mediated Drell-Yan (DY) \cite{Chen:2021vtu}
production.
The corrections are about $-2\%$ for most of the 
rapidity region. The conventional scale 
uncertainty at this order does not overlap 
with the NNLO one in the central rapidity region and 
barely overlaps in the higher rapidity region. 
A similar method has also been employed for 
$W^\pm$ bosons to obtain N3LO \cite{Chen:2022lwc} 
rapidity distributions and the charge asymmetry.
A relatively flat correction at N3LO 
was observed there as well, which amounts to about 
$-2.5\%$ corrections compared to the previous order.
Additionally, the scale uncertainty band at N3LO
is not covered by the previous order within the 
standard seven-point scale variation. 
This feature has also been observed 
in the inclusive production 
\cite{Duhr:2020seh,
Duhr:2020sdp,
Duhr:2021vwj} for these processes. 
It is thus interesting to investigate the threshold 
effects in these processes by resumming 
dominant threshold logarithms which are 
expected to change in the higher rapidity region.

Similar to the total inclusive cross section, 
the rapidity distribution receives similar threshold 
effects once one identifies the threshold variables 
properly. In the standard QCD framework, there are two 
different approaches in this context. In the 
Mellin-Fourier (M-F) approach \cite{Laenen:1992ey},
the double threshold behavior (corresponding to the
partonic threshold and the rapidity) is approximated 
by limiting partonic threshold variable to certain 
singular structures in the Fourier space. 
The $Z,W^\pm$ rapidity distributions 
have been studied at next-to-leading logarithmic (NLL) 
accuracy 
\cite{Mukherjee:2006uu,Sterman:2000pt,Bolzoni:2006ky} 
and also at next-to-next-to-leading logarithm
(NNLL) \cite{Bonvini:2010tp} using this approach. 
On the other hand in the double Mellin 
(M-M) approach 
\cite{Catani:1989ne,Ravindran:2006bu,Ravindran:2007sv,Westmark:2017uig,Banerjee:2017cfc,Banerjee:2018vvb,DASbbH,DASggH} 
both threshold variables 
corresponding to partonic threshold and rapidity
are consistently resummed retaining all the 
singular distributions correctly. Here one defines two threshold 
variables at the partonic level $z_1$ and $z_2$
in analogy to the inclusive case. 
These variables are straightforwardly related to the 
inclusive threshold variable ($z$) and the partonic 
rapidity. 
For convenience of the reader we will discuss the 
theoretical framework in the next section, 
although our presentation closely follows recent
article \cite{Banerjee:2017cfc}.
A numerical comparison between these two approaches 
has been first performed in 
\cite{Banerjee:2018vvb,Banerjee:2018mkm} and a better 
perturbative convergence has been observed in the 
M-M approach. 
A general framework has been put forward in 
\cite{Ahmed:2020amh} for the 
M-M approach following \cite{Ravindran:2006bu,Ravindran:2007sv}.
Later in \cite{Lustermans:2019cau}, 
the rapidity distribution has been improved beyond 
the double singular distributions (see also 
\cite{Ebert:2017uel} for resummation of $\pi^2$ 
terms) 
providing a general framework employing 
Soft-Collinear Effective Theory 
\cite{Bauer:2000ew,Bauer:2000yr,Bauer:2001ct,Bauer:2001yt,Bauer:2002nz,scetIain,Becher:2014oda}.

The purpose of this article is to perform a complete 
study of threshold effects at the second order 
in strong coupling, resumming large threshold 
logarithms corresponding to threshold variables 
$z_1, z_2$ and matching them to available NNLO results. 
We follow the M-M approach within the 
traditional QCD resummation framework
which correctly resums these large logarithms
in a systematic framework, 
and we study their impact on
the $Z,W^\pm$ rapidity distributions. 
Furthermore, we also provide analytical results 
and numerical estimates at the third order 
at SV as well as at N3LL accuracy.
The article is organized as follows:
in \sect{sec:theory} we lay out the 
necessary theoretical foundation for the 
resummation of rapidity distributions in 
the M-M approach. In \sect{sec:numerical} we 
present a detailed phenomenological study of the 
processes in the context of 13.6 TeV LHC. 
Finally, we conclude in \sect{sec:conclusion}
and collect all the analytic ingredients required 
up to N3LL in appendices (\ref{App:ANOMALOUS-DIMENSIONS} - \ref{App:SV-COEFFICIENTS}).
%%%Theoretical framework
\section{Theoretical Framework} \label{sec:theory}
The hadronic cross section for the rapidity distribution
of colorless particles can be formulated through the 
following master formula \cite{ellis_stirling_webber_1996},
\begin{align}\label{eq:rapidity-master}
        \frac{\df\sigma(\tau, y)}{\df Q~\df y } 
        =
        \sum_{a,b=q,\overline q, g}
        &\int_{0}^1 {\df  x_1}
        \int_{0}^1 {\df  x_2}~
        f_a\left({x_1},\muf\right)  
        f_b\left({x_2},\muf\right)
\nn & \times 
        \int_{0}^1 {\df  z_1}
        \int_{0}^1 {\df z_2}~
        \delta(x_1^0 - x_1 z_1)
        \delta(x_2^0 - x_2 z_2)~
\widehat{\sigma}_{d,ab} (z_1,z_2,\muf,\mur) \,.
\end{align} 
Here $\tau=Q^2/S = x_1^0 x_2^0 $ and  
$y = \frac{1}{2}\ln \left( x_1^0 / x_2^0\right)$
are the hadronic threshold variable and 
hadronic rapidity of the colorless final state 
respectively.
The partonic threshold variable on the other hand 
is given as $z= Q^2/\widehat{s}=z_1 z_2$. 
Here $\sqrt{S}\equiv E_{\rm CM}$ and $\widehat{s}$ are hadronic and partonic
center of mass energies respectively. 
For on-shell $Z,W^\pm$ production, $Q=m_V$ with 
$V=Z,W^\pm$.
The rapidity dependent partonic coefficient function 
($\widehat{\sigma}_{d,ab}$)
can be decomposed (up to a Born factor 
$\sigma_{_{0,V}}$)
in terms of singular SV piece and the regular piece,
\begin{align}\label{eq:partonic-decompose}
\widehat{\sigma}_{d,ab}(z_1,z_2,\muf,\mur)
=
\sigma_{_{0,V}}(Q) 
\Big( 
	\Delta_{d,ab}^{\rm sv}\left(z_1,z_2,\muf,\mur\right) 
	+ \Delta_{d,ab}^{\rm reg}\left(z_1,z_2,\muf,\mur\right)
\Big) \,.
\end{align}
Similar to the inclusive production, the SV part 
gets contributions only from the diagonal 
channel \textit{i.e.}\ in the present case 
$a,b = q, \bar{q}$ whereas the regular part gets 
contributions from all partonic channels.
In the threshold region, the SV part constitutes 
a significant contribution to the total cross section
due to large threshold logarithms.
Even away from the threshold, it is desirable to 
improve the theory predictions by resumming these
logarithms considering the importance of these 
processes.

The overall normalization constants are different 
for different processes. In particular,
the neutral DY production gets 
contributions from virtual photon $(\gamma^{*})$ 
and $Z$ boson
as well as their interference ($\gamma^{*} Z$). 
Their combined effect has been studied in detail 
in \cite{Banerjee:2018vvb}. Here we will mostly 
focus on the $Z$ and $W^\pm$ production.
For completeness, we present the 
explicit forms of the prefactors for different 
cases below,
\begin{align}\label{eq:normalization}
%%%% gamma
	{\sigma}_{0,\gamma^{*}}(Q) =&
	\frac{8 \pi \alpha^2}{3 n_c Q S} 
        \Bigg[ 
        e_l^2 e_q^2
        \Bigg]
        \,,
\nn
%%%% gamma-Z
{\sigma}_{0,\gamma^{*} Z}(Q) =&
	\frac{8 \pi \alpha^2}{3 n_c Q S} 
        \Bigg[ 
        \frac{2 Q^2(Q^2-m_Z^2)}{|P_{_Z}(Q)|^2 ~c_w^2 s_w^2}
        e_l e_q g_l^V g_q^V
        \Bigg]
        \,,
\nn
%%%% Z
{\sigma}_{0,Z}(Q) 
        =&
	\frac{8 \pi \alpha^2}{3 n_c Q S} 
        \Bigg[
	\frac{3 Q^4}{\alpha ~m_Z c_w^2 s_w^2}
        \Big((g_q^V)^2+(g_q^A)^2\Big) 
        \frac{\Gamma_Z {\cal B}_{Z}^{l}}{|P_{_Z}(Q)|^2 }
        \Bigg]\,,
\nn
%%%% W
{\sigma}_{0,W}(Q) =&
	\frac{8 \pi \alpha^2}{3 n_c Q S} 
        \Bigg[
	\frac{3Q^4 V_{qq'}^2}{\alpha ~m_W s_w^2 }
        \Big((g_q^{'V})^2+(g_q^{'A})^2\Big) 
        \frac{\Gamma_W {\cal B}_{W}^{l}}{|P_{_W}(Q)|^2}
        \Bigg]\,,
\end{align}
where, 
\begin{align}
g_a^A &= -\frac{1}{2} T_a^3 \,, & 
g_a^V &= \frac{1}{2} T_a^3  - s_w^2 e_a \,, &
\nn
g_a^{'A} &= -\frac{1}{2\sqrt{2}}\,, &
g_a^{'V} &= \frac{1}{2\sqrt{2}}\,.&
\nn
e_l &= -1 \,, &
e_u &= \frac{2}{3} \,, &
e_d &= - \frac{1}{3} \,.&
\end{align}
$V_{qq'}$ are the CKM matrix elements with 
$e_q + e_{q'} = \pm 1$.
Here $e_a$ is the electric charge and $T_a^3$ is 
the weak isospin number for leptons, 
$\Gamma_V$ is the total decay width of vector boson $V$ 
whereas ${\cal B}_{V}^{l}$ is its branching ratio 
to a pair of leptons. 
The propagator factor in \eq{eq:normalization} is 
given as,
\begin{align}
        P_{_V}(Q) &= Q^2 - m_V^2 + i ~m_V \Gamma_V.\
\end{align}
With the above normalization, the SV part takes the 
form 
$\Delta^{\rm sv}_d(z_1,z_2) = 
\delta(1-z_1)\delta(1-z_2) $
at the leading order. In terms of threshold 
contributions, the $Z$ and $W^\pm$ processes receive 
similar QCD corrections up to NNLO. 
At N3LO, the $Z$ boson production receives additional
charge weighted contributions whose numerical 
effect however is known to be negligible \cite{Ajjath:2020rci}.
In our analysis, these will be included nevertheless.

The double Mellin transformation on the partonic 
coefficient is performed in the following way taking the 
Mellin transformation with respect to both threshold 
variables $z_1$ and $z_2$ as,
\begin{align}\label{eq:mellin-partonic}
        \widetilde\Delta_{d,ab}(N_1,N_2) 
        =& 
        \int_0^1  \df z_1   z_1^{N_1-1}   
        \int_0^1  \df z_2   z_2^{N_2-1}   
        \Delta_{d,ab}^{\rm sv} (z_1,z_2) \,.
\end{align}
Here $N_i$ is the Mellin variable corresponding to 
$z_i$. In the Mellin space this can be organized in 
the following integral form 
\cite{Ravindran:2006bu,
Ravindran:2007sv,
Ahmed:2014uya} 
in terms of cusp anomalous dimensions $(A^q)$ and 
rapidity-dependent threshold soft anomalous 
dimensions $(D_d^q)$,
\begin{align}\label{eq:mellin-integral-form}
\widetilde{\Delta}_{d,ab}(N_1,N_2) 
= &
g_{_{d,0}}(\as) \exp
\Bigg(  
        \int_0^1 \df z_1 z_1^{N_1-1}   
        \int_0^1 \df z_2 z_2^{N_2-1}  
\nn&
\times
\bigg(
        \delta(\bar{z}_2)
        \Bigg[
        \frac{1}{\bar{z}_1} 
        \Bigg\{ \int_{\muf^2}^{Q^2 \bar{z}_1}
                {\df \eta^2 \over \eta^2}~ 
                A^q\left(\as(\eta^2)\right) 
% \nn&
                +D^q_d\left(\as\left(Q^2\bar{z}_1\right)\right) 
        \Bigg\}
        \Bigg]_+  
\nn&
        +
        \Bigg[
                \frac{1}{2\bar{z}_1\bar{z}_2} 
        \Bigg\{A^q(\as(z_{12})) 
        + \frac{\df D^q_{d}(\as(z_{12}))}{\df\ln z_{12}} 
        \Bigg\}
        \Bigg]_+
        + (z_1 \leftrightarrow z_2)
\bigg)
\Bigg)\,,
\end{align}
where $z_{12} = Q^2 \bar{z}_1 \bar{z}_2$ with $\bar{z}_i = 1-z_i$. 
We also redefine the strong coupling constant $\als$ in 
terms of $\as = \als/(4\pi)$ for convenience.
This partonic coefficient in
double Mellin space can be rearranged\footnote{Note that 
the Mellin transformation will produce some non-logarithmic contributions 
which are then combined with $g_{_{d,0}}$ to finally arrive at the 
coefficients $g_{_{d,0}}^V$.} 
in the following form to resum large logarithms 
consistently to all orders, 
\begin{align}\label{eq:resum-expression}
\widetilde{\Delta}_{d,ab}(N_1,N_2) 
= 
g^V_{_{d,0}}(\as,\mu_r,\mu_f) 
\exp 
\Bigg( 
        G_{d}(\as,\w,\mu_r, \mu_f)   
\Bigg)\,,
\end{align}
where 
$\w =  \as \beta_0 \ln(\Nbar1 \Nbar2)$, 
with $\Nbari = e^{\gamma_E} N_i, i=1,2$.
$\gamma_E \simeq 0.577$ is the Euler–Mascheroni constant.
The process-dependent coefficients 
($g^V_{_{d,0}}$) have the following perturbative expansion,
\begin{align}\label{eq:g0-expansion}
        g^V_{_{d,0}}(\as) = 1 + 
        \sum_{i=1}^{\infty} a_S^i ~g^V_{_{d,0i}} \,.
\end{align}
They are  collected up to the 
third order in the strong coupling 
in Appendix\ (\ref{App:g0}) for 
$Z,W^\pm$ processes.
The exponent, on the other hand, is universal 
and resums the large 
logarithms to all orders. It can be expanded 
in the strong coupling and the inclusion of 
successive terms defines the resummed order.
The explicit expansion takes the following 
form (suppressing the $\mur,\muf$ dependence and 
keeping in mind $\as \ln (N_1 N_2) \sim {\cal O}(1)$ 
in the threshold region), 
\begin{align}\label{eq:RESUM-EXPONENT}
G_d(\as,\w)
=
g^q_{d,1}(\w) ~\ln(\Nbar1 \Nbar2) 
+ \sum_{i=0}^\infty \as^i ~g^q_{d,i+2}(\w)\,.  
\end{align}

The resummed expression in \eq{eq:resum-expression}
captures the leading singular behavior to all orders 
provided the anomalous dimensions are known to 
sufficiently higher orders. However, this still 
lacks the non-singular (regular) contributions.
One way to include the effect of those regular 
terms into the prediction is  to match the resummed 
cross section corresponding to 
\eq{eq:resum-expression} with 
known fixed order results through some matching 
prescription. Since the fixed order results also 
contain the singular terms, one needs to remove them 
in order to avoid double-counting with the resummed 
expression in \eq{eq:resum-expression}.
The final matched expression then takes the following 
form, 
\begin{align}\label{eq:master-matched}
\frac{\df\sigma^{\text{matched}} }{ \df Q \df y } 
=  
\frac{\df \sigma^{\text{f.o.}} }{\df Q \df y } 
+& \sigma_{_{0,V}}(Q) 
\sum_{a,b=q,\bar{q}}
\int_{c_{1} - i\infty}^{c_1 + i\infty} 
        \frac{d N_{1}}{2\pi i}
\int_{c_{2} - i\infty}^{c_2 + i\infty} 
        \frac{d N_{2}}{2\pi i} 
e^{y(N_{2}-N_{1})}
\left(\sqrt{\tau}\right)^{-N_{1}-N_{2}} 
\nn&
\widetilde f_{a}(N_{1}) 
\widetilde f_{b}(N_{2}) 
%\nonumber\\&
%\hspace{-0.1cm}\times
\Big[ 
        \widetilde{\Delta}_{d,ab}(N_1,N_2) 
        -\widetilde{\Delta}^{\rm f.o.}_{d,ab}(N_1,N_2)  
\Big] \,.
\end{align}
The symbol `\text{f.o.}'\ in the second term 
($\widetilde{\Delta}^{\rm f.o.}_{d,ab}$) in 
the second line above means the function is 
truncated to a fixed order in order to avoid 
double counting the singular contributions which are 
already present in the first term 
$\left(\frac{\df \sigma^{\text{f.o.}} }{\df Q \df y } \right)$
in \eq{eq:master-matched}. 
$\widetilde f_j(N_i) \equiv \int_0^1 dz_i z_i^{N_i-1}f_j(z_i)$
are the PDFs in the Mellin space.
Ideally one needs Mellin space PDFs which can be 
used to perform the Mellin inversion together with 
the Mellin space partonic coefficient function. 
Such Mellin space PDF can be obtained with the publicly 
available code P{\sc egasus} \cite{Vogt:2004ns}. 
However, for practical purposes
one can use $x$-space PDF 
\cite{Catani:1989ne,Kulesza:2002rh,Catani:2003zt} 
with its derivative through \texttt{LHAPDF6} 
\cite{Buckley:2014ana} interface. 

Some comments are now essential on the implementation
of the double Mellin inversion. Let us first 
recall the case of inclusive production where only 
a single Mellin inversion ($N$) is needed for the 
inclusive threshold variable $z$. The straight-forward
Mellin inversion is divergent which is intrinsically 
related to the strong coupling which diverges near 
the Landau pole 
where the theory is not perturbative anymore. 
Thus, one needs to be careful while performing the Mellin 
inversion. One way to proceed further is to invoke so called 
`Minimal Prescription' (MP) \cite{Catani:1996yz}.
The idea is to choose the contour 
in such a way that all the poles remain at the left 
of the contour except the Landau pole which remains 
at the right of the contour. 
In the single Mellin 
inversion case, one can project the complex contour 
integration of the Mellin variable ($N$) onto a real 
variable ($r$)  using $N=c+r \exp(i \phi)$ with the
constant $c$ which appears in the contour.
One can then take suitable choices for $(c,\phi)$
according to the MP
to characterize the contour
and take the imaginary part of the integrand 
$h(N)$\footnote{See for example sec.\ (3.1) of \cite{Vogt:2004ns}.}
with appropriate normalization. 
The latter is possible as the integrand depends only on a 
single Mellin variable ($N$) such that the 
complex conjugate of the integrand is same as the 
original integrand with its argument complex 
conjugated \textit{i.e.} $h^{*}(N) = h(N^{*})$. 

The case with the double Mellin inversion is, on the 
other hand, more involved, as the integrand now 
depends on two independent Mellin variables 
$(N_1,N_2)$. Applying the same method on each of the 
Mellin variables does not work anymore  
as \textit{e.g.} $h^{*}(N_1,N_2) \neq h(N_1^{*}, N_2)$.
This is true for both the terms in the second line 
of \eq{eq:master-matched}. 
On top of this, one has to be careful with the Landau 
pole problem which now appears in the resummed 
exponent when $\w \to 1$. 
As before, one can apply MP and 
choose the contours in such a way that all the poles 
in $N_1, N_2$ remain on the left of the contours 
except the Landau pole.
In practice this is not 
straight-forward as the Landau pole now depends on 
two Mellin variables. One can still fix the contour
for one Mellin variable 
(say $N_1= c_1 + r_1 \exp(i \phi_1)$) 
in the similar way as the single Mellin inversion, 
however the Landau pole is not restricted anymore 
to the real axis of the second Mellin variable 
($N_2=c_2+r_2\exp(i\phi_2)$). 
Therefore, with a choice of $c_1, c_2$, 
the angle corresponding to $N_2$ has to be 
fixed according to $N_1$.
In fact for larger $r_1$ in order to satisfy MP,
$\phi_2$ has to be set close to $\pi$.
A suitable choice for 
$\phi_2$ can be taken 
\cite{Catani:1989ne,Westmark:2017uig} as 
$\phi_2 = 
\max\left(
        \pi/2
        -1/2\arg(\frac{1}{N_1} \exp(\frac{1}{\as \bt0} - 2 \gamma_E))
        ,3\pi/4 \right)
$ which is also fine for higher values of $r_1$.
The stability of the inverted results then needs to
be checked by changing the constants $c_1, c_2$ 
around the default choices which will be discussed in 
the next section.
%%%Numerical Results%%%%%%%%%%%%%%%%%%%%%%%%%%%%%
\section{Numerical Results}\label{sec:numerical}
%%%%%%%%%%%%%%%%%%%%%%%%%%%%%%%%%%%%%%%%%%%%%%%%%
With the setup introduced in the previous section,
we now turn to study the effects of threshold 
resummation on the rapidity distributions of $Z,W^\pm$
bosons at the LHC.
As stated earlier, the neutral DY rapidity 
distribution gets contributions from $Z$, 
$\gamma^{*}$ as well as their interference.  
The experimentally interesting region is near the 
$Z$ boson resonance where the rapidity distribution 
is studied by typically integrating the di-lepton 
invariant mass near the resonance. Note that the 
interference term vanishes when one 
sets $Q=m_Z$. A detailed study on matched DY rapidity 
distribution at NNLL has been performed previously
\cite{Banerjee:2018vvb}
keeping in mind the invariant mass region 
$80<Q<120$ GeV. For the $Z,W^\pm$ rapidity 
distributions, we simply set $Q=m_V$ to study 
the behavior at the resonance.

Now we set up all the parameters for our numerical
study as follows: 
we consider the $13.6$ TeV LHC collider 
with \texttt{MSHT20} \cite{Bailey:2020ooq} as our 
default PDF set which we use through the 
\texttt{LHAPDF6} \cite{Buckley:2014ana} interface 
(version 6.5.3). 
The strong coupling $(\als)$ is taken through the 
\texttt{LHAPDF6} whereas at the third order 
we evolved it using 
the renormalization group equation (RGE) of 
strong coupling with the four-loop 
QCD beta function 
(see App.\ \ref{App:ANOMALOUS-DIMENSIONS}).
For the evolution we use the initial condition 
$\als(m_Z) = 0.118$. 
An essential ingredient of this study is the 
fixed-order rapidity distribution for which 
we use the publicly available code \texttt{Vrap-0.9}
\cite{Anastasiou:2003ds}. The flexibility of this code 
allows us to study $Z, \gamma^{*}$ channels 
separately in addition to $W^\pm$ rapidities 
at NNLO. 
For the resummation in the double 
Mellin space, we use our in-house code which 
we linked to the \texttt{LHAPDF6} interface
as well. As explained in the previous section, 
the Mellin inversion in \eq{eq:master-matched} 
requires certain choices of the contour 
in order to avoid the Landau pole where 
the strong coupling constant diverges.
For this purpose, we chose the contours 
of the double Mellin integrations in such a way that 
all the poles remain to the left of the 
contours except the Landau pole.
We find suitable choices (inspired by the single 
Mellin case in \cite{Vogt:2004ns})
for the contour to be $c_1 = c_2 = 1.9$ and $\phi_1 = 3\pi/4$
whereas $\phi_2$ is then fixed by $N_1$ as stated earlier.
We also checked the stability of these choices by 
varying $c_1,c_2$ slightly below and above, and 
find a good agreement.
We interfaced our code to the {\tt Cuba} library 
\cite{Hahn:2004fe,Hahn:2014fua}
and use the built-in {\tt Vegas} integrator 
for the final integration. 
Furthermore, we validate our code against an 
independent implementation in Mathematica with a
toy PDF and found excellent agreement between 
two codes.

Our default central scale choices are 
$(\mur^c, \muf^c) = (1,1)m_V$ and for the scale 
uncertainty estimation, we typically vary them 
simultaneously within the range $[1/2,2]$ and 
with the constraint $1/2 \leq \mur/\muf \leq 2$ 
to invoke the seven-point uncertainty.
The envelope of the scale uncertainty band 
is then obtained from the maximum absolute deviations 
from the central scale choices.
We follow the $G_\mu$ scheme to fix the EW parameters
which translates to fixing $m_W, m_Z$ and Fermi 
constant $G_F$ as input while $\alpha ,s_w$ are 
now derived quantities. This choice is known to 
minimize the EW corrections \cite{Dittmaier:2009cr}.
Below we summarize all the parameters 
\cite{ParticleDataGroup:2020ssz} for the 
numerical study,
\begin{align}
        E_{\rm CM} &= 13.6 \text{ TeV},&
        % \alpha_{\rm QED}^{-1}(m_Z) &= 127.955,&
        G_F &= 1.6639 \times 10^{-5},&
        n_f &= 5, & 
        % s_w^2 &= 0.223, & 
        \nn
        m_Z &= 91.1876 \text{ GeV},&
        \Gamma_Z &=2.4952 \text{ GeV},&
        {\cal B}_Z^l &=0.03366,&
        \nn
        m_W &= 80.379 \text{ GeV},&
        \Gamma_W &=2.085 \text{ GeV},& 
        {\cal B}_W^l &=0.1086,&
        \nn
        % |V_{ud}| &= 0.9737,&
        % |V_{us}| &= 0.2245,&
        |V_{ud}| &= 0.9737,&
        |V_{us}| &= 0.2245,&
        |V_{ub}| &= 0.00382,&
        \nn
        |V_{cd}| &= 0.221,&
        |V_{cs}| &= 0.987,&
        |V_{cb}| &= 0.041 \,. &
\end{align}

In order to estimate the higher order effects in 
the rapidity distribution,
we define the following quantities' 
\textit{viz.}
the $K$-factor and $R$-factor 
\cite{Das:2019bxi,Das:2020gie,Das:2020pzo} 
corresponding to the 
fixed order and resummed order effects,
\begin{align}\label{eq:K-R-FACTORS}
        K_{ij} = 
        {\left[\frac{\df \sigma}{\df Q \df y}\right]_{{\rm N}i{\rm LO}} }\bigg/
        {\left[\frac{\df \sigma}{\df Q \df y}\right]_{{\rm N}j{\rm LO}} }\,,
        R_{ij} = 
        {\left[\frac{\df \sigma}{\df Q \df y}\right]_{{\rm N}i{\rm LO}+ {\rm N}i{\rm LL}} }\bigg/
        {\left[\frac{\df \sigma}{\df Q \df y}\right]_{{\rm N}j{\rm LO}+ {\rm N}j{\rm LL}} }\,. 
\end{align}
In addition, to get an estimate on the effect of the
SV cross section as well as the matched cross section
against fixed order (FO),
we define the following two perturbative ratios,
\begin{align}\label{eq:SF-RF-FACTORS}
        SF_{ij} = 
        {\left[\frac{\df \sigma}{\df Q \df y}\right]_{{\rm N}i{\rm LOsv}} }\bigg/
        {\left[\frac{\df \sigma}{\df Q \df y}\right]_{{\rm N}j{\rm LO}} }\,,
        RF_{ij} = 
        {\left[\frac{\df \sigma}{\df Q \df y}\right]_{{\rm N}i{\rm LO}+ {\rm N}i{\rm LL}} }\bigg/
        {\left[\frac{\df \sigma}{\df Q \df y}\right]_{{\rm N}j{\rm LO}} } \,.
\end{align}
Here we point out that although there are some 
intrinsic ambiguities concerning 
the estimation of theoretical scale uncertainties,
the traditional procedure\footnote{There 
is no such \textit{first principle}
of the scale uncertainty estimation, 
however there are studies \cite{Cacciari:2011ze,Bonvini:2020xeo} 
to estimate scale uncertainties 
through Baysian analysis as well.} 
is to vary the $\mur,\muf$
scales with the seven-point variation described before.
However, for the quantities defined as the ratio 
\text{e.g.} the $K$-factor or the $R$-factor 
or the charge asymmetry $A_W(y)$ (see 
\eq{eq:charge-asymmetry}) in case of $W^{\pm}$
productions there are ambiguities on how the scale 
should be treated both in the denominators and the 
numerators.
We study the following two standard methods:
\begin{itemize}
        \item \textbf{Correlated error:}
        here we evaluate the ratio at the same scale 
        for both the numerator and the denominator, and 
        finally we calculate the seven-point scale 
        uncertainty from all such seven ratios.

        \item \textbf{Uncorrelated error:}
        here we evaluate the seven-point uncertainty 
        of the numerator and denominator 
        independently with the constraint 
        $1/2 \leq \mu/\mu' \leq 2$ for any scale 
        combination $\mu,\mu' \in \muf, \mur$ 
        of the numerator and denominator. Essentially 
        this amounts to taking $31$ combinations from 
        different scales in the numerator and denominator
        to estimate the scale uncertainty. 
\end{itemize}
Estimation of intrinsic PDF uncertainties depends 
on the PDF set being used as they use different
methods to estimate uncertainties.
To assess the 
intrinsic PDF uncertainties from 
\texttt{LHAPDF6} interface, we define the following 
quantity, 
\begin{align}\label{eq:PDF-ERROR}
        \delta {\rm PDF} 
        = 1 \pm
        {\delta\left[\frac{\df \sigma}{\df Q \df y}\right] } 
        \bigg/
        {\left[\frac{\df \sigma}{\df Q \df y}\right]_{0} }\,. 
        % {[\df \sigma/dy]_{\rm N2LO}}, 
\end{align}
The intrinsic PDF uncertainty 
$\big({\delta\big[\frac{\df \sigma}{\df Q \df y}\big]} \big)$
corresponds to all the subsets and can also be 
asymmetric depending on the PDF set, 
while the denominator is evaluated at the $0$-th set
for each of these PDF sets. 
We will treat the above \eq{eq:PDF-ERROR} only at the 
central scale choice.
The reason we divided this  by the central set is 
linked with fact that different PDF sets can 
give different central values and hence a relative 
error is more justified. 
It is also interesting to compare these central 
predictions for different PDF groups along with their 
seven-point scale uncertainties to estimate how they 
behave in different rapidity regions. 
To do this we define another quantity as the 
ratio of the cross sections for different PDFs as,
\begin{align}\label{eq:rPDF}
        r{\rm PDF}
        =
        \left[\frac{\df \sigma}{\df Q \df y}\right]_{{\rm 0}}
        \bigg/
        \left[\frac{\df \sigma}{\df Q \df y}\right]_{{\rm 0,MSHT}} \,,
\end{align} 
where the ratio is taken with respect to the default 
PDF choice {\tt MSHT}. The subscript `0' indicates 
that we are considering only the central subset 
both in the numerator and denominator.
%%% DY-analysis%%%%%%%%%%%%%%%%%%%%%%%%%%%%%%%%%%%%%%%%%%%%%%%%%%%%
\subsection{DY rapidity distribution}\label{sec:dy-analysis}
%%%%%%%%%%%%%%%%%%%%%%%%%%%%%%%%%%%%%%%%%%%%%%%%%%%%%%%%%%%%%%%%%%
%% DY RAPIDITY
\begin{figure}[ht]
        \centering{
\includegraphics[width=7.4cm,height=5.6cm]{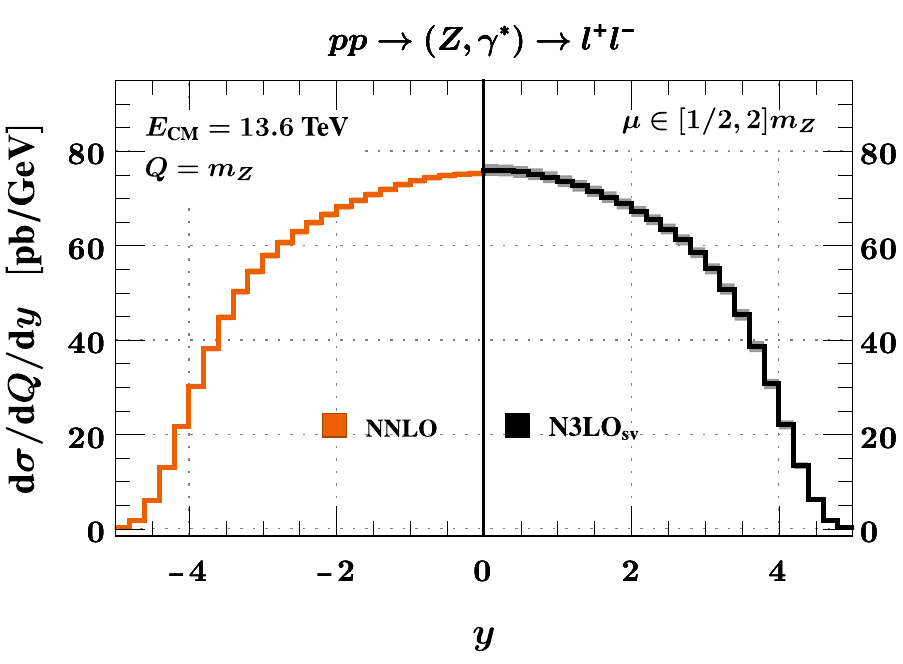}
\hspace{0.05cm}
\includegraphics[width=7.4cm,height=5.6cm]{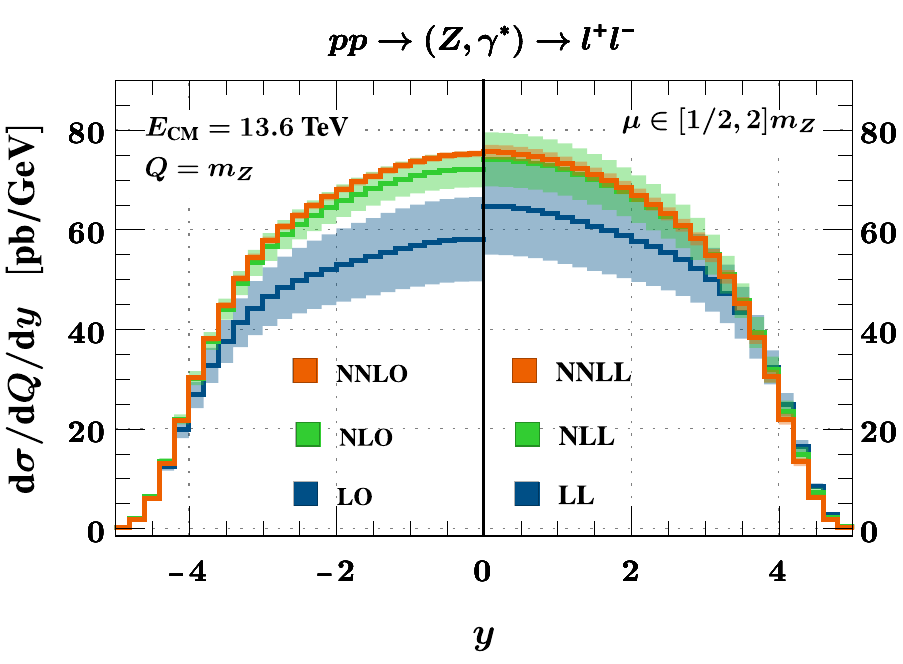}
}
\caption{DY rapidity distribution with 
scale uncertainty at $13.6$ TeV LHC. 
The left figure compares NNLO with N3LOsv whereas
the right figure compares FO against matched results.
Contributions from both $Z$ and $\gamma^{*}$ have 
been taken into account.}
\label{fig:DY-RAPIDITY}
\end{figure} 
%% DY SF,RF factors
\begin{figure}[ht!]
        \centering{
\includegraphics[width=7.4cm,height=5.6cm]{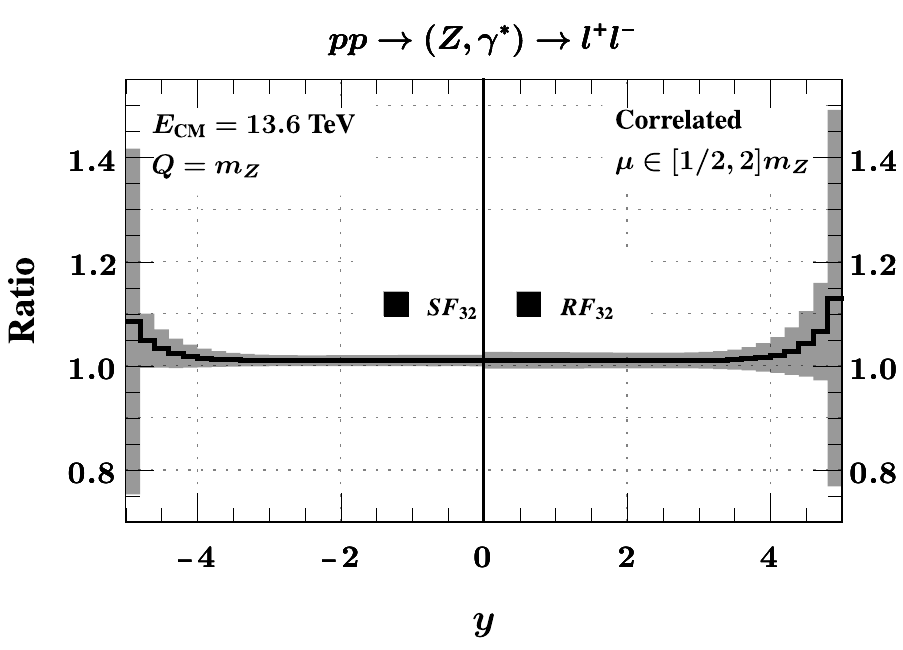}
\hspace{0.05cm}
\includegraphics[width=7.4cm,height=5.6cm]{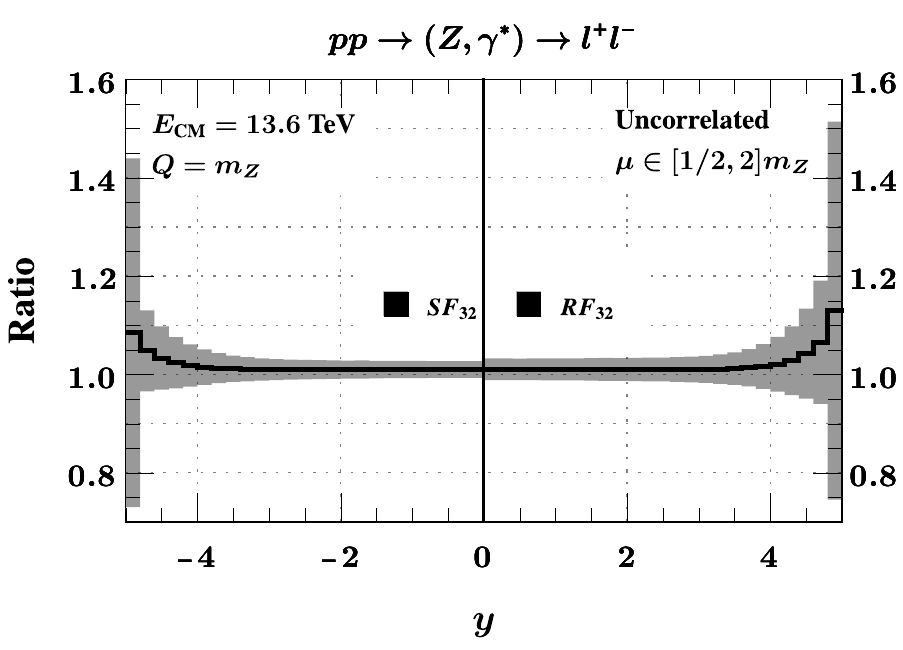}
}
\caption{The $SF$-factors and $RF$-factors as defined 
in \eq{eq:SF-RF-FACTORS} with seven-point scale 
uncertainties (left: correlated, right: uncorrelated)
included for $13.6$ TeV LHC. }
\label{fig:DY-SFRF-RATIO}
\end{figure}
We first revisit the case of neutral DY rapidity 
distribution. We set the invariant mass of the DY 
production as $Q=m_Z$. With this choice, the 
interference term vanishes, however, the 
virtual photon and $Z$ boson contributions remain.
The latter is dominant in this invariant mass region.
A detailed study has been already performed in 
\cite{Banerjee:2018vvb} up to NNLL matched to NNLO 
result. Here we estimate the new resummed effect in 
addition to the SV effects at third order in the 
double Mellin space. 
In \fig{fig:DY-RAPIDITY}, we compare 
the FO results to the matched resummed result up to 
the second order, as well as to the third order SV 
result.\footnote{Throughout this section and the 
following sections we use the 
notation(for convenience) N$i$LL to indicate the matched result 
where the resummed contribution has been matched 
to the corresponding FO result according to 
\eq{eq:master-matched}.} We observe a better 
perturbative convergence in the case of matched result.
We also present the ratios taken against the NNLO results 
in \fig{fig:DY-SFRF-RATIO}. 
For the third order study we use NNLO PDFs, but we 
use the strong coupling evolving it 
with the four-loop beta function in QCD.
On the fixed order side, we find $1\%$ corrections 
at the central rapidity region over the NNLO 
contribution while the seven-point scale 
uncertainties are around $1.5\%$ in the 
central region and increase to around $5\%$
at higher rapidities ($y=4$). 
After matching this third order SV results 
with the new N3LL results we find an increment of 
$0.5\%$  
over the previous NNLL order for a wide range of rapidity. 
The scale uncertainties in the resummed case 
are in general larger than those of the corresponding
fixed order. At NNLL they amount to around 
$2\%$ in the central region which is almost 
double that of the fixed order case.  
In addition to using the NNLO PDF, we use the 
available approximate N3LO PDFs \cite{McGowan:2022nag} 
from the \texttt{MSHT} 
group. It is recommended however to vary only the 
renormalization scale at the third order, since the 
effect of factorization scale variation is taken care 
of in the PDF fit. Using this approximate third 
order PDF  set we found the correction to be around 
$1.3\%$
in the central region 
while the scale uncertainty in the SV case becomes 
$1\%$ in the central region and around $2\%$ 
for higher rapidities. We find a similar 
behavior in the matched case.
This reduces the scale uncertainties 
dramatically in the higher rapidity region compared 
to using NNLO PDFs which is mostly 
due to the variation of the 
renormalization scale only. 
In the following sections, however, we refrain 
from using this approximate PDF set 
and use the NNLO one 
instead. This will also allow us to gauge the 
perturbative correction at the third order 
using the same PDFs
at NNLO and N3LO level. On the 
other hand we expect a similar behavior  as 
for the DY case
with approximate N3LO PDF for 
other processes studied here.

%%% Z-analysis%%%%%%%%%%%%%%%%%%%%%%%%%%%%%%%%%%%%%%%%%%%%%%%%%%%%
\subsection{$Z$ boson rapidity distribution}\label{sec:z-analysis}
%%%%%%%%%%%%%%%%%%%%%%%%%%%%%%%%%%%%%%%%%%%%%%%%%%%%%%%%%%%%%%%%%%
%% Z Rapidity
\begin{figure}[ht!]
        \centering{
\includegraphics[width=7.4cm,height=5.6cm]{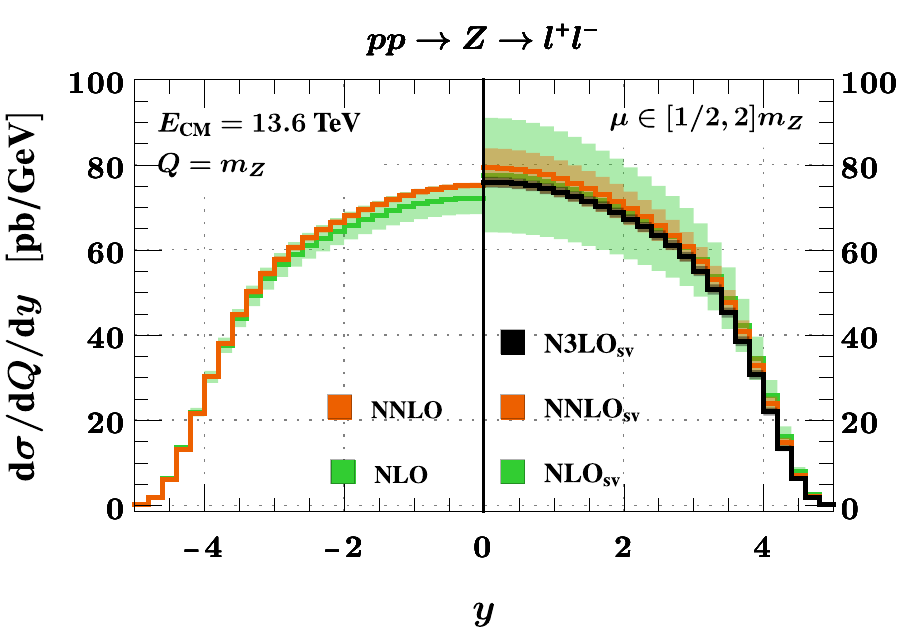}
\hspace{0.05cm}
\includegraphics[width=7.4cm,height=5.6cm]{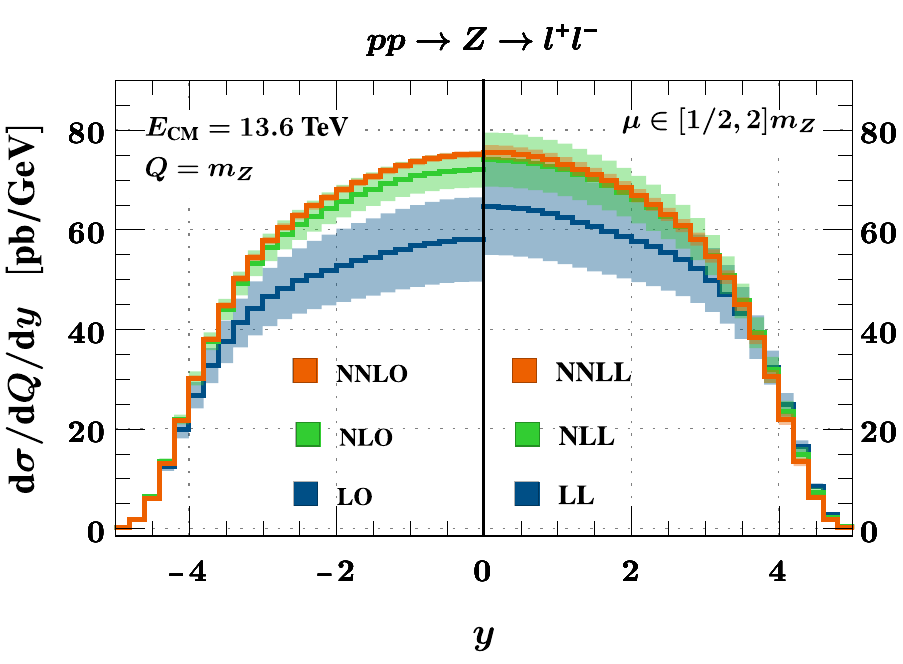}
        }
\caption{$Z$ boson rapidity distribution with 
scale uncertainty at $13.6$ TeV LHC. The left figure
compares fixed order to the SV contribution while 
the right figure compares it to matched calculations.}
\label{fig:Z-RAPIDITY}
\end{figure} 
The DY and $Z$ rapidities are comparable to each other 
particularly at the $Z$ resonance. 
To quantify, we find the photon contribution 
at the $Z$ resonance below $0.2\%$ in all rapidity 
regions at NNLO and also at NNLL. 
To explicitly 
study the $Z$ boson rapidity, we completely remove 
this small contribution coming from the virtual photon 
at the resonance. 
In \fig{fig:Z-RAPIDITY} (left), we present the 
comparison of the SV approximation to the fixed order.
We find, in general, higher cross sections at all 
rapidities compared to the fixed order at NLO and NNLO.
The scale uncertainty is however much higher and 
amounts to $5-8\%$ at NNLO compared to 
$1-3\%$ at the fixed order. 
In the right figure we present the FO and matched 
distributions up to the second order. As before 
we observe a better perturbative convergence 
in the case of the matched result while the 
scale uncertainty does not improve compared to 
the fixed order case, an observation which is 
also seen in \cite{Banerjee:2018mkm} in the case of 
DY rapidities. Going from LL to NNLL, we find a 
$14\%$ correction at NLL and $2\%$ at NNLL. 

%% Z SF,RF factors
\begin{figure}[ht!]
        \centering{
\includegraphics[width=7.4cm,height=5.6cm]{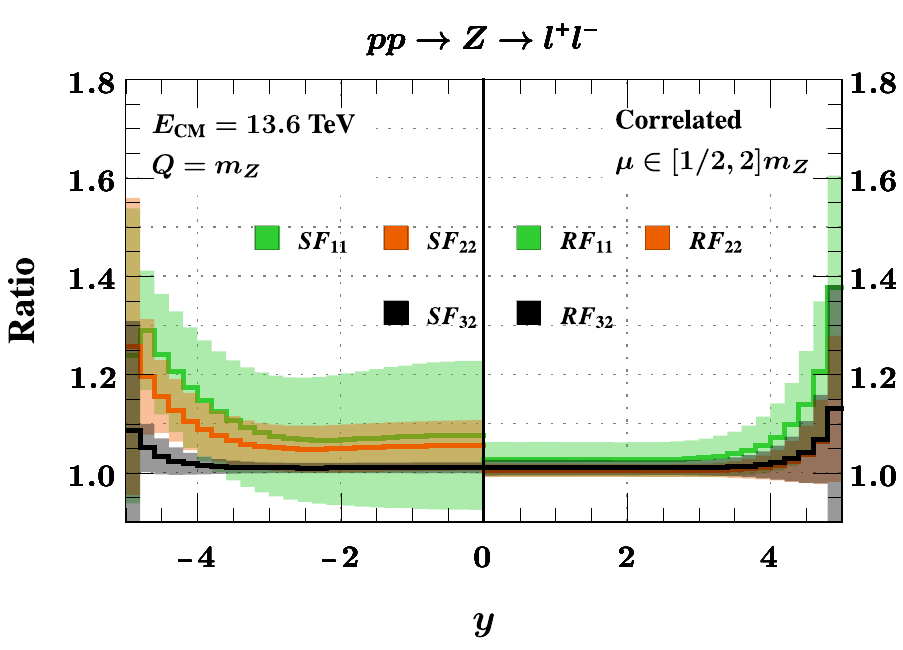}
\hspace{0.05cm}
\includegraphics[width=7.4cm,height=5.6cm]{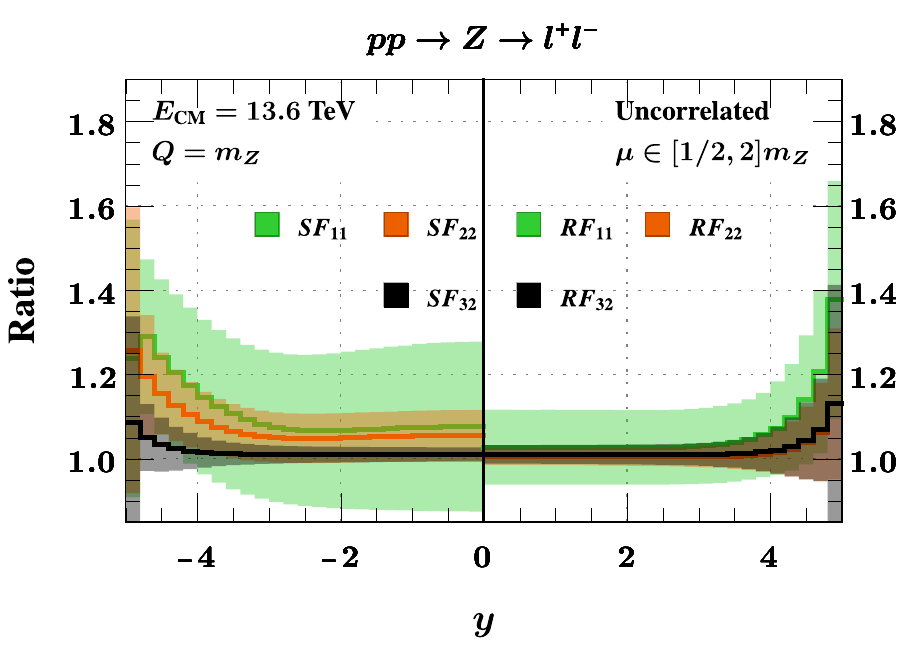}
}
\caption{The $SF$-factors and $RF$-factors as defined 
in \eq{eq:SF-RF-FACTORS} with seven-point scale 
uncertainties (left: correlated, right: uncorrelated)
included for $13.6$ TeV LHC. }
\label{fig:Z-SFRF-RATIO}
\end{figure}
In order to compare the SV approximation, we have 
studied the $SF$ ratios against the fixed order 
in \fig{fig:Z-SFRF-RATIO}. It is evident from the 
central predictions that SV leads to higher 
cross section at all orders. At NLO, 
we observe that the SV correction amounts to $5\%$ of 
the fixed order cross section over the central 
rapidity region
with $14\%$ correlated and $18\%$ uncorrelated 
error. The correlated and uncorrelated errors 
are larger in the higher rapidity region, amounting 
to $10\%$ and $15\%$ respectively at $y=4$.
These are reduced to $4-6\%$ at NNLO for the correlated 
case,
and to $5-8\%$ in the case of uncorrelated error.
For the $SF_{32}$ ratio, the uncertainties are 
$1-3\%$ in the correlated case and around $2-5\%$ 
in the uncorrelated case.
The scale uncertainties on the resummed case, 
however, are reduced to below $3-7\%$ in the 
correlated case, and to $8-10\%$ in the case of 
uncorrelated errors already at NLL level ($RF_{11}$).
These are further improved to $1-4\%$ (correlated)
and $2-6\%$ (uncorrelated) at NNLL.
At N3LLsv level ($RF_{32}$), both correlated and 
uncorrelated uncertainties do not improve on the 
previous order and remain at a similar level 
as seen from the right panels of 
\fig{fig:Z-SFRF-RATIO}. 
%
%% Z KR factors
\begin{figure}[ht!]
        \centering{
\includegraphics[width=7.4cm,height=5.6cm]{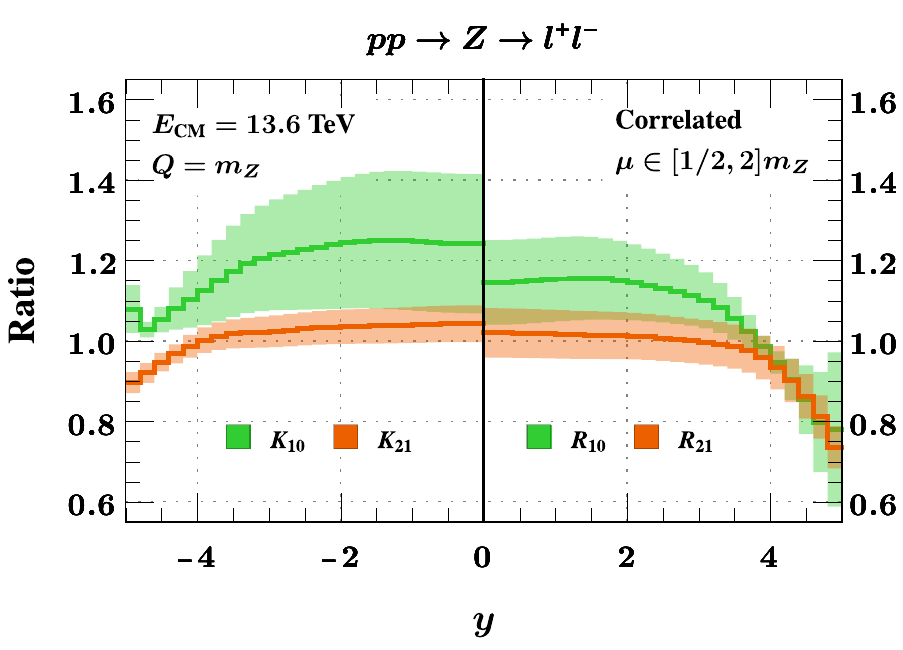}
\hspace{0.05cm}
\includegraphics[width=7.4cm,height=5.6cm]{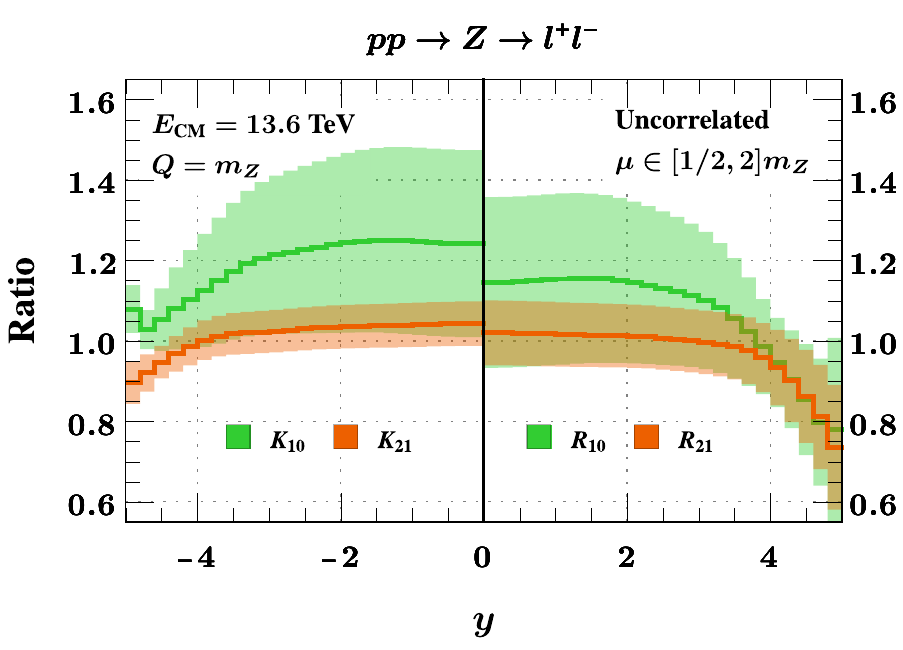}
}
\caption{The $K$-factors and $R$-factors as defined 
in \eq{eq:K-R-FACTORS} with seven-point scale 
uncertainties (left: correlated, right: uncorrelated)
included for $13.6$ TeV LHC. }
\label{fig:Z-KR-RATIO}
\end{figure}
In order to estimate the higher order contribution 
from the FO and resummed cases, we study 
ratios of successive orders and present them 
in \fig{fig:Z-KR-RATIO}.
We show both the correlated and uncorrelated
errors for these ratios. The differences can be seen 
in the higher rapidity region in terms of the 
$K$-factor and the $R$-factor. The fixed order 
$K$ factors do not overlap 
in the higher rapidity region and 
barely overlap in the central rapidity region for 
both, the correlated and uncorrelated cases. The $R$ factors 
on the other hand completely fall within the lower order 
scale uncertainty in the uncorrelated case.
The quality of the perturbative convergence is clearly
visible.
Even the scale uncertainty is reduced at the NLL 
level compared to the FO counterpart, although the same 
does not hold at the next order.
Here we also point out that the remarkably 
improved scale uncertainty at NNLO is a consequence 
of large cancellations of contributions from 
different channels at different scales, a feature 
for DY type production processes and also seen in the 
total inclusive cross section.

%%% Z-Intrinsic PDF Uncertainty
\begin{figure}
        \centering{
\includegraphics[width=7.4cm,height=5.6cm]{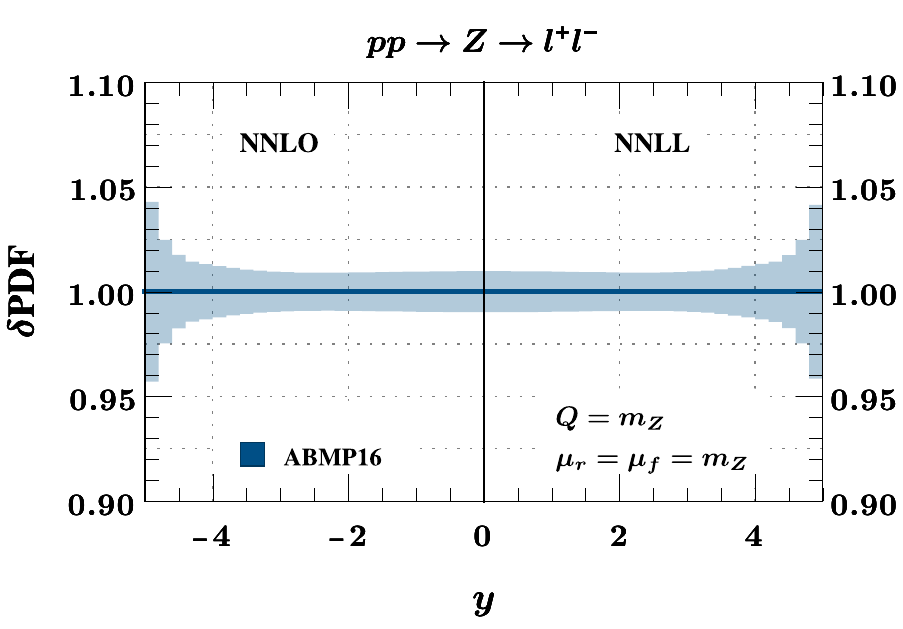}
\hspace{0.05cm}
\includegraphics[width=7.4cm,height=5.6cm]{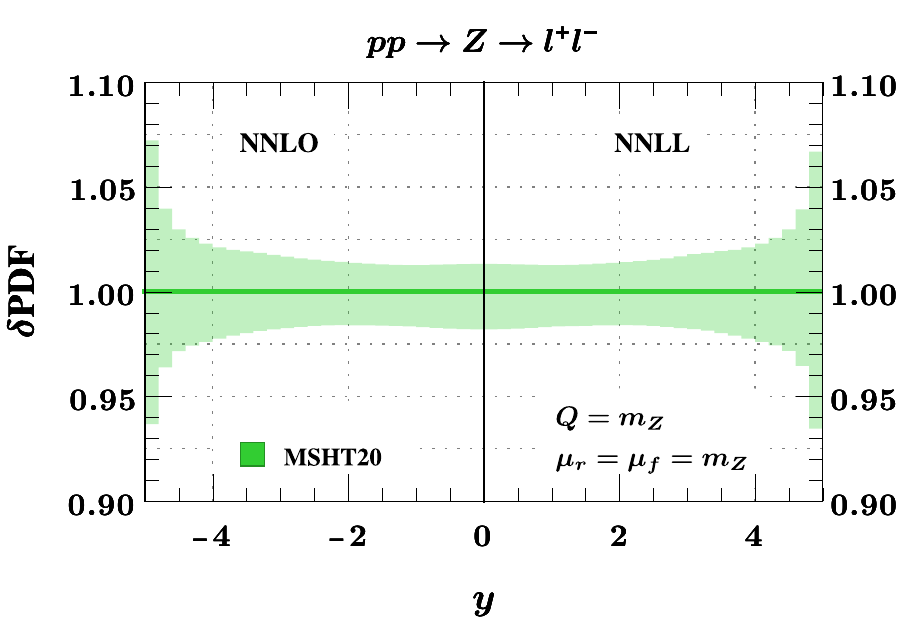}
\includegraphics[width=7.4cm,height=5.6cm]{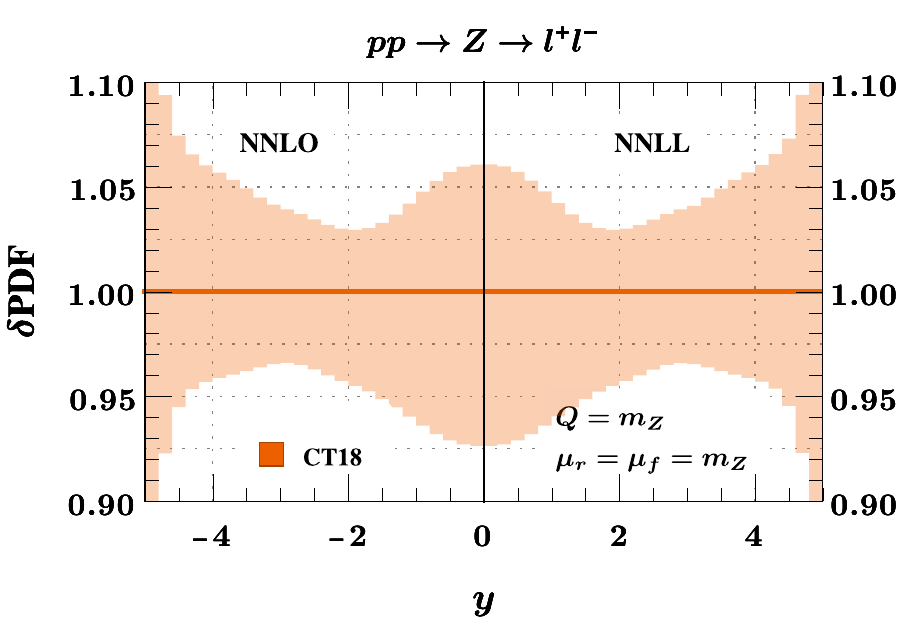}
\hspace{0.05cm}
\includegraphics[width=7.4cm,height=5.6cm]{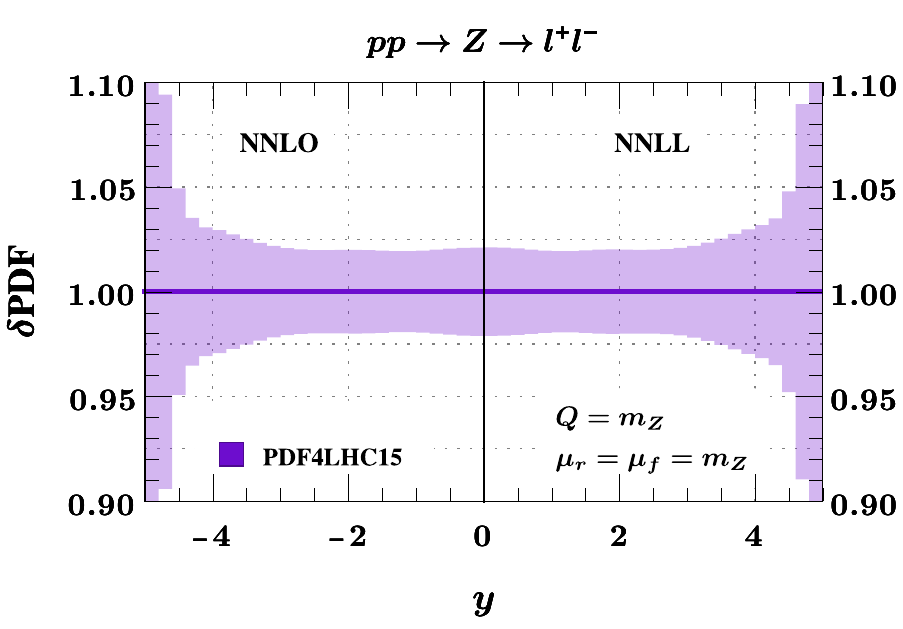}
}
\caption{Intrinsic PDF uncertainty defined 
in \eq{eq:PDF-ERROR} for $Z$ boson 
rapidity for $13.6$ TeV LHC at second order. Left panel: the fixed order 
case. Right panel: matched case.}
\label{fig:Z-PDF-UNCERTAINTY} 
\end{figure} 

So far we have used the {\tt MSHT} PDF sets as our 
default parton distributions. However, it is also 
instructive to have an estimation of using different 
PDF sets which use different methods in their 
PDF fits. Ideally the results from different PDF sets 
should be consistent within the intrinsic PDF errors.
For this study we chose a handful of standard sets as 
example, however we stress that a similar analysis 
could be done with other available PDFs as well.
For this study, we have chosen apart from 
the {\tt MSHT20}, the {\tt ABMP16},
{\tt CT18}, and {\tt PDF4LHC30} sets. 
In terms of the 
PDF uncertainty, we observe in the central rapidity 
region a smaller PDF error.
On the other hand we observe a different
behavior for the {\tt CT18} set. In fact this set provides 
larger uncertainties throughout the whole rapidity 
region including the central rapidity region.
The {\tt ABMP} PDF set provides the lowest errors among these 
choices which amount to $1\%$ in the central region and 
$2\%$ to the higher rapidity region. 
The {\tt MSHT} and {\tt PDF4LHC30} sets provide 
$1.3\%$ and $2\%$ uncertainties respectively in the central rapidity
region and $7\%$ and $10\%$  respectively in the higher rapidities. 
For \texttt{CT18}, the PDF uncertainty is around 
$6\%$ at the central region which reduces to $3\%$ 
in the intermediate region $y=3$ and then again 
increases in the higher rapidities.
The resummed results show slightly better 
PDF uncertainties for all the cases in particular 
in the higher rapidities, although the uncertainty 
remains similar to the FO case in the central region.

\begin{figure}
        \centering{
\includegraphics[width=7.4cm,height=5.6cm]{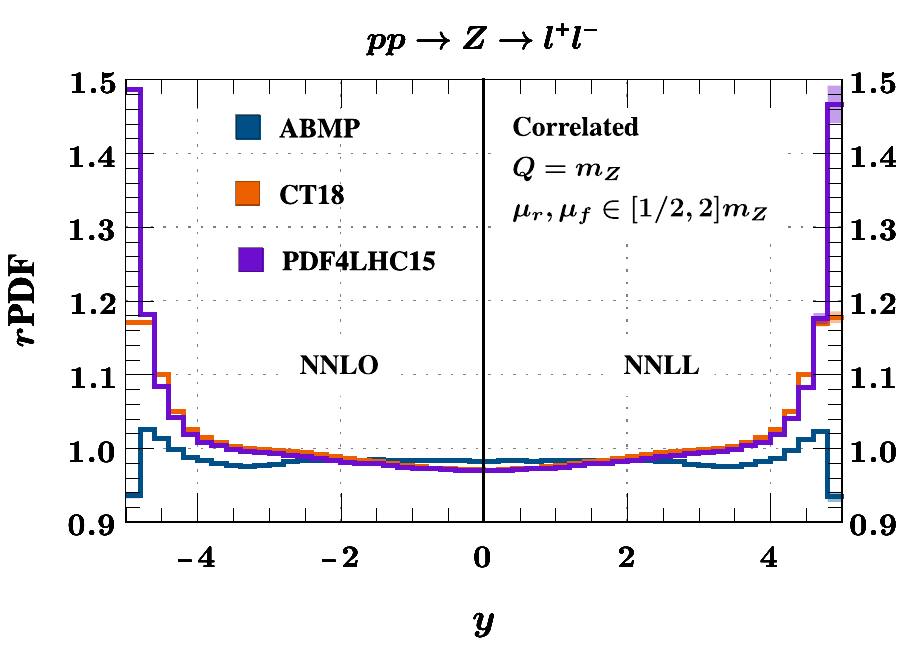}
\hspace{0.05cm}
\includegraphics[width=7.4cm,height=5.6cm]{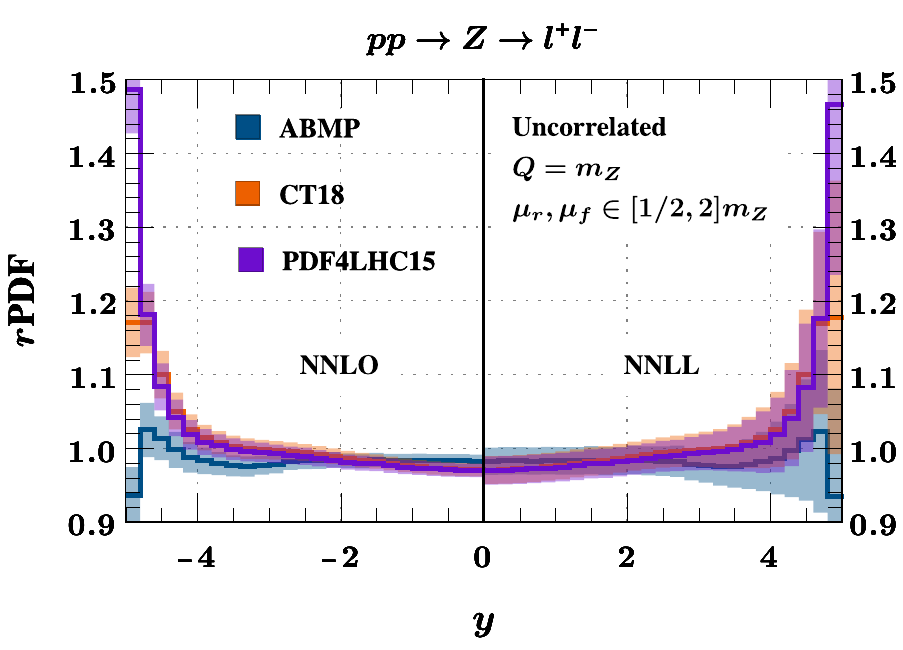}
}
\caption{Ratio for central scale predictions taken 
with MSHT PDF sets as defined in \eq{eq:rPDF}
with scale uncertainties for $Z$ boson rapidities 
at FO and for resummed cases.
(Left: correlated, Right: uncorrelated)
included for $13.6$ TeV LHC. }
\label{fig:Z-PDF-RATIO}
\end{figure}

We also observe that the central predictions are different
for different sets and hence we also compared them 
against our default choice {\tt MSHT20} and plotted them 
for 
both correlated and uncorrelated errors from seven-point 
uncertainties in \fig{fig:Z-PDF-RATIO}. 
The deviation of the central scale prediction could be 
as large as $10\%$ except in the last two bins where 
it becomes $50\%$ for \texttt{PDF4LHC} and $20\%$
for \texttt{CT18}. For most of the rapidity 
region, they are consistent within their PDF 
uncertainties. The uncorrelated uncertainty
becomes $0.9\%$ in the central region while it becomes 
around $4\%$ at higher rapidities for all the ratios 
at NNLO. The resummed scale uncertainty at NNLL is
almost double to that of the NNLO in the central region
whereas in the higher rapidities it becomes as large 
as $10\%$ for all the ratios considered here.
The uncorrelated errors for 
other sets agree within the errors over a large rapidity 
range except the higher rapidity region where they 
are not covered by the scale uncertainties.
These studies suggest that even in the resummed case,
these PDF fits are very different both 
qualitatively and quantitatively
in particular in the higher rapidity region. 

%% Z PDF RATIO

%%% W-analysis%%%%%%%%%%%%%%%%%%%%%%%%%%%%%%%%%%%%%%%%%%%%%%%%%%%%
\subsection{$W$ boson rapidity distribution}\label{sec:w-analysis}
%%%%%%%%%%%%%%%%%%%%%%%%%%%%%%%%%%%%%%%%%%%%%%%%%%%%%%%%%%%%%%%%%%
% {\bf $W^-$ Analysis:}
%% W- RAPIDITY
\begin{figure}[ht!]
        \centering{
\includegraphics[width=7.4cm,height=5.6cm]{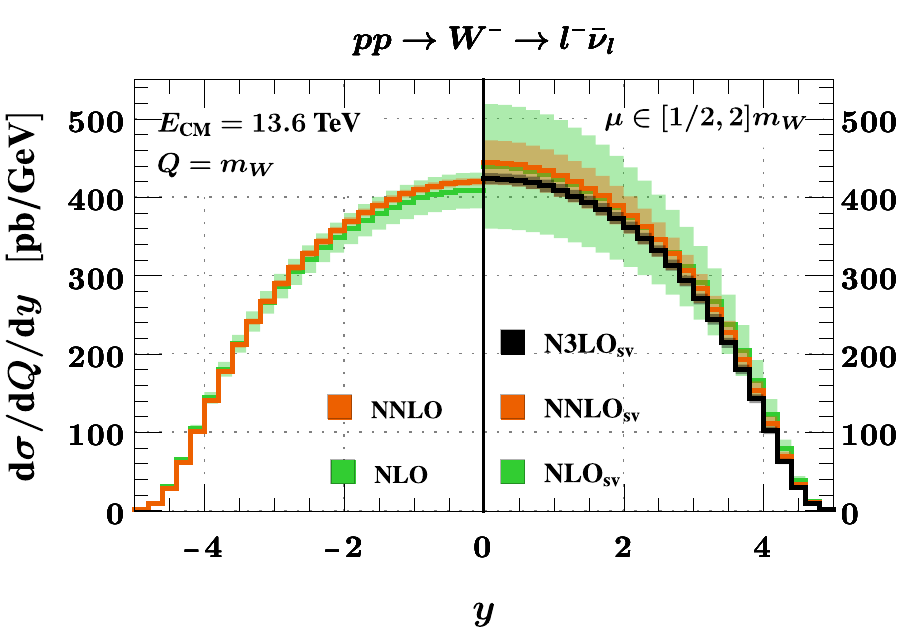}
\hspace{0.05cm}
\includegraphics[width=7.4cm,height=5.6cm]{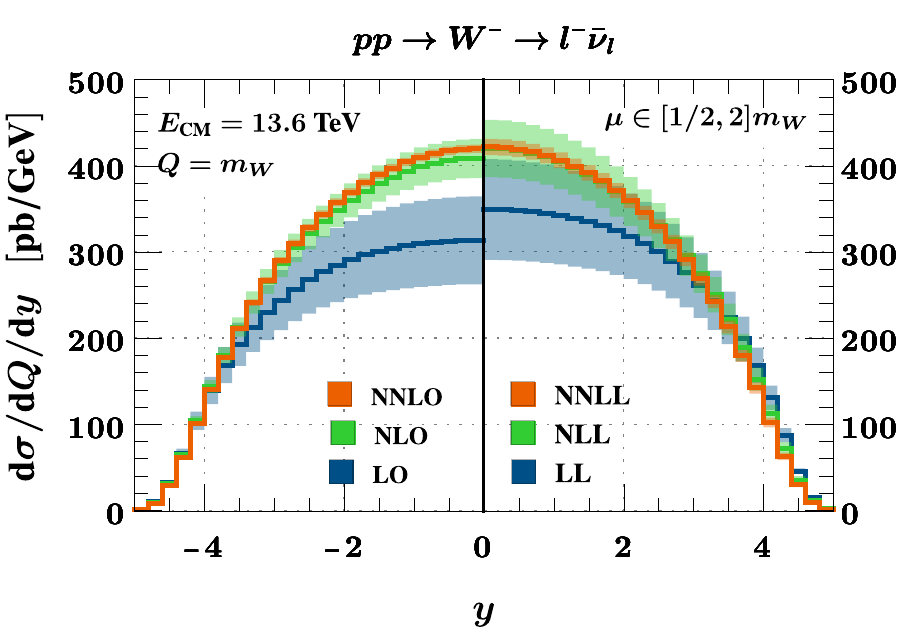}
        }
\caption{$W^{-}$ boson rapidity distribution with 
scale uncertainty at $13.6$ TeV LHC. 
The left panel is for the fixed order 
and the right panel is for SV/resummed calculations.}
\label{fig:Wm-RAPIDITY}
\end{figure} 
%% W- SF,RF factors
\begin{figure}[ht!]
        \centering{
\includegraphics[width=7.4cm,height=5.6cm]{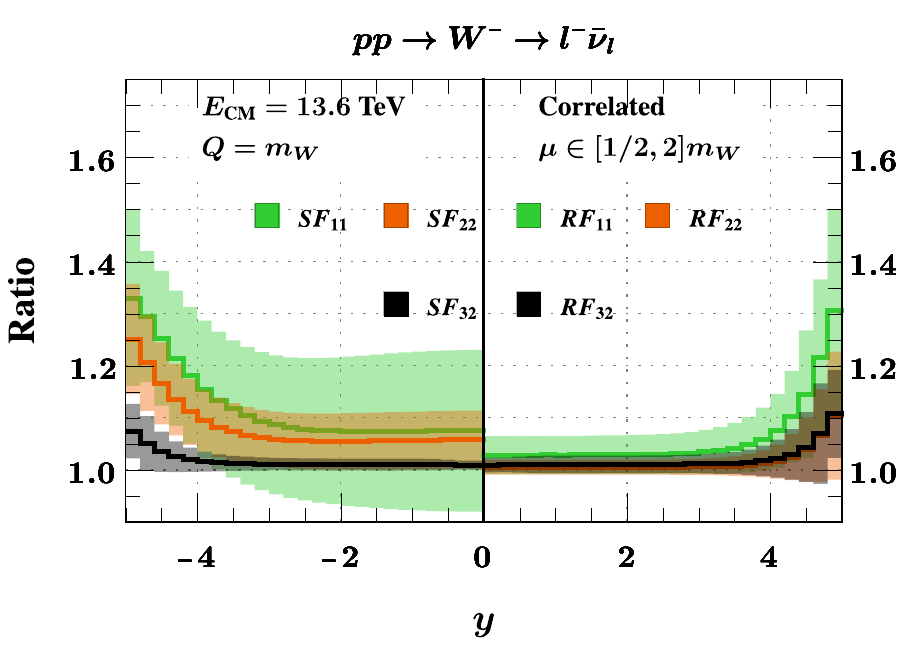}
\hspace{0.05cm}
\includegraphics[width=7.4cm,height=5.6cm]{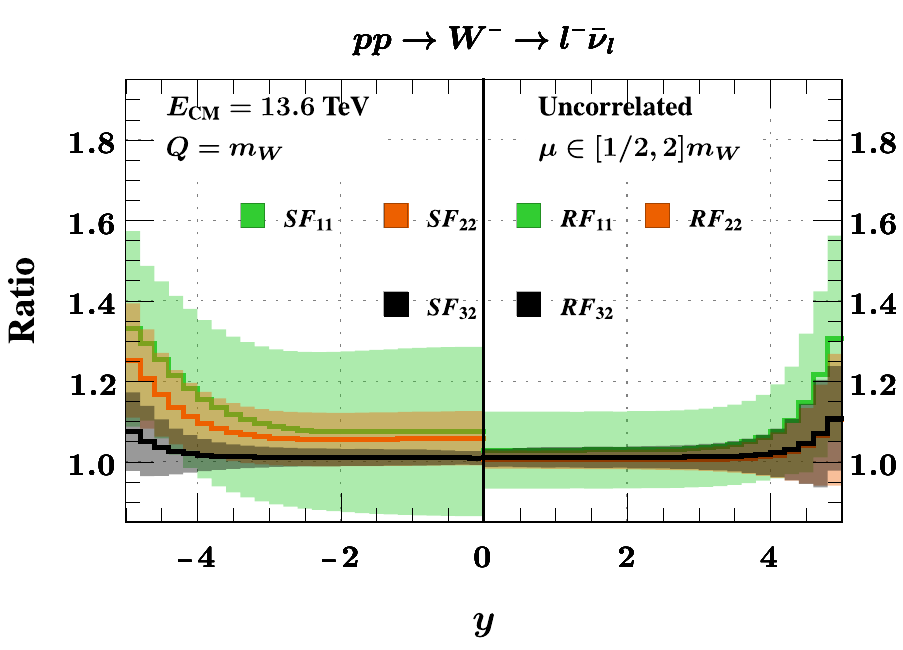}
}
\caption{The $SF$-factors and $RF$-factors as defined 
in \eq{eq:SF-RF-FACTORS} with seven-point scale 
uncertainties (left: correlated, right: uncorrelated)
included for $13.6$ TeV LHC. }
\label{fig:Wm-SFRF-RATIO}
\end{figure}

\noindent
{\bf $W^-$ boson rapidity:\\}
We now turn to the $W^\pm$ rapidities. 
In \fig{fig:Wm-RAPIDITY}, we present the 
comparison between the fixed order results and 
the SV results for the $W^{-}$ rapidity. 
We observe higher cross sections 
at all rapidities in the case of the SV. 
The scale uncertainty at N3LOsv becomes $1.6\%$ 
compared to $1.2\%$ at fixed order at NNLO
at $y=0$.  
In the right plot of \fig{fig:Wm-RAPIDITY},
we present the new matched results up to 
NNLL accuracy. We observe a better perturbative 
convergence in the resummed case where 
NLL receives a correction of $20\%$ over LL with a
scale uncertainty of $8\%$. The corresponding fixed
order NLO gets a correction of $30\%$ over the LO.
The rapidity distribution stabilizes at NNLL level
where it gets contributions below $0.5\%$ with a scale 
uncertainty of $2.3\%$ in the central rapidity region.
At the higher rapidity region, 
the cross section decreases compared to the previous 
order by $8\%$ at NNLO and $16\%$ at NNLL. 
%% W- KR factors
\begin{figure}[ht!]
        \centering{
\includegraphics[width=7.4cm,height=5.6cm]{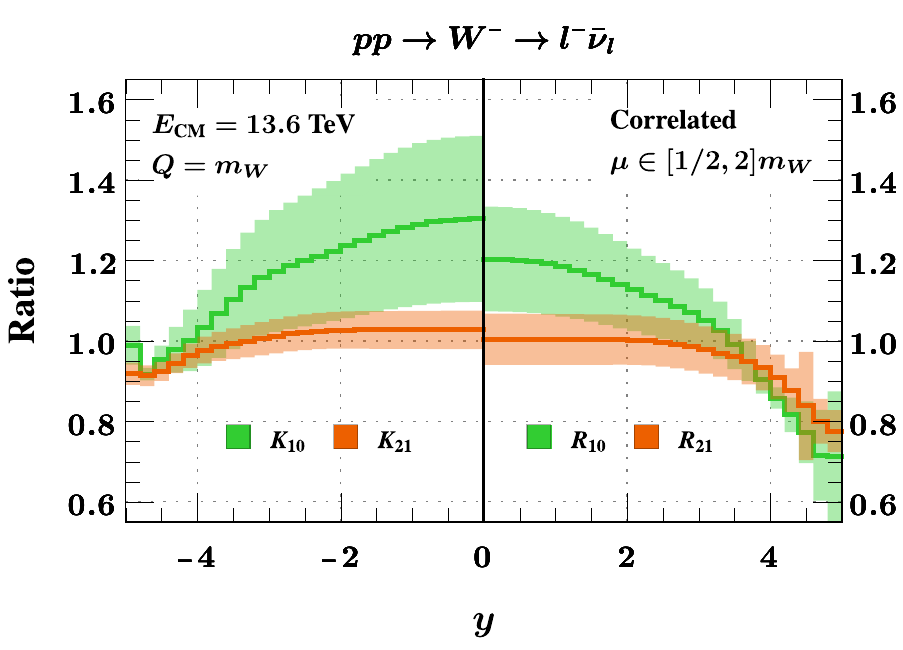}
\hspace{0.05cm}
\includegraphics[width=7.4cm,height=5.6cm]{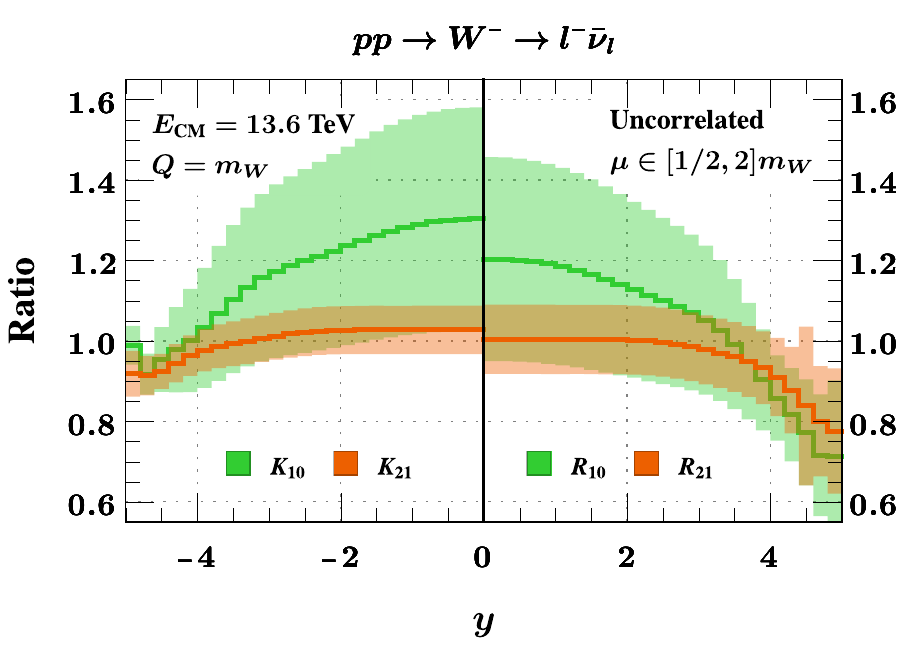}
}
\caption{The $K$-factors and $R$-factors as defined 
in \eq{eq:K-R-FACTORS} with seven-point scale 
uncertainties (left: correlated, right: uncorrelated)
included for $13.6$ TeV LHC. }
\label{fig:Wm-KR-RATIO}
\end{figure}
%% W- PDF SCALE ratio
\begin{figure}
        \centering{
\includegraphics[width=7.4cm,height=5.6cm]{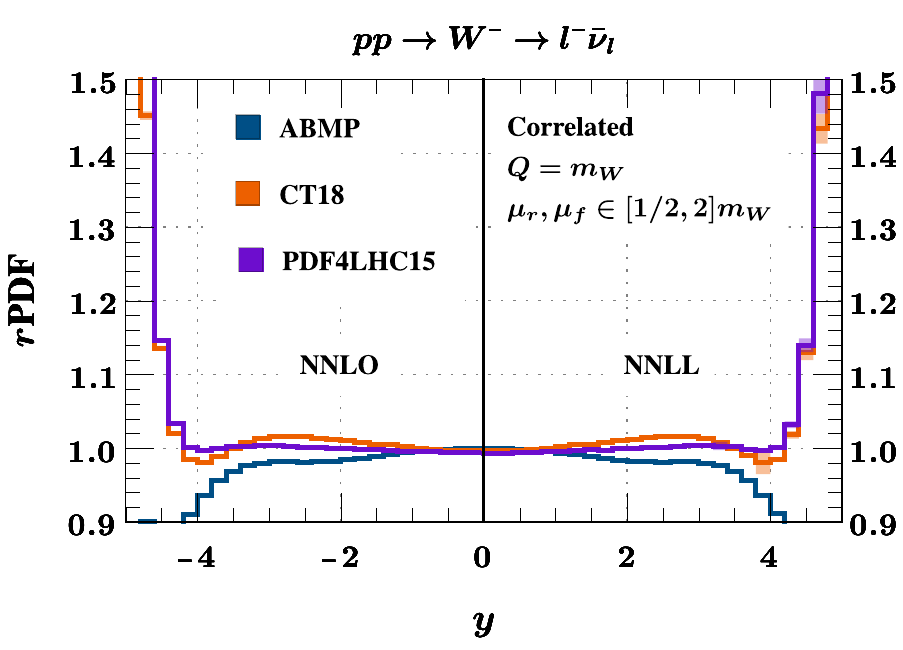}
\hspace{0.05cm}
\includegraphics[width=7.4cm,height=5.6cm]{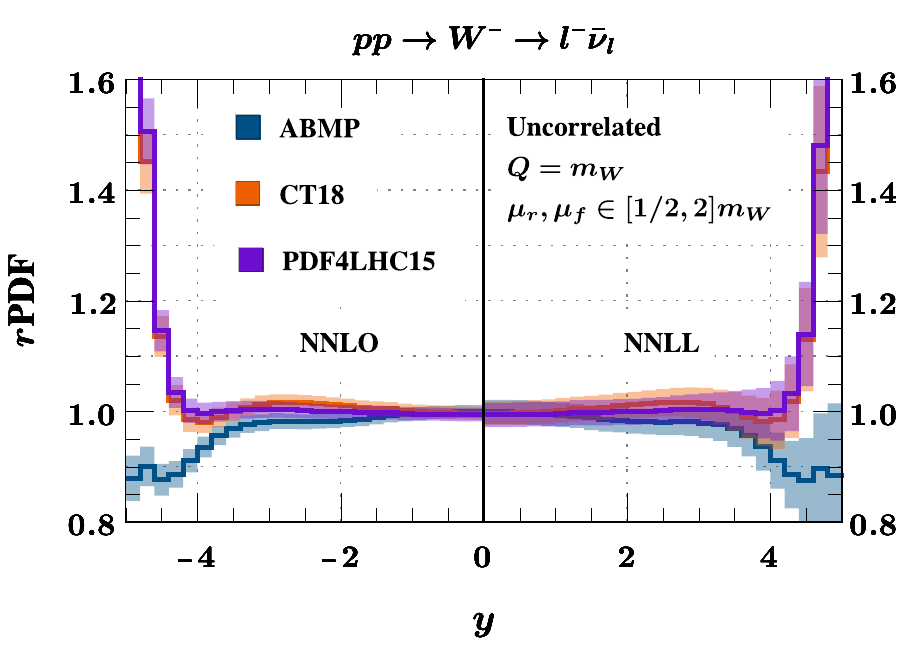}
}
\caption{Ratio for central scale predictions taken with 
MSHT PDF sets as defined in \eq{eq:rPDF}
with scale uncertainties for $Z$ boson rapidities 
at FO and for resummed cases.
(left: correlated, right: uncorrelated)
included for $13.6$ TeV LHC. }
\label{fig:Wm-PDF-RATIO}
\end{figure}

We further plotted the SV and the resummed 
contributions 
against the corresponding FO at NLO and NNLO in 
\fig{fig:Wm-SFRF-RATIO} along with the correlated 
and uncorrelated uncertainties. At the third order, 
we take the ratio of SV and matched result against the 
NNLO contribution. Note that  we match 
the resummed case at the third order with the partial 
SV result at the same order as before.
The SV contribution at N3LO 
is below $1\%$ in the central 
region and around $2-7\%$ beyond $y=4$.
At the matched level, the numbers are similar to the 
NNLL contribution i.e. around $0.5\%$ correction
except at the higher rapidities, where the N3LLsv
matched correction is still of the same order 
compared to large NNLL correction as stated before.
At the third order the correlated and uncorrelated 
uncertainties for $SF_{32}$ become $0.9\%$ and $1.7\%$
in the central region and around $5\%$ and $9\%$ 
respectively at higher rapidities.
For $RF_{32}$ the uncertainties are 
higher and amount to $1.4\%$ and $2.2\%$ in the 
central region and $7\%$  and $10\%$ at 
higher rapidities.

%%% W-Intrinsic PDF Uncertainty
\begin{figure}
        \centering{
\includegraphics[width=7.4cm,height=5.6cm]{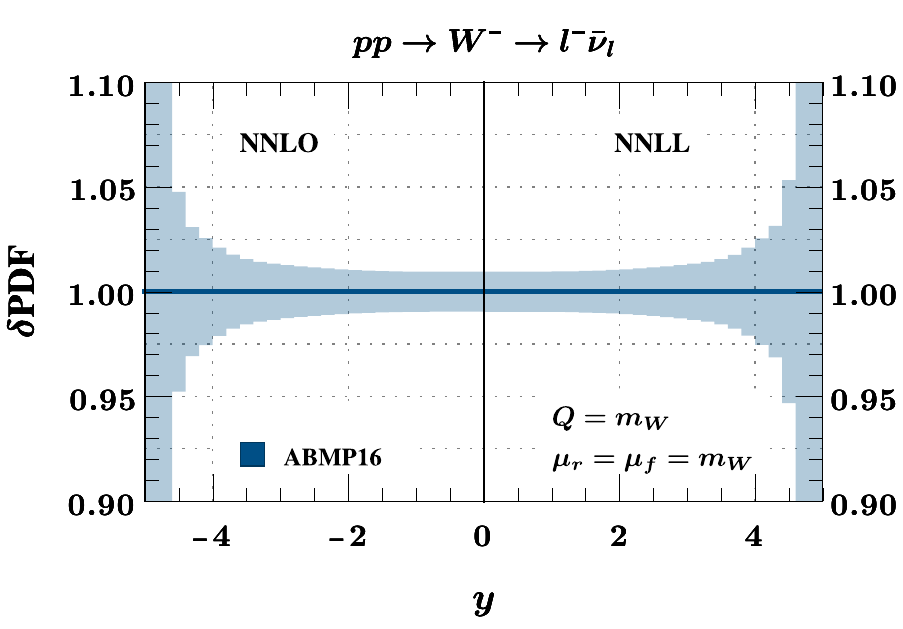}
\hspace{0.05cm}
\includegraphics[width=7.4cm,height=5.6cm]{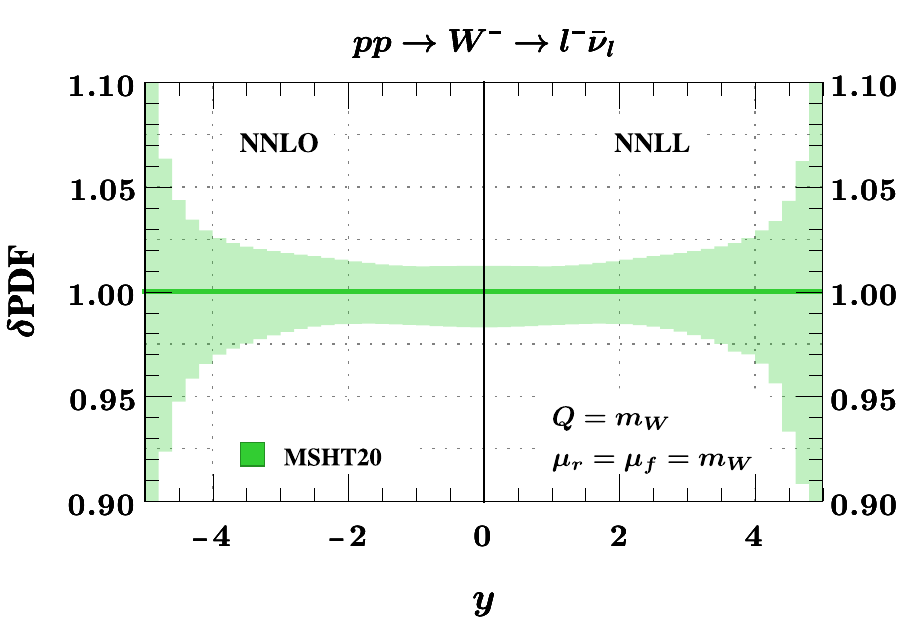}
\includegraphics[width=7.4cm,height=5.6cm]{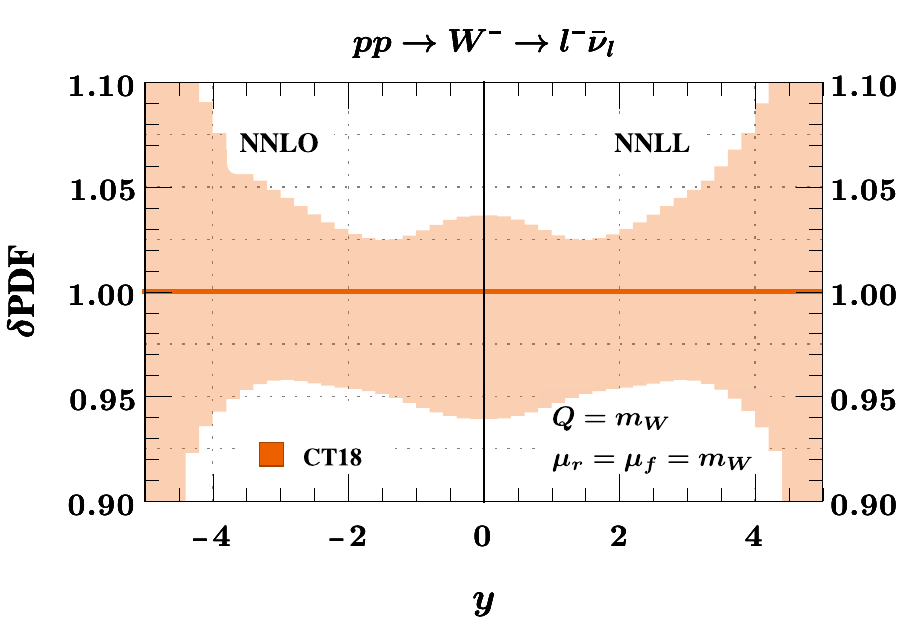}
\hspace{0.05cm}
\includegraphics[width=7.4cm,height=5.6cm]{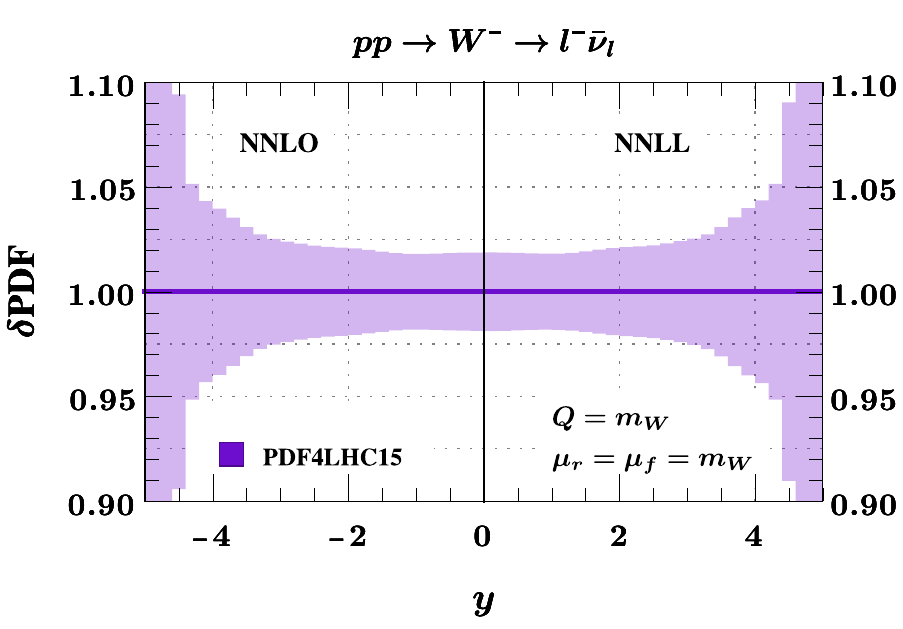}
}
\caption{Intrinsic PDF uncertainty defined 
in \eq{eq:PDF-ERROR} for $W^-$ boson 
rapidity for $13.6$ TeV LHC at second order. Left panel: the fixed order 
case. Right panel: matched case.} 
\label{fig:Wm-PDF-UNCERTAINTY}
\end{figure} 
In \fig{fig:Wm-KR-RATIO}, we further plotted 
the $K$ and $R$ factors to access the correction
at the FO and at the resummed orders respectively.
While the NLO $K$ factor defined against LO shows 
a correction as large as $30\%$, the NNLO 
$K$ factor defined with respect to NLO stabilizes
particularly in the central rapidity region. The matched 
distribution on the other hand gets around $20\%$
correction compared to LL which further reduces to 
$0.5\%$ at NNLL. The correlated and uncorrelated 
uncertainties for the $K_{21}$ factor 
tend to be around $4.5\%$ and $5.5\%$ 
respectively at NNLO in most of the rapidity region.
For the $R_{21}$ factor, the corresponding
uncertainties are $6\%$ and $8.5\%$ respectively.

We also studied the effect of different choices of 
PDFs and present it in \fig{fig:Wm-PDF-RATIO}.
We observe that within the  uncorrelated error, 
all the PDFs agree in the central rapidity region, whereas 
they start to deviate beyond $|y|=3.5$. 
Furthermore, we investigated the intrinsic uncertainty 
and found that except for the \texttt{CT18} PDF set,
all of them give less than $2\%$ PDF uncertainty 
in the central region. For \texttt{CT18} the 
uncertainty is around $3-4\%$ in the central region.
This is due to $90\%$ c.l.\ error in \texttt{CT18} 
compared to $68\%$ c.l.\ error in other PDFs.\footnote{
Thanks to S.\ Moch for pointing this out.}
The error reduces by $1.6$ times 
\cite{Accardi:2016ndt} when it is reduced 
to $68\%$ c.l.\ which is consistent with other PDFs.
In the high rapidity region, all the PDFs give error
more than $5\%$ as seen in \fig{fig:Wm-PDF-UNCERTAINTY}.\\

\noindent
{\bf $W^+$ boson rapidity:\\}
%%%%%%%%%%%%%%%%%%%%%%%%%%%%%%%%%%%%%%%%%%%%%%%%%%%%%%
%%%W+ Analysis
%%%%%%%%%%%%%%%%%%%%%%%%%%%%%%%%%%%%%%%%%%%%%%%%%%%%%%
%% W+ RAPIDITY
\begin{figure}[ht!]
        \centering{
\includegraphics[width=7.4cm,height=5.6cm]{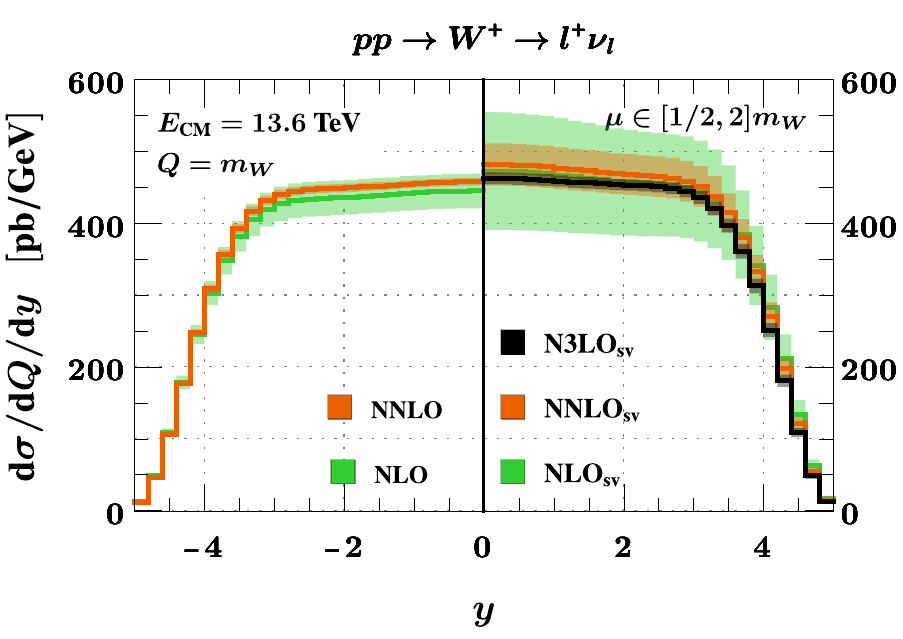}
\hspace{0.05cm}
\includegraphics[width=7.4cm,height=5.6cm]{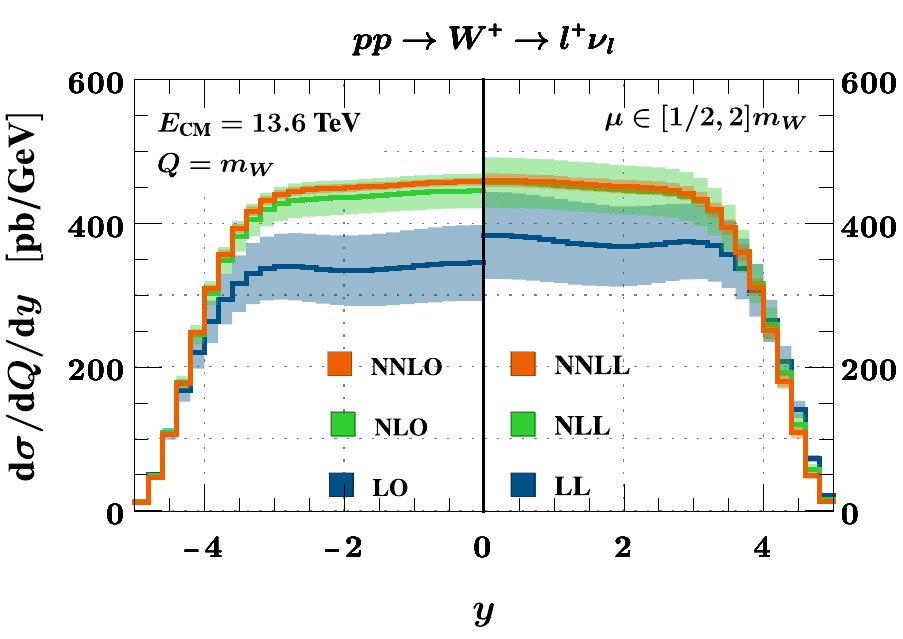}
        }
\caption{Same as \fig{fig:Wm-RAPIDITY} but for 
$W^+$ rapidity.}
\label{fig:Wp-RAPIDITY}
\end{figure} 
%% W+ KR factors
\begin{figure}[ht!]
        \centering{
\includegraphics[width=7.4cm,height=5.6cm]{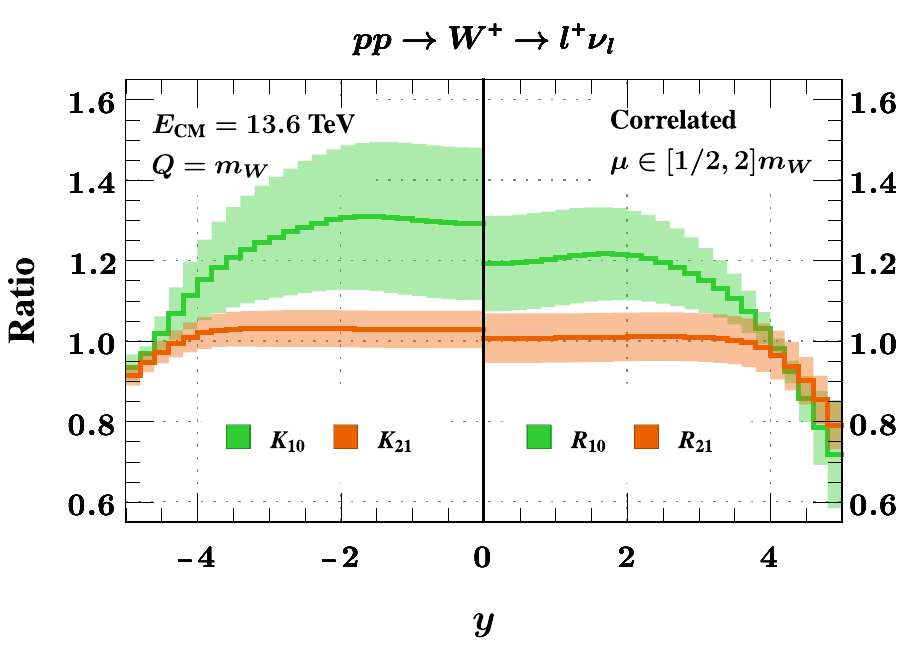}
\hspace{0.05cm}
\includegraphics[width=7.4cm,height=5.6cm]{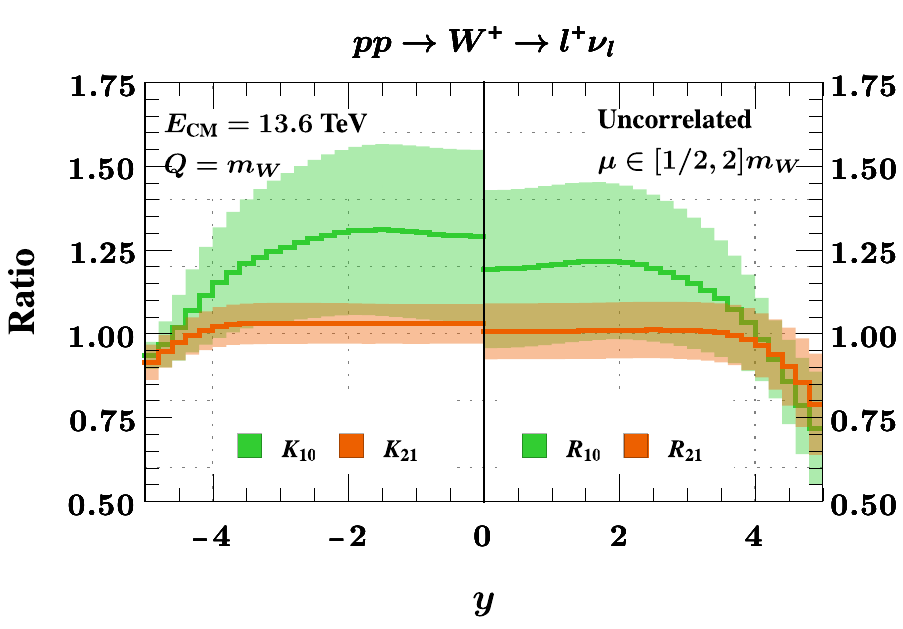}
}
\caption{Same as \fig{fig:Wm-KR-RATIO} but for 
$W^+$ rapidity.}
\label{fig:Wp-KR-RATIO}
\end{figure}
%% W+ SF,RF factors
\begin{figure}[ht!]
        \centering{
\includegraphics[width=7.4cm,height=5.6cm]{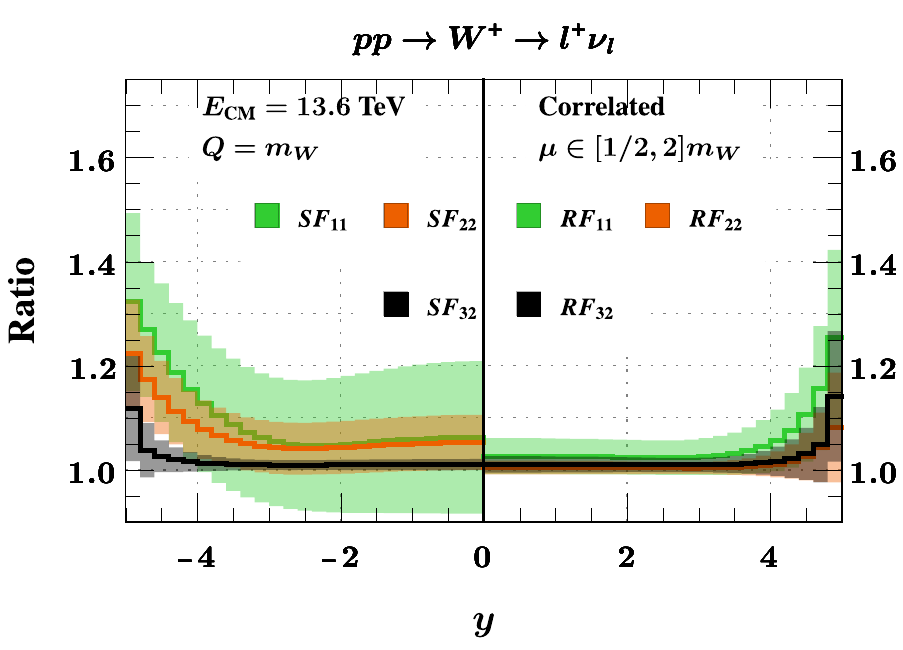}
\hspace{0.05cm}
\includegraphics[width=7.4cm,height=5.6cm]{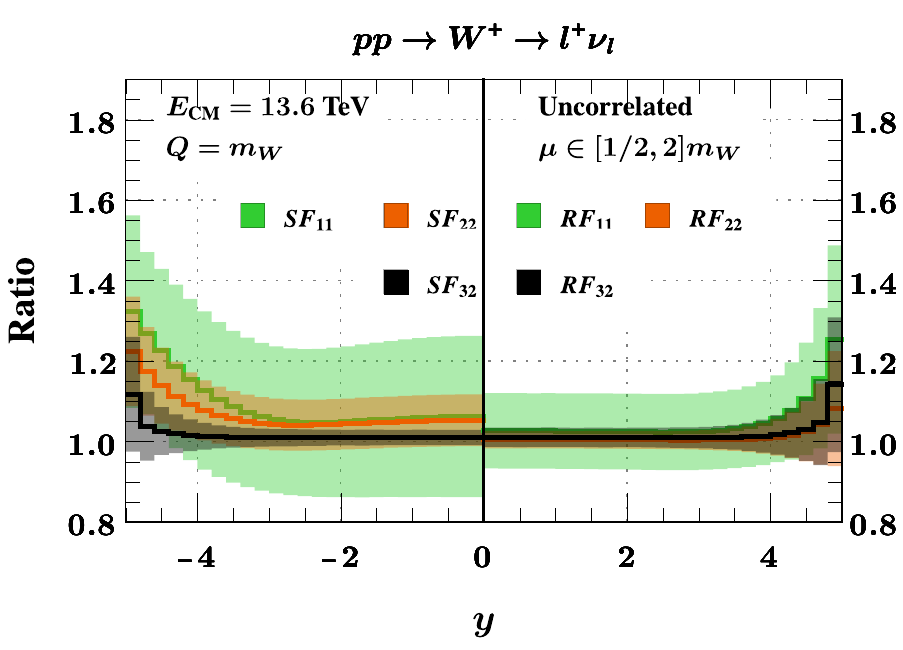}
}
\caption{Same as \fig{fig:Wm-SFRF-RATIO} 
but for $W^+$ rapidity.}
\label{fig:Wp-SFRF-RATIO}
\end{figure}
We now study the $W^+$ boson rapidity and in 
\fig{fig:Wp-RAPIDITY} we compare the 
FO distributions against SV and matched predictions.
While in the FO case, it gets a contribution
of around $3\%$ at NNLO compared to the previous 
order, in the resummed case the correction only amounts
to around $0.7\%$ in the central rapidity region 
showing the stability of the resummed prediction.
On the other hand, in the higher rapidity region 
the picture changes as the cross section itself 
is lower at NNLO compared to NLO and at NNLL it further
reduces amounting to $-10\%$ correction at $y=4.4$
compared to $-4\%$ at NNLO. This is also evident 
from the ratio plots in \fig{fig:Wp-KR-RATIO}.
The third order SV correction amounts to about 
$1\%$ increment for most of the rapidity region 
except in the higher rapidities where 
the correction could reach to about $3\%$ at $y=4.4$
as seen in \fig{fig:Wp-SFRF-RATIO}. 
Although their ($W^\pm$) cross sections are different, 
the behavior of the
$K$-factor or the $R$-factor are very similar 
as they receive similar QCD corrections, the only 
difference being their fluxes.

\begin{figure}
        \centering{
\includegraphics[width=7.4cm,height=5.6cm]{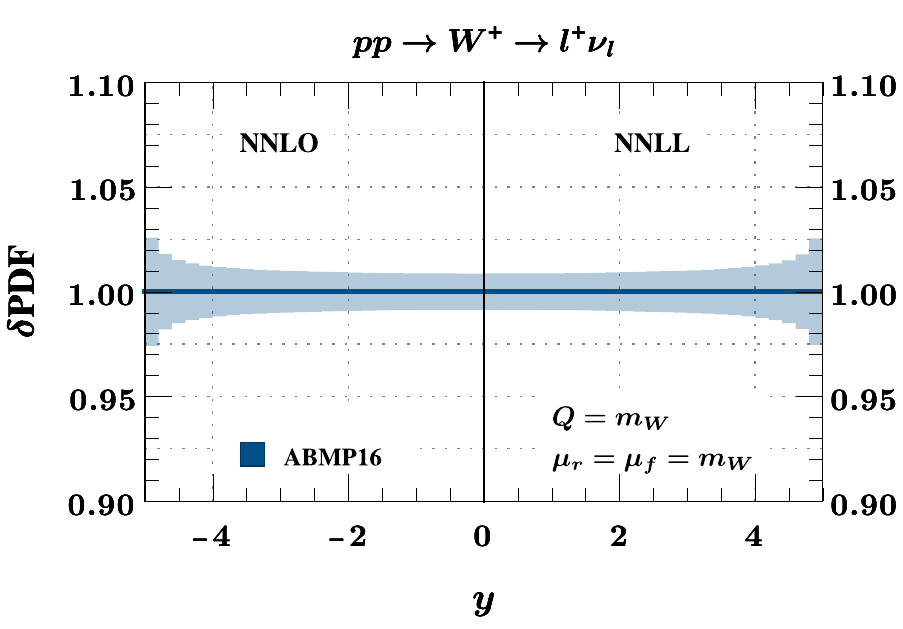}
\hspace{0.05cm}
\includegraphics[width=7.4cm,height=5.6cm]{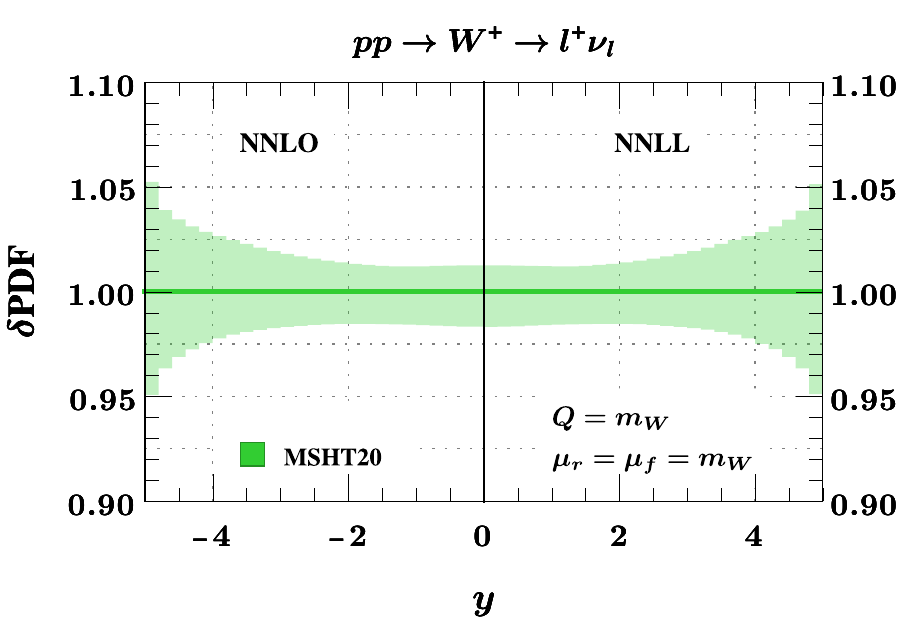}
\includegraphics[width=7.4cm,height=5.6cm]{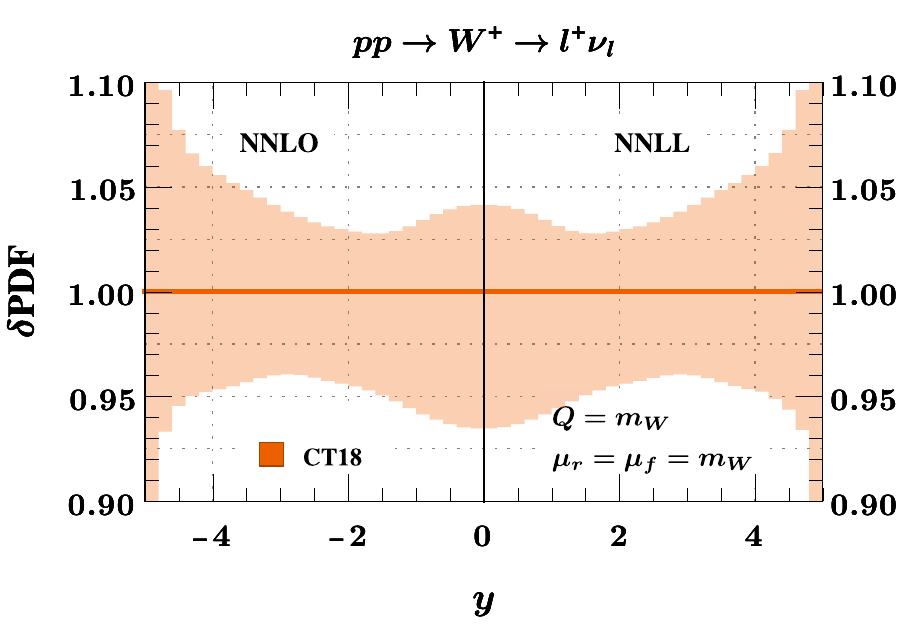}
\hspace{0.05cm}
\includegraphics[width=7.4cm,height=5.6cm]{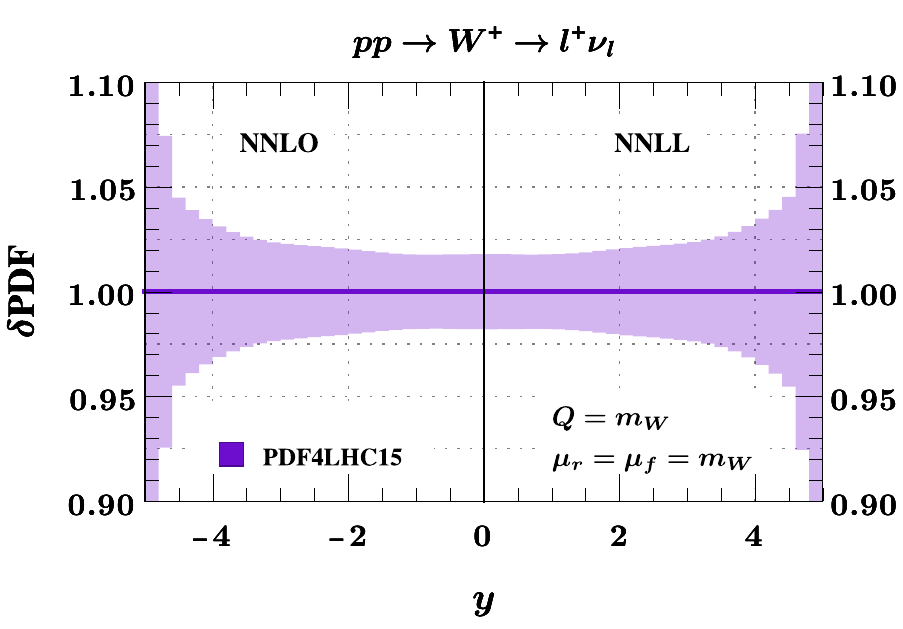}
}
\caption{Same as \fig{fig:Wm-PDF-UNCERTAINTY} 
but for $W^+$ rapidity.}
\label{fig:Wp-PDF-UNCERTAINTY}
\end{figure}

In the central rapidity region, the NLO $K$-factor
can become as large as $1.3$, on the other hand 
at the NNLO level the $K$-factor becomes stable 
and varies in the range $1.01-1.05$.
This shows that the higher order QCD corrections 
are already stable at NNLO.
%% W+ PDF SCALE ratio
\begin{figure}
        \centering{
\includegraphics[width=7.4cm,height=5.6cm]{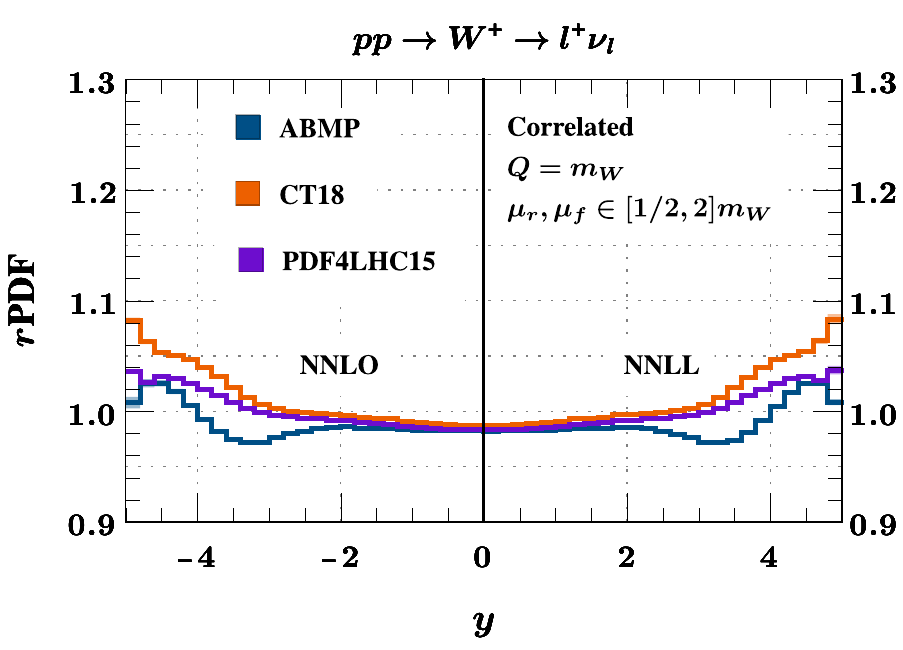}
\hspace{0.05cm}
\includegraphics[width=7.4cm,height=5.6cm]{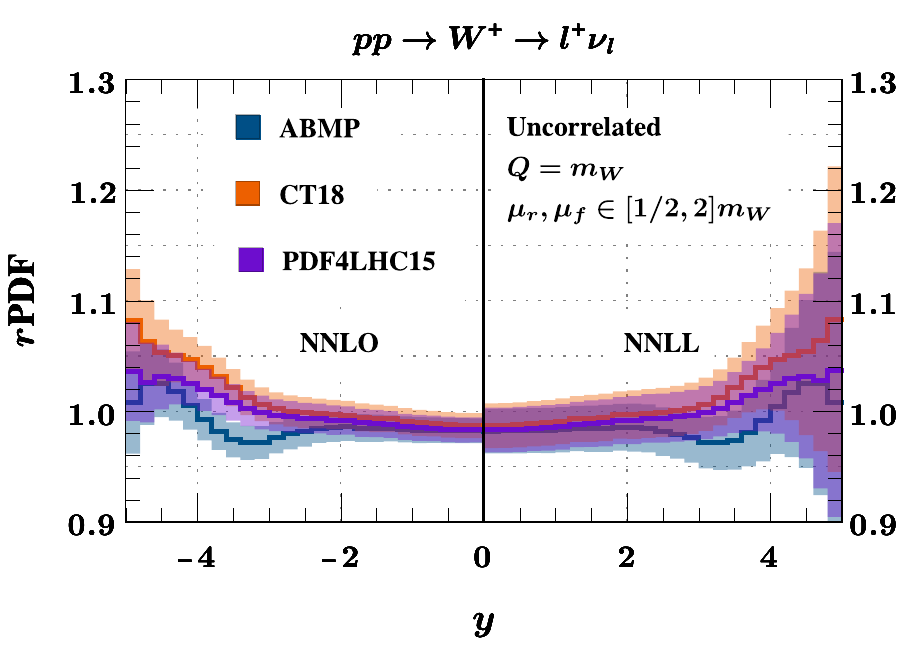}
}
\caption{Same as \fig{fig:Wm-PDF-RATIO} 
but for $W^+$ rapidity.}
\label{fig:Wp-PDF-RATIO}
\end{figure}
From \fig{fig:Wp-PDF-RATIO}, it is more clear that 
the different PDFs start to differ already from 
$|y|=2.0$. In the higher rapidity region, the 
differences are covered by the uncorrelated scale 
uncertainty which could amount to around $10\%$ 
at $y=4.4$.
In \fig{fig:Wp-PDF-UNCERTAINTY}, we also show the 
corresponding intrinsic PDF uncertainties, and we find 
a similar trend as before for the case of $W^-$.
Interestingly, the large rapidity region is less 
affected compared to the $W^-$ case and amounts to 
below $10\%$ for all the sets with \texttt{ABMP} 
giving the smallest errors throughout below $2\%$.
%%% W-Intrinsic PDF Uncertainty

\noindent
{\bf \\$W$ charge asymmetry:\\}
%% W+- Charge Asymmetry
\begin{figure}
        \centering{
\includegraphics[width=7.4cm,height=5.6cm]{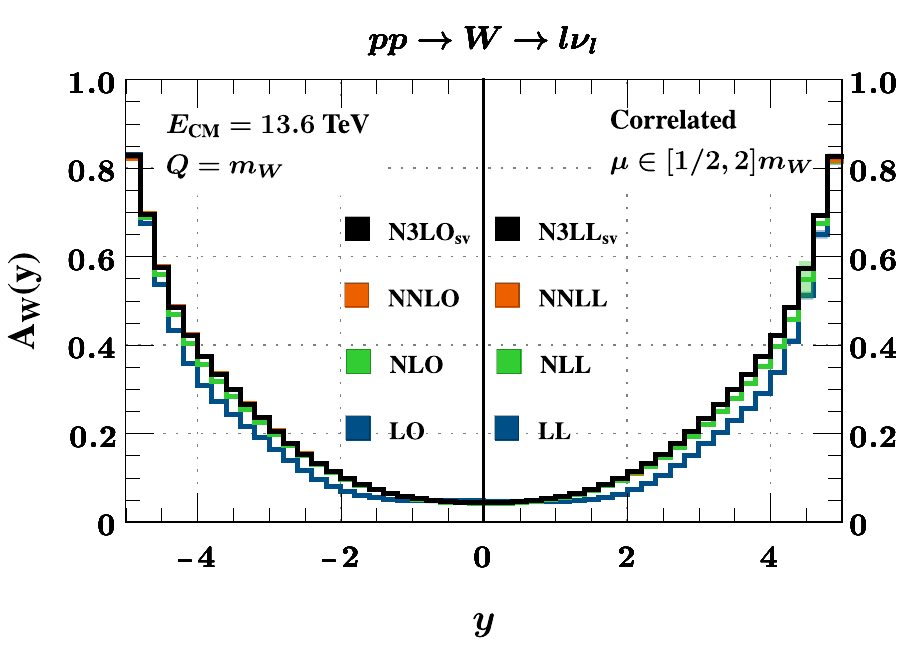}
\hspace{0.05cm}
\includegraphics[width=7.4cm,height=5.6cm]{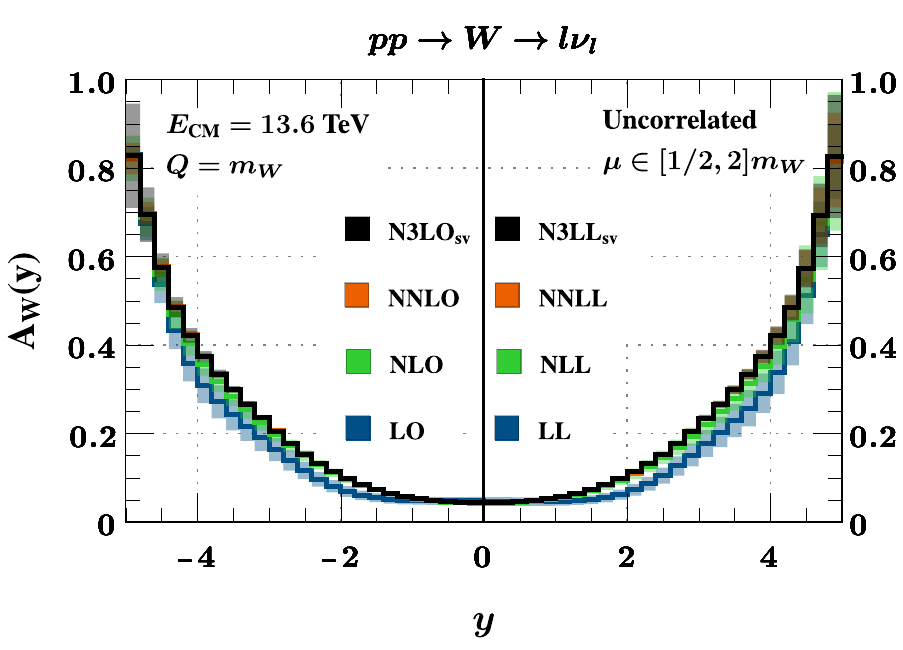}
}
\caption{The $W$ boson charge asymmetry as 
defined in \eq{eq:charge-asymmetry} with correlated
and uncorrelated errors are shown. The left panels are 
for the fixed order contribution whereas the right panels 
show the matched prediction. At the third order 
the matching is done with fixed order SV results.}
\label{fig:Wpm-CHARGE-ASYMMETRY}
\end{figure} 
Finally, with the available fixed order and matched 
results for both $W^-$ and $W^+$ rapidities, we 
study the charge asymmetry at the LHC.
The $W$ boson charge asymmetry at the LHC is defined 
as, 
\begin{align}\label{eq:charge-asymmetry}
        A_W(y) 
        = 
        % \frac{\df \sigma_{W^+}/\df y - \df \sigma_{W^-} / \df y}{\df \sigma_{W^+} / \df y + \df \sigma_{W^-} / \df y}
        {
        \bigg(
        \frac{\df \sigma_{W^+}}{\df Q \df y} 
        - 
        \frac{\df \sigma_{W^-} }{ \df Q \df y} 
        \bigg)}\bigg/{
        \bigg(
        \frac{\df \sigma_{W^+}}{\df Q \df y} 
        + 
        \frac{\df \sigma_{W^-} }{ \df Q \df y}
        \bigg)} \,.
\end{align}
At the LHC, the $W^+$ cross section is larger than 
the $W^-$ one due to the fluxes as the $W^+$ boson is 
produced through $u\bar{d}$ annihilation compared 
to the $d\bar{u}$ in the case of $W^-$.
We present the charge asymmetry for the on-shell $W$
bosons at different rapidities.
The $W$ charge asymmetry is stable against 
the QCD corrections at the LHC. The asymmetry does 
not vanish at $y=0$. Typically, the asymmetry 
increases at the higher rapidity region due to 
different behavior of $u$ and $d$ quarks in the 
region. Due to the CP invariance of the QCD 
cross section, both $W^+$ and $W^-$ rapidity 
distributions are forward-backward symmetric 
which is also translated to the charge asymmetry.

In the case of correlated errors in 
\fig{fig:Wpm-CHARGE-ASYMMETRY}, we observe
a stable result already at NLO which shapes 
the distribution over the large rapidity range
with a scale uncertainty around $2\%$ in 
the central rapidities and gets improved to below 
$0.5\%$ in the higher rapidities.
At NNLO the corrections are around $1\%$ 
compared to NLO for the 
central and higher rapidities whereas 
they become larger for moderate rapidities amounting 
to around $5\%$ at $y=3$. The 
scale uncertainty becomes half of the NLO 
throughout the region. The third order SV effects are 
relatively flat on top of that and amount below $0.5\%$
corrections compared to NNLO with correlated 
uncertainties around $2\%$ at the central 
region and below $1\%$ for higher rapidities.
The NLL scale uncertainty, however is much larger
in the central region around $4\%$ which gets 
better below $1\%$ at higher rapidities.
Similar to the FO case, we observe larger 
corrections at the moderate rapidities which 
amount to around $6.5\%$ at $y=3$ compared to the 
previous order, whereas in the both rapidity ends,
the corrections are of $3\%$.
The correlated scale uncertainties are comparable 
at NNLL and at N3LLsv amounting to around $2\%$
in the central region and below $1\%$ in the 
higher rapidities. While NNLL gets a correction
of about $3.2\%$, the N3LLsv get a further improvement 
with a $2\%$ correction compared to NNLL.
However, in the higher rapidity region at 
N3LLsv the corrections are below $0.2\%$.

The uncorrelated error on the other hand gets reduced
from $8\%$ to around $2\%$ going from NLO to NNLO 
whereas at N3LOsv it amounts to around $3\%$
in the central region. 
At higher rapidities 
$(y=4.4)$ the uncertainties become $6.6\%$ at NLO,
$3\%$ at NNLO and $5\%$ at N3LOsv. It is evident
that the scale uncertainties do not improve at the 
N3LOsv level where the missing regular piece as well 
as the remaining channels are equally important.
We also performed similar analyses at the resummed 
level where we see that at NLL the uncorrelated 
scale uncertainty is around $10\%$ which gets better 
to $3\%$ at NNLL and $3.1\%$ at N3LLsv in the 
central rapidity region. At higher rapidities,
the corresponding uncorrelated uncertainties are 
$14\%$, $7.1\%$, and $7.4\%$ respectively.

%%%Conclusions%%%%%%%%%%%%%%%%%%%%%%%%%%%%%%
\section{Conclusions}\label{sec:conclusion}
%%%%%%%%%%%%%%%%%%%%%%%%%%%%%%%%%%%%%%%%%%%%
We have studied the rapidity distributions of massive 
gauge bosons including threshold effects at NNLL 
and beyond. Our approach employs the double Mellin 
formalism which correctly takes into account all the 
double singular terms in the double threshold limits
corresponding to the inclusive threshold and the rapidity.
This is in contrast to the Mellin-Fourier approach 
for rapidity resummation which approximates one of 
the threshold variable to a certain singular structure 
while resums the other, effectively making it similar 
to a single threshold case. 
The double Mellin resummation formalism closely follows 
the inclusive case and the associated anomalous dimensions 
are related to the inclusive one. We have derived all the analytic
ingredients needed for resummation up to the third order 
in the strong coupling. We also performed a complete 
matched study at the NNLL level where the effects from 
subleading regular terms are also taken into account.
In the central region we find that the resummed 
distributions amount to a further $0.5\%$ correction 
compared to the fixed order NNLO whereas in 
the higher rapidity region, the corrections could 
be as large as $3-7\%$ depending on the rapidity 
value. The conventional scale uncertainty however 
does not improve compared to the fixed order but 
amounts to around $2-9\%$ compared to $1-4\%$ in 
the case of NNLO. Although the seven point 
scale uncertainty failed to capture the higher 
order effects at the fixed order NLO, 
the resummed result on the other hand provides a 
nice overlap starting from LL to N3LL.
In the large rapidity region, we found that the 
scale uncertainty in N3LLsv remains at a similar 
level like NNLL. In order to access the 
threshold effects, we studied  a number of 
ratios in detail and the associated correlated 
and uncorrelated uncertainties. We also studied 
the truncated threshold effect in the Mellin space 
in order to compare to the complete fixed order and 
found that the SV approximation contributes 
positively to all rapidities, however, it suffers 
from large scale uncertainties.

We observed significant changes while comparing the 
predictions for the central scale among a few PDF
sets. The differences in the higher rapidity 
regions are particularly striking. In this region 
the two threshold variables take opposite extreme 
values with one of them being very small and the 
other very large. For the $W$ production, we 
additionally studied the charge asymmetry at the 
LHC at different rapidity points and found that 
it receives larger correction in moderate 
rapidity regions both in the fixed order 
case and in resummed case. Our study can be easily 
extended to the complete matched N3LL level with all 
the necessary resummed ingredients collected in the 
appendix. However, as stated previously, an 
important ingredient of this study involves 
matching to the fixed order to include the 
missing regular piece which are particularly 
important at this level. Some partial 
N3LO results have become available in the literature 
\cite{Chen:2021vtu,Chen:2022lwc} very recently, 
and it will be interesting to include these effects 
into the analyses in the future.
Nevertheless, we 
believe our resummed NNLL(N3LL) results 
supplemented with the third order 
double Mellin space SV results are also competitive.

Our resummed results will also be useful in 
understanding and constraining the PDFs at 
large momentum fraction, a region not well 
constraint by the current results. In the large 
rapidity region there are competitive 
contributions from small-$x$ enhancement of gluon 
and sea-quark. Thus, in addition to double 
threshold improvements (which incorporates the 
corrections from annihilation diagrams) presented 
in this article, it will be interesting to include 
the small-$x$ resummation 
\cite{Altarelli:2008aj,Caola:2010kv,Bonvini:2016wki} 
to achieve further improvements. 
At this accuracy level, it will be also interesting 
to study power corrections from subleading 
logarithms and beyond, which are often important 
at the higher orders 
\cite{DelDuca:2017twk,Ebert:2018lzn,Beneke:2018gvs,Bahjat-Abbas:2019fqa,Ajjath:2020lwb,Ajjath:2021pre,Alekhin:2021xcu}
which we leave for future studies.
%%%%%%%%%%%%%%%%%%%%%%%%%%%%%%%%%%%%%%%%
\section*{Acknowledgements}
%%%%%%%%%%%%%%%%%%%%%%%%%%%%%%%%%%%%%%%%
G.D.\ is indebted to A.\ Vogt for 
enlightening discussion related to 
the Mellin inversion and to S.\ Moch and 
V.\ Ravindran for many clarifications 
at different stages of this work. 
G.D.\ is grateful to G.\ Bell, M.\ Czakon, 
M.\ C.\ Kumar, and S.\ Moch for carefully 
reading the manuscript and providing insightful 
comments.
The computation has been performed on 
the TTP/ITP HPC cluster at KIT within the P3H
network and G.D. thanks F.\ Lange for related help.
Part of the analytical computation has been performed
using symbolic manipulation system F{\sc orm} 
\cite{Vermaseren:2000nd,Ruijl:2017dtg}.
The research of G.D. was supported by the
Deutsche Forschungsgemeinschaft 
(DFG, German Research Foundation)
under grant  396021762 - TRR 257
(\textit{Particle Physics Phenomenology 
after Higgs discovery.}).

\appendix
\section{Anomalous Dimensions}\label{App:ANOMALOUS-DIMENSIONS}
In this section we collect all the anomalous dimensions appearing in this work. 
The QCD beta function is defined as
\begin{align}
        \frac{\df \as}{\df \ln \mu^2}
        = \beta(\as)
        = - \sum_{n=0}^{\infty} 
        \as^{n+2}\beta_n  \,.
\end{align}
We note down the first four coefficients needed up to
N3LL accuracy, 
\cite{Gross:1973id,Politzer:1973fx,Caswell:1974gg,Jones:1974mm,Egorian:1978zx,Tarasov:1980au,Larin:1993tp,vanRitbergen:1997va,Czakon:2004bu,Baikov:2016tgj,Herzog:2017ohr,Luthe:2017ttg},
\begin{align}
\begin{autobreak}
        \beta_{0} =  
\frac{11}{3} \ca 
- \frac{4}{3} \tf \nf \,,
\end{autobreak}
\nn
% %%
\begin{autobreak}
        \beta_{1} = 
\frac{34}{3} \cas 
- \frac{20}{3} \ca \tf \nf
- 4 \cf \tf \nf \,,
\end{autobreak}
\nn
% %%
\begin{autobreak}
        \beta_{2} =  
\frac{2857}{54} \cat 
- \frac{1415}{27} \cas \tf \nf
- \frac{205}{9} \cf \ca \tf \nf 
+ 2 \cfs \tf \nf 
+ \frac{44}{9} \cf \tfs \nfs 
+ \frac{158}{27} \ca \tfs \nfs  \,,
\end{autobreak}
% \end{align}
%%
\nn
% \begin{align}
\begin{autobreak}
        \beta_{3} =
\caf    \bigg( 
        \frac{150653}{486} 
        - \frac{44}{9} \z3 
        \bigg)   
+ \frac{d_{AA}^{(4)}}{N_A} \bigg(  
        - \frac{80}{9} 
        + \frac{704}{3} \z3 
        \bigg)
+  \cat \tf \nf  
        \bigg(  
        - \frac{39143}{81} 
        + \frac{136}{3} \z3 
        \bigg)
        + \cas \cf \tf \nf 
        \bigg( 
        \frac{7073}{243} 
        - \frac{656}{9} \z3 
        \bigg)
+ \ca \cfs \tf \nf 
        \bigg(  
        - \frac{4204}{27} 
        + \frac{352}{9} \z3 
        \bigg)
+ \frac{d_{FA}^{(4)}}{N_A} \nf 
        \bigg(  
        \frac{512}{9} 
        - \frac{1664}{3} \z3 
        \bigg)
+ 46 \cft \tf \nf 
+  \cas \tfs \nfs 
        \bigg( 
        \frac{7930}{81} 
        + \frac{224}{9} \z3 
        \bigg)
+  \cfs \tfs \nfs 
        \bigg( 
        \frac{1352}{27} 
        - \frac{704}{9} \z3 
        \bigg)
+  \ca \cf \tfs \nfs 
        \bigg( 
        \frac{17152}{243} 
        + \frac{448}{9} \z3 
        \bigg)
+ \frac{d_{FF}^{(4)}}{N_A} \nfs 
        \bigg( 
        - \frac{704}{9} 
        + \frac{512}{3} \z3 
        \bigg)
        + \frac{424}{243} \ca \tft \nft 
        + \frac{1232}{243}\cf \tft \nft \,, 
\end{autobreak}
\end{align}
where the quartic Casimirs are defined as,
\begin{align}
        \frac{d_{AA}^{(4)}}{N_A}\equiv
        \frac{d_A^{abcd} d_A^{abcd} }{n_c^2-1}
        &= \frac{n_c^2(n_c^2 + 36)}{24}
        =\frac{135}{8}\,,
        \nn
        \frac{d_{FA}^{(4)}}{N_A}\equiv
        \frac{d_F^{abcd} d_A^{abcd} }{n_c^2-1}
        &=\frac{n_c(n_c^2+6)}{48}
        =\frac{15}{16}\,,
        \nn
        \frac{d_{FF}^{(4)}}{N_A}\equiv
        \frac{d_F^{abcd} d_F^{abcd} }{n_c^2-1}
        &=
        \frac{(n_c^4 - 6 n_c^2 + 18)}{96n_c^2}
        =\frac{5}{96}\,,
\end{align}
with $N_A=n_c^2-1$ and $n_c=3$ for QCD.
All the anomalous dimensions ($F = A, D$) 
for rapidity can be expanded in strong coupling as,
\begin{align}
        F_{d}^{q}
        = \sum_{i=1}^{\infty}
        \as^i F_{d,i}^{q}\,.
\end{align}
The cusp anomalous dimensions ($A$) are known 
up to fourth order completely analytically
\cite{Henn:2019swt,Huber:2019fxe,vonManteuffel:2020vjv} 
and are collected below, 
\begin{align}
\begin{autobreak}
A_1^q = C_{F}
\Bigg\{ 4
\Bigg\},
\end{autobreak}
\nn
\begin{autobreak}
A_2^q = C_{F}
\Bigg\{ \nf    \bigg(
- \frac{40}{9} \bigg)
+ \Ca    \bigg( \frac{268}{9}
- 8  \z2 \bigg)
\Bigg\},
\end{autobreak}
\nn
\begin{autobreak}
A_3^q = C_{F}
\Bigg\{ \nf^2    \bigg(
- \frac{16}{27} \bigg)
+ \Cf  \nf    \bigg(
- \frac{110}{3}
+ 32  \z3 \bigg)
+ \Ca  \nf    \bigg(
- \frac{836}{27}
- \frac{112}{3}  \z3
+ \frac{160}{9}  \z2 \bigg)
+ \Ca^2    \bigg( \frac{490}{3}
+ \frac{88}{3}  \z3
- \frac{1072}{9}  \z2
+ \frac{176}{5}  \z2^2 \bigg)
\Bigg\},
\end{autobreak}
\nn
\begin{autobreak}
A_4^q = C_{F}
\Bigg\{
        \nf^3    \bigg(
- \frac{32}{81}
+ \frac{64}{27}  \z3 \bigg)
+ \Cf  \nf^2    \bigg( \frac{2392}{81}
- \frac{640}{9}  \z3
+ \frac{64}{5}  \z2^2 \bigg)
+ \Ca  \nf^2    \bigg( \frac{923}{81}
+ \frac{2240}{27}  \z3
- \frac{608}{81}  \z2
- \frac{224}{15}  \z2^2 \bigg)
+ \Cf^2  \nf    \bigg( \frac{572}{9}
- 320  \z5
+ \frac{592}{3}  \z3 \bigg)
+ \Ca^2  \nf    \bigg(
- \frac{24137}{81}
+ \frac{2096}{9}  \z5
- \frac{23104}{27}  \z3
+ \frac{20320}{81}  \z2
+ \frac{448}{3}  \z2  \z3
- \frac{352}{15}  \z2^2 \bigg)
+ \Ca  \Cf  \nf    \bigg(
- \frac{34066}{81}
+ 160  \z5
+ \frac{3712}{9}  \z3
+ \frac{440}{3}  \z2
- 128  \z2  \z3
- \frac{352}{5}  \z2^2 \bigg)
+ \Ca^3    \bigg( \frac{84278}{81}
- \frac{3608}{9}  \z5
+ \frac{20944}{27}  \z3
- 16  \z3^2
- \frac{88400}{81}  \z2
- \frac{352}{3}  \z2  \z3
+ \frac{3608}{5}  \z2^2
- \frac{20032}{105}  \z2^3 \bigg)
\Bigg\}
+ \nf  \dFF    \bigg(
% + \nf  \dFFoNF    \bigg(
- \frac{1280}{3}  \z5
- \frac{256}{3}  \z3
+ 256  \z2 \bigg)
+ \dFA     \bigg( \frac{3520}{3}  \z5
% + \dFAoNF     \bigg( \frac{3520}{3}  \z5
+ \frac{128}{3}  \z3
- 384  \z3^2
- 128  \z2
- \frac{7936}{35}   \z2^3 \bigg)
\,,
\end{autobreak}
%\end{align}
\end{align}
% where the quartic casimirs are defined as
% \begin{align}
% % \frac{d_A^{abcd}d_A^{abcd}}{N_A} &= \frac{n_c^2 (n_c^2 + 36)}{24},
% % \frac{d_A^{abcd}d_F^{abcd}}{N_A} = \frac{n_c (n_c^2 + 6)}{48}, \nn\\
% \frac{d_{FA}^{(4)}}{N_F} &\equiv
% \frac{d_F^{abcd}d_A^{abcd}}{n_c} 
% =  \frac{(n_c^2-1)(n_c^2+6)}{48}
% = \frac{5}{2},
% \nn
% \frac{d_{FF}^{(4)}}{N_F} &\equiv
% \frac{d_F^{abcd}d_F^{abcd}}{n_c} 
% = \frac{(n_c^2-1)(n_c^4 - 6 n_c^2 + 18)}{96n_c^3}
% = \frac{5}{36},
% \end{align}
with $N_F = n_c = 3$ for QCD.
The threshold non-cusp anomalous dimensions ($D$) 
for rapidity are maximally non-abelian up to 
third order 
($D_{d}^{q} = C_F/C_A ~D_{d}^{g}$) and 
are related to the inclusive one (see \text{e.g.}\ \cite{Ravindran:2006bu}).
Up to third order they are given as, 
\begin{align}
D^q_{d,1} 
=&~ 
\CF \Bigg\{0\Bigg\},
\nn
D^q_{d,2} 
=&~
\CF \Bigg\{\NF   
\bigg(
        \frac{112}{27}
        - \frac{8}{3} \z2
\bigg)
+ \CA   
\bigg(
        - \frac{808}{27}
        + 28 \z3
        + \frac{44}{3} \z2
\bigg)
\Bigg\},
\nn
D^q_{d,3} 
=&~
\CF \Bigg\{\NF^2  
\bigg(
        - \frac{1856}{729} 
        - \frac{32}{27} \z3 
        + \frac{160}{27} \z2 
\bigg)
+ \CA \NF   
\bigg(
        \frac{62626}{729}
        + \frac{208}{15} \z2^2
        - \frac{536}{9} \z3
        - \frac{7760}{81} \z2
\bigg)

\nn&
+ \CF \NF   
\bigg(
        \frac{1711}{27}
        - \frac{32}{5} \z2^2
        - \frac{304}{9} \z3
        - 8 \z2
\bigg)
+ \CA^2   
\bigg(
        - \frac{297029}{729}
        - \frac{616}{15} \z2^2
        - 192 \z5
\nn&
        + \frac{14264}{27} \z3
        + \frac{27752}{81} \z2
        - \frac{176}{3} \z2 \z3
\bigg) \Bigg\}.
\end{align}
\section{The process dependent coefficient $g_{_{d,0}}^V$}\label{App:g0}
Here we collect all the process-dependent 
coefficients for the process at each order
as defined in \eq{eq:g0-expansion}.
\input{outg0DYR.tex} 
Note that up to the second order in the strong 
coupling both $Z, W^\pm$ receive similar 
SV corrections. Thus, up to the overall normalization
factor the $g^V_0$ coefficients are same for 
both these processes.
At the third order the $Z$ 
production process receives additional 
contributions.
This corresponds to the term $n_{fv}$ 
proportional to 
the charge weighted sum \cite{Larin:1996wd,Vermaseren:2005qc} 
of quark
flavors with $\frac{d^{abc} d_{abc}}{N_A} \equiv 
\frac{(n_c^2-4)}{16 n_c} = 5/48$ in QCD. 
This term is absent for $W$ production case. 
\section{Universal double-Mellin resummed exponent}\label{App:GN}
Here we collect the universal resummed exponent 
$\left(g^{q}_{d,i}(\w)\right)$
in the double-Mellin variables as introduced in 
\eq{eq:RESUM-EXPONENT} up to 
N3LL accuracy.
Recalling $\w = \as \bt0 \ln (\Nbar1 \Nbar2) $, 
and in addition, defining 
$\widetilde{\omg} = 1-\w \,,~ \wminv = 1/(1-\w)\,,$ and 
\begin{align}
\boldsymbol{A}^{'}_{i} &= A_{i}^{q}/\beta_0^{i}\,,&
\boldsymbol{D}^{'}_{i} &= D_{d,i}^{q}/\beta_0^{i}\,,&
\boldsymbol{\beta}^{'}_{i} &= \beta_{i}/\beta_0^{i+1}\,,&
\nn
\boldsymbol{g}^{'q}_{d,1(2)}(\w)&= g^q_{d,1(2)}(\w)\,,&
\boldsymbol{g}^{'q}_{d,3}(\w)&= g^q_{d,3}(\w)/\bt0\,,&
\boldsymbol{g}^{'q}_{d,4}(\w)&= g^q_{d,4}(\w)/\bt0^2\,,
\end{align}
we present the coefficients up to N3LL as,
\input{outgnDYR.tex}

\section{Soft-virtual coefficients in double-Mellin space}\label{App:SV-COEFFICIENTS}
For completeness, here we have collected all the 
large $N_i$ coefficients up to third order in strong coupling. 
We write down the perturbative expansion as follows,
\begin{align}
        \widetilde{\Delta}_{d,ab}^{\rm f.o.}(N_1, N_2)
        =
        1+\sum_{i=1}^{\infty} 
        \as^i
        \widetilde{\Delta}_{d,ab}^{{\rm f.o.}(i)} \,.
\end{align}
Defining $\LNoNtb=\ln (\Nbar1 \Nbar2)$, we write out the 
coefficients up to third order as follows,
\input{outsvnDYR.tex}
where the non-logarithmic constant terms 
($g^V_{_{d,0i}}$) are collected before in \app{App:g0}.
Note that the double threshold logarithmic structures 
are same for both $Z,W^\pm$ production up to third 
order.
%%%References
\bibliographystyle{JHEP}
\bibliography{references}
\end{document}

%% file: outg0DYR.tex
\begin{align} \label{eq:G0}
\begin{autobreak} 
\g01DY = 
  \CF    \bigg\{ 
- 16
+ 16  \z2
+ \bigg(-6\bigg)  \Lfr
+ \bigg(6\bigg)  \Lqr \bigg\} ,  
\end{autobreak} 
\\ 
\begin{autobreak} 
\g02DY = 
  \CF  \NF    \bigg\{ \frac{127}{6}
+ \frac{8}{9}  \z3
- \frac{64}{3}  \z2
+ \bigg(
- \frac{34}{3}
+ \frac{16}{3}  \z2\bigg)  \Lqr
+ \bigg(\frac{2}{3}
+ \frac{16}{3}  \z2\bigg)  \Lfr
+ \bigg(-2\bigg)  \Lfr^2
+ \bigg(2\bigg)  \Lqr^2 \bigg\}      
+ \CF^2    \bigg\{ \frac{511}{4}
- 60  \z3
- 198  \z2
+ \frac{552}{5}  \z2^2
+ \bigg(
- 93
+ 48  \z3
+ 72  \z2 \bigg)  \Lqr
+ \bigg(93
- 48  \z3
- 72  \z2\bigg)  \Lfr
+ \bigg(-36\bigg)  \Lqrfr
+ \bigg(18\bigg)  \Lfr^2
+ \bigg(18\bigg)   \Lqr^2 \bigg\}      
+ \CA  \CF    \bigg\{ 
- \frac{1535}{12}
+ \frac{604}{9}  \z3
+ \frac{376}{3}  \z2
- \frac{92}{5}  \z2^2
+ \bigg(
- \frac{17}{3}
+ 24  \z3
- \frac{88}{3}  \z2\bigg)  \Lfr
+ \bigg(\frac{193}{3}
- 24  \z3
- \frac{88}{3}  \z2\bigg)  \Lqr
+ \bigg(-11\bigg)  \Lqr^2
+ \bigg(11\bigg)  \Lfr^2 \bigg\} ,  
\end{autobreak} 
\\ 
\begin{autobreak} 
\g03DY = 
  \CF  \nfv  \frac{d^{abc}d_{abc}}{N_A} \bigg\{ 128
- \frac{2560}{3}  \z5
+ \frac{448}{3}  \z3
+ 320  \z2
- \frac{64}{5}  \z2^2 \bigg\}
%   \CF  \nfv  \Nfour    \bigg\{ 8
% - \frac{160}{3}  \z5
% + \frac{28}{3}  \z3
% + 20  \z2
% - \frac{4}{5}  \z2^2 \bigg\}
%%      
+ \CF  \NF^2    \bigg\{ 
- \frac{7081}{243}
+ \frac{16}{81}  \z3
+ \frac{1072}{27}  \z2
+ \frac{448}{135}  \z2^2
+ \bigg(
- \frac{68}{9}
+ \frac{32}{9}  \z2\bigg)  \Lqr^2
+ \bigg(
- \frac{8}{9}\bigg)  \Lfr^3
+ \bigg(\frac{4}{9}
+ \frac{32}{9}  \z2\bigg)  \Lfr^2
+ \bigg(\frac{8}{9} \bigg)  \Lqr^3
+ \bigg(\frac{34}{9}
+ \frac{32}{9}  \z3
- \frac{160}{27}  \z2\bigg)  \Lfr
+ \bigg(\frac{220}{9}
- \frac{64}{27}  \z3
- \frac{608}{27}  \z2\bigg)  \Lqr \bigg\}      
+ \CF^2  \NF    \bigg\{ 
- \frac{421}{3}
- \frac{608}{9}  \z5
+ \frac{9448}{27}  \z3
+ \frac{9064}{27}  \z2
- \frac{256}{3}  \z2   \z3
- \frac{36208}{135}  \z2^2
+ \bigg(
- 92
+ 32  \z3
+ 48  \z2\bigg)  \Lqr^2
+ \bigg(
- \frac{275}{3}
+ \frac{256}{3}  \z3
+ 40  \z2
+ \frac{272}{5}  \z2^2\bigg)  \Lfr
+ \bigg(20
- 32  \z3
- 48  \z2\bigg)  \Lfr^2
+ \bigg(72
+ \bigg(
-12\bigg)  \Lfr
+ \bigg(-12\bigg)  \Lqr\bigg)  \Lqrfr
+ \bigg(230
- \frac{496}{3}  \z3
- 272  \z2
+ \frac{464}{5}  \z2^2\bigg)   \Lqr
+ \bigg(12\bigg)  \Lfr^3
+ \bigg(12\bigg)  \Lqr^3 \bigg\}      
+ \CF^3    \bigg\{ 
- \frac{5599}{6}
+ 1328  \z5
- 460  \z3
+ 32  \z3^2
+ \frac{2936}{3}  \z2
- 400  \z2   \z3
- \frac{5972}{5}  \z2^2
+ \frac{169504}{315}  \z2^3
+ \bigg(
- \frac{1495}{2}
+ 480  \z5
+ 992  \z3
+ 720  \z2
- 704  \z2  \z3
- \frac{1968}{5}  \z2^2\bigg)  \Lfr
+ \bigg(
- 270
+ 288  \z3
+ 144  \z2\bigg)   \Lfr^2
+ \bigg(
- 270
+ 288  \z3
+ 144  \z2\bigg)  \Lqr^2
+ \bigg(540
- 576  \z3
- 288  \z2
+ \bigg(
-108\bigg)  \Lqr
+ \bigg(108\bigg)  \Lfr\bigg)  \Lqrfr
+ \bigg(\frac{1495}{2}
- 480  \z5
- 992  \z3
- 720  \z2
+ 704  \z2  \z3
+ \frac{1968}{5}  \z2^2\bigg)  \Lqr
+ \bigg(-36\bigg)  \Lfr^3
+ \bigg(36\bigg)  \Lqr^3 \bigg\}      
+ \CA  \CF  \NF    \bigg\{ \frac{110651}{243}
- 8  \z5
- \frac{24512}{81}  \z3
- \frac{44540}{81}  \z2
+ \frac{880}{9}  \z2   \z3
+ \frac{1156}{135}  \z2^2
+ \bigg(
- \frac{3052}{9}
+ \frac{3440}{27}  \z3
+ \frac{7504}{27}  \z2
- \frac{344}{15}   \z2^2\bigg)  \Lqr
+ \bigg(
- 40
- \frac{400}{9}  \z3
+ \frac{2672}{27}  \z2
- \frac{8}{5}  \z2^2\bigg)  \Lfr
+ \bigg(
- \frac{146}{9}
+ 16  \z3
- \frac{352}{9}  \z2\bigg)  \Lfr^2
+ \bigg(
- \frac{88}{9}\bigg)  \Lqr^3
+ \bigg(\frac{88}{9}\bigg)  \Lfr^3
+ \bigg(\frac{850}{9}
- 16  \z3
- \frac{352}{9}  \z2\bigg)  \Lqr^2 \bigg\}      
+ \CA  \CF^2    \bigg\{ \frac{74321}{36}
- \frac{5512}{9}  \z5
- \frac{51508}{27}  \z3
+ \frac{592}{3}  \z3^2
- \frac{66544}{27}  \z2
+ \frac{3680}{3}  \z2  \z3
+ \frac{258304}{135}  \z2^2
- \frac{123632}{315}  \z2^3
+ \bigg(
- \frac{3439}{2}
+ 240  \z5
+ \frac{5368}{3}  \z3
+ 1552  \z2
- 352  \z2  \z3
- \frac{2912}{5}  \z2^2\bigg)  \Lqr
+ \bigg(
- 420
+ 288  \z3
+ \bigg(66\bigg)  \Lfr
+ \bigg(66\bigg)  \Lqr\bigg)  \Lqrfr
+ \bigg(
- 131
+ 32  \z3
+ 264  \z2\bigg)   \Lfr^2
+ \bigg(551
- 320  \z3
- 264  \z2\bigg)  \Lqr^2
+ \bigg(\frac{2348}{3}
- 240  \z5
- \frac{4048}{3}  \z3
- 100  \z2
+ 352  \z2  \z3
- \frac{1136}{5}  \z2^2\bigg)  \Lfr
+ \bigg(-66\bigg)  \Lfr^3
+ \bigg(-66\bigg)  \Lqr^3 \bigg\}      
+ \CA^2  \CF    \bigg\{ 
- \frac{1505881}{972}
- 204  \z5
+ \frac{139345}{81}  \z3
- \frac{400}{3}  \z3^2
+ \frac{130295}{81}  \z2
- \frac{7228}{9}  \z2  \z3
- \frac{23357}{135}  \z2^2
+ \frac{7088}{63}  \z2^3
+ \bigg(
- \frac{2429}{9}
+ 88  \z3
+ \frac{968}{9}  \z2\bigg)  \Lqr^2
+ \bigg(
- \frac{242}{9}\bigg)  \Lfr^3
+ \bigg(\frac{242}{9}\bigg)  \Lqr^3
+ \bigg(\frac{493}{9}
- 88  \z3
+ \frac{968}{9}  \z2\bigg)  \Lfr^2
+ \bigg(\frac{1657}{18}
- 80  \z5
+ \frac{3104}{9}  \z3
- \frac{8992}{27}  \z2
+ 4  \z2^2\bigg)  \Lfr
+ \bigg(\frac{3082}{3}
+ 80  \z5
- \frac{22600}{27}  \z3
- \frac{20720}{27}   \z2
+ \frac{1964}{15}  \z2^2\bigg)  \Lqr \bigg\} \,.
\end{autobreak} 
\end{align}

%% file: outgnDYR.tex
\begin{align} 
\begin{autobreak} 
\GNDY1 = 
  \Ap1    \bigg\{ 1
-   \zlmmo
%+ \winv   \zlmmo \bigg\} ,  
+ (\w)^{-1}   \zlmmo \bigg\} ,  
\end{autobreak} 
\\ 
\begin{autobreak} 
\GNDY2 = 
  \Dp1    \bigg\{ \zlmmo \bigg\}      
+ \Ap2    \bigg\{ 
- \zlmmo
- \w \bigg\}      
+ \Ap1    \bigg\{ \bigg(\zlmmo
+ \frac{1}{2}  \zlmmto
+ \w\bigg)  \btp1
+ \bigg(\w\bigg)  \Lfr
+ \bigg(\zlmmo\bigg)  \Lqr \bigg\} ,  
\end{autobreak} 
\\ 
\begin{autobreak} 
\GNDY3 = 
  \Dp2    \bigg\{ 1
- \wminv \bigg\}      
+ \Dp1    \bigg\{ \bigg(
- 1
+ \wminv\bigg)  \Lqr
+ \bigg(\wminv  \zlmmo
+ \w  \wminv\bigg)  \btp1 \bigg\}      
+ \Ap3    \bigg\{ 
- \frac{1}{2}
+ \frac{1}{2}  \wminv
- \frac{1}{2}  \w \bigg\}      
+ \Ap2    \bigg\{ \bigg(1
- \wminv\bigg)  \Lqr
+ \bigg(
- \wminv  \zlmmo
- \w  \wminv
- \frac{1}{2}  \w^2  \wminv\bigg)   \btp1
+ \bigg(\w\bigg)  \Lfr \bigg\}      
+ \Ap1    \bigg\{ 
- \z2
+ \z2  \wminv
+ \bigg(
- \frac{1}{2}
+ \frac{1}{2}  \wminv\bigg)  \Lqr^2
+ \bigg(\zlmmo
+ \w   \wminv
- \frac{1}{2}  \w^2  \wminv\bigg)  \btp2
+ \bigg(\frac{1}{2}  \wminv  \zlmmto
+ \w  \wminv  \zlmmo
+ \frac{1}{2}  \w^2  \wminv\bigg)  \btp1^2
+ \bigg(
- \frac{1}{2}  \w\bigg)  \Lfr^2
+ \bigg(\bigg(\wminv  \zlmmo
+ \w  \wminv\bigg)   \Lqr\bigg)  \btp1 \bigg\} ,  
\end{autobreak} 
\\ 
\begin{autobreak} 
\GNDY4 = 
  \Dp3    \bigg\{ \frac{1}{2}
- \frac{1}{2}  \wminv^2 \bigg\}      
+ \Dp2    \bigg\{ \bigg(
- 1
+ \wminv^2\bigg)  \Lqr
+ \bigg(\wminv^2  \zlmmo
+ \w  \wminv^2
- \frac{1}{2}  \w^2   \wminv^2\bigg)  \btp1 \bigg\}      
+ \Dp1    \bigg\{ \z2
- \z2  \wminv^2
+ \bigg(\frac{1}{2}
- \frac{1}{2}  \wminv^2\bigg)  \Lqr^2
+ \bigg(
- \frac{1}{2}  \wminv^2   \zlmmto
+ \frac{1}{2}  \w^2  \wminv^2\bigg)  \btp1^2
+ \bigg(
- \frac{1}{2}  \w^2  \wminv^2\bigg)  \btp2
+ \bigg(\bigg(
- \wminv^2  \zlmmo\bigg)  \Lqr\bigg)  \btp1 \bigg\}      
+ \Ap4    \bigg\{ 
- \frac{1}{6}
+ \frac{1}{6}  \wminv^2
- \frac{1}{3}  \w \bigg\}      
+ \Ap3    \bigg\{ \bigg(\frac{1}{2}
- \frac{1}{2}  \wminv^2\bigg)  \Lqr
+ \bigg(
- \frac{1}{2}  \wminv^2  \zlmmo
- \frac{1}{2}  \w   \wminv^2
- \frac{1}{4}  \w^2  \wminv^2
+ \frac{1}{3}  \w^3  \wminv^2\bigg)  \btp1
+ \bigg(\w\bigg)  \Lfr \bigg\}      
+ \Ap2    \bigg\{ 
- \z2
+ \z2  \wminv^2
+ \bigg(
- \frac{1}{2}
+ \frac{1}{2}  \wminv^2\bigg)  \Lqr^2
+ \bigg(\frac{1}{2}   \wminv^2  \zlmmo
+ \frac{1}{2}  \wminv^2  \zlmmto
+ \frac{1}{2}  \w  \wminv^2
- \frac{1}{4}  \w^2   \wminv^2
- \frac{1}{3}  \w^3  \wminv^2\bigg)  \btp1^2
+ \bigg(
- \w\bigg)  \Lfr^2
+ \bigg(\frac{1}{3}  \w^3  \wminv^2\bigg)   \btp2
+ \bigg(\bigg(\wminv^2  \zlmmo
+ \w  \wminv^2
- \frac{1}{2}  \w^2  \wminv^2\bigg)  \Lqr\bigg)  \btp1 \bigg\}      
+ \Ap1    \bigg\{ 
- \frac{2}{3}  \z3
+ \frac{2}{3}  \z3  \wminv^2
+ \bigg(
- \frac{1}{12}
+ \frac{1}{2}  \zlmmo
+ \frac{1}{12}   \wminv^2
+ \frac{1}{3}  \w\bigg)  \btp3
+ \bigg(\frac{1}{6}
- \frac{1}{6}  \wminv^2\bigg)  \Lqr^3
+ \bigg(
- \frac{1}{6}  \wminv^2   \zlmmte
+ \frac{1}{2}  \w^2  \wminv^2  \zlmmo
+ \frac{1}{3}  \w^3  \wminv^2\bigg)  \btp1^3
+ \bigg(\z2
- \z2  \wminv^2\bigg)  \Lqr
+ \bigg(
- \z2  \wminv^2  \zlmmo
+ \bigg(
- \frac{1}{2}  \wminv^2  \zlmmo\bigg)   \Lqr^2
+ \bigg(
- \frac{1}{2}  \wminv^2  \zlmmo
- \frac{1}{2}  \w  \wminv^2
+ \w  \wminv^2  \zlmmo
+ \frac{3}{4}  \w^2  \wminv^2
- \w^2  \wminv^2  \zlmmo
- \frac{2}{3}  \w^3  \wminv^2\bigg)  \btp2
+ \bigg(
- \frac{1}{2}   \w\bigg)  \Lfr^2\bigg)  \btp1
+ \bigg(\frac{1}{3}  \w\bigg)  \Lfr^3
+ \bigg(\bigg(
- \frac{1}{2}  \wminv^2  \zlmmto
+ \frac{1}{2}  \w^2   \wminv^2\bigg)  \Lqr\bigg)  \btp1^2
+ \bigg(\bigg(
- \frac{1}{2}  \w^2  \wminv^2\bigg)  \Lqr\bigg)  \btp2 \bigg\} ,
\end{autobreak} 
\end{align}

%% file: outsvnDYR.tex
\begin{align} 
\begin{autobreak} 
\gSVN1 = 
  \LNoNtb^2    \bigg\{ \bigg(2\bigg)  \CF \bigg\}      
+ \LNoNtb    \bigg\{ \bigg(\bigg(-4\bigg)  \Lqr
+ \bigg(4\bigg)  \Lfr\bigg)  \CF \bigg\} 
+ \g01DY \,,     
% + \bigg(16  \z2
% + \bigg(-6\bigg)  \Lfr
% + \bigg(6\bigg)  \Lqr
% - 16\bigg)  \CF ,  
\end{autobreak} 
\\ 
\begin{autobreak} 
\gSVN2 =
  \LNoNtb^4    \bigg\{ \bigg(2\bigg)  \CF^2 \bigg\}      
+ \LNoNtb^3    \bigg\{ \bigg(\bigg(-8\bigg)  \Lqr
+ \bigg(8\bigg)  \Lfr\bigg)  \CF^2
+ \bigg(
- \frac{4}{9}\bigg)  \CF  \NF
+ \bigg(\frac{22}{9}\bigg)  \CA   \CF \bigg\}      
+ \LNoNtb^2    \bigg\{ \bigg(
- 4  \z2
+ \bigg(
- \frac{22}{3}\bigg)  \Lqr
+ \frac{134}{9}\bigg)  \CA  \CF
+ \bigg(32  \z2
+ \bigg(-16\bigg)   \Lqrfr
+ \bigg(-12\bigg)  \Lfr
+ \bigg(8\bigg)  \Lqr^2
+ \bigg(8\bigg)  \Lfr^2
+ \bigg(12\bigg)  \Lqr
- 32\bigg)  \CF^2
+ \bigg(\bigg( \frac{4}{3}\bigg)  \Lqr
- \frac{20}{9}\bigg)  \CF  \NF \bigg\}      
+ \LNoNtb    \bigg\{ \bigg(
- 28  \z3
+ \bigg(
- 8  \z2
+ \frac{268}{9}\bigg)  \Lfr
+ \bigg(8  \z2
- \frac{268}{9}\bigg)  \Lqr
+ \bigg(
- \frac{22}{3}\bigg)  \Lfr^2
+ \bigg(\frac{22}{3}\bigg)  \Lqr^2
+ \frac{808}{27}\bigg)  \CA  \CF
+ \bigg(\bigg(
- 64  \z2
+ 64\bigg)  \Lqr
+ \bigg(64  \z2
- 64\bigg)  \Lfr
+ \bigg(-24\bigg)  \Lqr^2
+ \bigg(-24\bigg)  \Lfr^2
+ \bigg(48\bigg)  \Lqrfr\bigg)  \CF^2
+ \bigg(\bigg(
- \frac{40}{9}\bigg)  \Lfr
+ \bigg(
- \frac{4}{3}\bigg)  \Lqr^2
+ \bigg(\frac{4}{3}\bigg)  \Lfr^2
+ \bigg(\frac{40}{9}\bigg)  \Lqr
- \frac{112}{27}\bigg)   \CF  \NF \bigg\}   
+\g02DY\,,   
% + \bigg(
% - \frac{92}{5}  \z2^2
% + \frac{376}{3}  \z2
% + \frac{604}{9}  \z3
% + \bigg(
% - \frac{88}{3}  \z2
% - 24  \z3
% + \frac{193}{3}\bigg)   \Lqr
% + \bigg(
% - \frac{88}{3}  \z2
% + 24  \z3
% - \frac{17}{3}\bigg)  \Lfr
% + \bigg(-11\bigg)  \Lqr^2
% + \bigg(11\bigg)  \Lfr^2
% - \frac{1535}{12}\bigg)  \CA  \CF
% + \bigg(\frac{552}{5}  \z2^2
% - 198  \z2
% - 60  \z3
% + \bigg(
% - 72  \z2
% - 48  \z3
% + 93\bigg)  \Lfr
% + \bigg(72  \z2
% + 48  \z3
% - 93\bigg)  \Lqr
% + \bigg(-36\bigg)  \Lqrfr
% + \bigg(18\bigg)  \Lqr^2
% + \bigg(18\bigg)  \Lfr^2
% + \frac{511}{4}\bigg)  \CF^2
% + \bigg(
% - \frac{64}{3}  \z2
% + \frac{8}{9}  \z3
% + \bigg(\frac{16}{3}  \z2
% - \frac{34}{3}\bigg)  \Lqr
% + \bigg(\frac{16}{3}   \z2
% + \frac{2}{3}\bigg)  \Lfr
% + \bigg(-2\bigg)  \Lfr^2
% + \bigg(2\bigg)  \Lqr^2
% + \frac{127}{6}\bigg)  \CF  \NF ,  
\end{autobreak} 
\\ 
\begin{autobreak} 
\gSVN3 = 
  \LNoNtb^6    \bigg\{ \bigg(\frac{4}{3}\bigg)  \CF^3 \bigg\}      
+ \LNoNtb^5    \bigg\{ \bigg(\bigg(-8\bigg)  \Lqr
+ \bigg(8\bigg)  \Lfr\bigg)  \CF^3
+ \bigg(
- \frac{8}{9}\bigg)  \CF^2  \NF
+ \bigg(\frac{44}{9}\bigg)  \CA   \CF^2 \bigg\}      
+ \LNoNtb^4    \bigg\{ \bigg(
- 8  \z2
+ \bigg(
- \frac{220}{9}\bigg)  \Lqr
+ \bigg(\frac{88}{9}\bigg)  \Lfr
+ \frac{268}{9}\bigg)  \CA  \CF^2
+ \bigg(32  \z2
+ \bigg(-32\bigg)  \Lqrfr
+ \bigg(-12\bigg)  \Lfr
+ \bigg(12\bigg)  \Lqr
+ \bigg(16\bigg)  \Lqr^2
+ \bigg(16\bigg)   \Lfr^2
- 32\bigg)  \CF^3
+ \bigg(\bigg(
- \frac{16}{9}\bigg)  \Lfr
+ \bigg(\frac{40}{9}\bigg)  \Lqr
- \frac{40}{9}\bigg)  \CF^2  \NF
+ \bigg(
- \frac{44}{27}\bigg)  \CA  \CF  \NF
+ \bigg(\frac{4}{27}\bigg)  \CF  \NF^2
+ \bigg(\frac{121}{27}\bigg)  \CA^2  \CF \bigg\}      
+ \LNoNtb^3    \bigg\{ \bigg(
- \frac{88}{9}  \z2
+ \bigg(
- \frac{484}{27}\bigg)  \Lqr
+ \frac{3560}{81}\bigg)  \CA^2  \CF
+ \bigg(
- \frac{64}{9}  \z2
+ \bigg(
- \frac{136}{9}\bigg)  \Lfr
+ \bigg(\frac{8}{3}\bigg)  \Lfr^2
+ \bigg(\frac{16}{3}\bigg)  \Lqrfr
+ \bigg(\frac{136}{9}\bigg)  \Lqr
+ \bigg(-8\bigg)  \Lqr^2
- \frac{68}{27}\bigg)  \CF^2  \NF
+ \bigg(\frac{16}{9}  \z2
+ \bigg(\frac{176}{27}\bigg)  \Lqr
- \frac{1156}{81}\bigg)   \CA  \CF  \NF
+ \bigg(\frac{352}{9}  \z2
- 56  \z3
+ \bigg(
- 32  \z2
+ \frac{940}{9}\bigg)  \Lfr
+ \bigg(32  \z2
- \frac{940}{9}\bigg)  \Lqr
+ \bigg(
- \frac{88}{3}\bigg)  \Lqrfr
+ \bigg(
- \frac{44}{3}\bigg)  \Lfr^2
+ \bigg(44\bigg)  \Lqr^2
+ \frac{560}{27}\bigg)  \CA   \CF^2
+ \bigg(\bigg(
- 128  \z2
+ 128\bigg)  \Lqr
+ \bigg(128  \z2
- 128\bigg)  \Lfr
+ \bigg(\bigg(-32\bigg)  \Lfr
+ \bigg(32 \bigg)  \Lqr
+ 96\bigg)  \Lqrfr
+ \bigg(
- \frac{32}{3}\bigg)  \Lqr^3
+ \bigg(\frac{32}{3}\bigg)  \Lfr^3
+ \bigg(-48\bigg)  \Lqr^2
+ \bigg(
-48\bigg)  \Lfr^2\bigg)  \CF^3
+ \bigg(\bigg(
- \frac{16}{27}\bigg)  \Lqr
+ \frac{80}{81}\bigg)  \CF  \NF^2 \bigg\}      
+ \LNoNtb^2    \bigg\{ \bigg(
- \frac{504}{5}  \z2^2
+ \frac{4976}{9}  \z2
+ \frac{1208}{9}  \z3
+ \bigg(
- 200  \z2
+ 64   \z3
+ \frac{5822}{27}\bigg)  \Lqr
+ \bigg(
- \frac{104}{3}  \z2
- 64  \z3
+ \frac{514}{27}\bigg)  \Lfr
+ \bigg(
- 32  \z2
+ \frac{478}{9}\bigg)  \Lqr^2
+ \bigg(
- 32  \z2
+ \frac{1270}{9}\bigg)  \Lfr^2
+ \bigg(64  \z2
+ \bigg(\frac{88}{3}\bigg)  \Lqr
+ \bigg( \frac{88}{3}\bigg)  \Lfr
- \frac{1748}{9}\bigg)  \Lqrfr
+ \bigg(
- \frac{88}{3}\bigg)  \Lqr^3
+ \bigg(
- \frac{88}{3}\bigg)  \Lfr^3
- \frac{8893}{18}\bigg)  \CA  \CF^2
+ \bigg(\frac{88}{5}  \z2^2
- \frac{536}{9}  \z2
- 88  \z3
+ \bigg(\frac{88}{3}  \z2
- \frac{3560}{27}\bigg)  \Lqr
+ \bigg(\frac{242}{9}\bigg)  \Lqr^2
+ \frac{15503}{81}\bigg)  \CA^2  \CF
+ \bigg(\frac{1104}{5}  \z2^2
- 396  \z2
- 120  \z3
+ \bigg(
- 256  \z2
+ \bigg(-144\bigg)  \Lqr
+ \bigg(144\bigg)  \Lfr
+ 184\bigg)  \Lqrfr
+ \bigg(
- 144  \z2
- 96   \z3
+ 186\bigg)  \Lfr
+ \bigg(128  \z2
- 92\bigg)  \Lqr^2
+ \bigg(128  \z2
- 92\bigg)  \Lfr^2
+ \bigg(144  \z2
+ 96  \z3
- 186\bigg)  \Lqr
+ \bigg(-48\bigg)  \Lfr^3
+ \bigg(48\bigg)  \Lqr^3
+ \frac{511}{2}\bigg)  \CF^3
+ \bigg(
- \frac{704}{9}  \z2
+ \frac{160}{9}  \z3
+ \bigg(\frac{32}{3}  \z2
- \frac{52}{27}\bigg)  \Lfr
+ \bigg(32  \z2
- \frac{992}{27}\bigg)  \Lqr
+ \bigg(\bigg(
- \frac{16}{3}\bigg)  \Lqr
+ \bigg(
- \frac{16}{3}\bigg)  \Lfr
+ \frac{248}{9}\bigg)  \Lqrfr
+ \bigg(
- \frac{196}{9}\bigg)  \Lfr^2
+ \bigg(
- \frac{52}{9}\bigg)  \Lqr^2
+ \bigg(\frac{16}{3}\bigg)  \Lqr^3
+ \bigg(\frac{16}{3}\bigg)  \Lfr^3
+ \frac{536}{9}\bigg)  \CF^2  \NF
+ \bigg(\frac{80}{9}   \z2
+ \bigg(
- \frac{16}{3}  \z2
+ \frac{1156}{27}\bigg)  \Lqr
+ \bigg(
- \frac{88}{9}\bigg)  \Lqr^2
- \frac{4102}{81}\bigg)  \CA  \CF   \NF
+ \bigg(\bigg(
- \frac{80}{27}\bigg)  \Lqr
+ \bigg(\frac{8}{9}\bigg)  \Lqr^2
+ \frac{200}{81}\bigg)  \CF  \NF^2 \bigg\}      
+ \LNoNtb    \bigg\{ \bigg(
- \frac{88}{5}  \z2^2
+ \frac{176}{3}  \z2  \z3
- \frac{6392}{81}  \z2
- \frac{12328}{27}  \z3
+ 192  \z5
+ \bigg(
- \frac{176}{5}  \z2^2
+ \frac{1072}{9}  \z2
+ 176  \z3
- \frac{31006}{81}\bigg)  \Lqr
+ \bigg(\frac{176}{5}   \z2^2
- \frac{1072}{9}  \z2
+ \frac{88}{3}  \z3
+ \frac{490}{3}\bigg)  \Lfr
+ \bigg(
- \frac{88}{3}  \z2
+ \frac{3560}{27}\bigg)   \Lqr^2
+ \bigg(\frac{88}{3}  \z2
- \frac{3560}{27}\bigg)  \Lfr^2
+ \bigg(
- \frac{484}{27}\bigg)  \Lqr^3
+ \bigg(\frac{484}{27}\bigg)   \Lfr^3
+ \frac{297029}{729}\bigg)  \CA^2  \CF
+ \bigg(
- \frac{16}{5}  \z2^2
+ \frac{824}{81}  \z2
+ \frac{904}{27}  \z3
+ \bigg(
- \frac{160}{9}  \z2
+ \frac{8204}{81}\bigg)  \Lqr
+ \bigg(
- \frac{16}{3}  \z2
+ \frac{1156}{27}\bigg)  \Lfr^2
+ \bigg(\frac{16}{3}   \z2
- \frac{1156}{27}\bigg)  \Lqr^2
+ \bigg(\frac{160}{9}  \z2
- \frac{112}{3}  \z3
- \frac{836}{27}\bigg)  \Lfr
+ \bigg(
- \frac{176}{27}\bigg)  \Lfr^3
+ \bigg(\frac{176}{27}\bigg)  \Lqr^3
- \frac{62626}{729}\bigg)  \CA  \CF  \NF
+ \bigg(\frac{32}{5}  \z2^2
- \frac{1792}{27}  \z2
+ \frac{304}{9}  \z3
+ \bigg(
- \frac{1408}{9}  \z2
+ \frac{320}{9}  \z3
+ 144\bigg)  \Lfr
+ \bigg(
- \frac{128}{3}  \z2
+ \frac{268}{3}\bigg)  \Lqr^2
+ \bigg(\frac{128}{3}  \z2
+ 12\bigg)  \Lfr^2
+ \bigg(\frac{1408}{9}  \z2
- \frac{320}{9}  \z3
- 144 \bigg)  \Lqr
+ \bigg(\bigg(16\bigg)  \Lqr
+ \bigg(16\bigg)  \Lfr
- \frac{304}{3}\bigg)  \Lqrfr
+ \bigg(-16\bigg)  \Lqr^3
+ \bigg(-16\bigg)   \Lfr^3
+ 3\bigg)  \CF^2  \NF
+ \bigg(
- 448  \z2  \z3
+ \frac{12928}{27}  \z2
+ 448  \z3
+ \bigg(
- \frac{1008}{5}  \z2^2
+ \frac{9952}{9}  \z2
+ \frac{3928}{9}  \z3
- \frac{3503}{3}\bigg)  \Lfr
+ \bigg(\frac{1008}{5}  \z2^2
- \frac{9952}{9}   \z2
- \frac{3928}{9}  \z3
+ \frac{3503}{3}\bigg)  \Lqr
+ \bigg(
- \frac{560}{3}  \z2
+ 96  \z3
- 84\bigg)  \Lfr^2
+ \bigg(
- 96  \z2
- 192  \z3
+ \bigg(-88\bigg)  \Lqr
+ \bigg(-88\bigg)  \Lfr
+ \frac{1912}{3}\bigg)  \Lqrfr
+ \bigg(\frac{848}{3}  \z2
+ 96  \z3
- \frac{1660}{3}\bigg)  \Lqr^2
+ \bigg(88\bigg)  \Lqr^3
+ \bigg(88\bigg)  \Lfr^3
- \frac{12928}{27}\bigg)  \CA   \CF^2
+ \bigg(\frac{32}{9}  \z3
+ \bigg(
- \frac{400}{81}\bigg)  \Lqr
+ \bigg(
- \frac{80}{27}\bigg)  \Lfr^2
+ \bigg(
- \frac{16}{27}\bigg)   \Lqr^3
+ \bigg(
- \frac{16}{27}\bigg)  \Lfr
+ \bigg(\frac{16}{27}\bigg)  \Lfr^3
+ \bigg(\frac{80}{27}\bigg)  \Lqr^2
+ \frac{1856}{729}\bigg)  \CF   \NF^2
+ \bigg(\bigg(
- \frac{2208}{5}  \z2^2
+ 792  \z2
+ 240  \z3
- 511\bigg)  \Lqr
+ \bigg(\frac{2208}{5}  \z2^2
- 792  \z2
- 240  \z3
+ 511\bigg)  \Lfr
+ \bigg(
- 288  \z2
- 192  \z3
+ 372\bigg)  \Lqr^2
+ \bigg(
- 288  \z2
- 192  \z3
+ 372\bigg)  \Lfr^2
+ \bigg(576  \z2
+ 384  \z3
+ \bigg(-216\bigg)  \Lfr
+ \bigg( 216\bigg)  \Lqr
- 744\bigg)  \Lqrfr
+ \bigg(-72\bigg)  \Lqr^3
+ \bigg(72\bigg)  \Lfr^3\bigg)  \CF^3 \bigg\}  
+ \g03DY\,,    
\end{autobreak} 
\end{align}

%% file: WZrap.bbl
\providecommand{\href}[2]{#2}\begingroup\raggedright\begin{thebibliography}{100}

\bibitem{Altarelli:1978id}
G.~Altarelli, R.~K. Ellis and G.~Martinelli, \emph{{Leptoproduction and
  Drell-Yan Processes Beyond the Leading Approximation in Chromodynamics}},
  \href{https://doi.org/10.1016/0550-3213(78)90067-6}{\emph{Nucl. Phys. B}
  {\bfseries 143} (1978) 521}.

\bibitem{Altarelli:1979ub}
G.~Altarelli, R.~K. Ellis and G.~Martinelli, \emph{{Large Perturbative
  Corrections to the Drell-Yan Process in QCD}},
  \href{https://doi.org/10.1016/0550-3213(79)90116-0}{\emph{Nucl. Phys. B}
  {\bfseries 157} (1979) 461}.

\bibitem{Fiaschi:2022wgl}
J.~Fiaschi, F.~Giuli, F.~Hautmann, S.~Moch and S.~Moretti,
  \emph{{$\mathbf{Z^\prime}$-boson dilepton searches and the high-$\mathbf{x}$
  quark density}},  \href{https://arxiv.org/abs/2211.06188}{{\ttfamily
  2211.06188}}.

\bibitem{Das:2016pbk}
G.~Das, C.~Degrande, V.~Hirschi, F.~Maltoni and H.-S. Shao, \emph{{NLO
  predictions for the production of a spin-two particle at the LHC}},
  \href{https://doi.org/10.1016/j.physletb.2017.05.007}{\emph{Phys. Lett. B}
  {\bfseries 770} (2017) 507}
  [\href{https://arxiv.org/abs/1605.09359}{{\ttfamily 1605.09359}}].

\bibitem{Berger:1988tu}
E.~L. Berger, F.~Halzen, C.~S. Kim and S.~Willenbrock, \emph{{WEAK BOSON
  PRODUCTION AT TEVATRON ENERGIES}},
  \href{https://doi.org/10.1103/PhysRevD.40.83}{\emph{Phys. Rev. D} {\bfseries
  40} (1989) 83}.

\bibitem{Martin:1988aj}
A.~D. Martin, R.~G. Roberts and W.~J. Stirling, \emph{{Improved Parton
  Distributions and W, Z Production at p anti-p Colliders}},
  \href{https://doi.org/10.1142/S0217732389001313}{\emph{Mod. Phys. Lett. A}
  {\bfseries 4} (1989) 1135}.

\bibitem{Hamberg:1990np}
R.~Hamberg, W.~L. van Neerven and T.~Matsuura, \emph{{A complete calculation of
  the order $\alpha_s^{2}$ correction to the Drell-Yan $K$ factor}},
  \href{https://doi.org/10.1016/0550-3213(91)90064-5}{\emph{Nucl. Phys. B}
  {\bfseries 359} (1991) 343}.

\bibitem{Harlander:2002wh}
R.~V. Harlander and W.~B. Kilgore, \emph{{Next-to-next-to-leading order Higgs
  production at hadron colliders}},
  \href{https://doi.org/10.1103/PhysRevLett.88.201801}{\emph{Phys. Rev. Lett.}
  {\bfseries 88} (2002) 201801}
  [\href{https://arxiv.org/abs/hep-ph/0201206}{{\ttfamily hep-ph/0201206}}].

\bibitem{Ahmed:2014cla}
T.~Ahmed, M.~Mahakhud, N.~Rana and V.~Ravindran, \emph{{Drell-Yan Production at
  Threshold to Third Order in QCD}},
  \href{https://doi.org/10.1103/PhysRevLett.113.112002}{\emph{Phys. Rev. Lett.}
  {\bfseries 113} (2014) 112002}
  [\href{https://arxiv.org/abs/1404.0366}{{\ttfamily 1404.0366}}].

\bibitem{Li:2014bfa}
Y.~Li, A.~von Manteuffel, R.~M. Schabinger and H.~X. Zhu, \emph{{N$^3$LO Higgs
  boson and Drell-Yan production at threshold: The one-loop two-emission
  contribution}}, \href{https://doi.org/10.1103/PhysRevD.90.053006}{\emph{Phys.
  Rev. D} {\bfseries 90} (2014) 053006}
  [\href{https://arxiv.org/abs/1404.5839}{{\ttfamily 1404.5839}}].

\bibitem{Catani:2014uta}
S.~Catani, L.~Cieri, D.~de~Florian, G.~Ferrera and M.~Grazzini,
  \emph{{Threshold resummation at N$^3$LL accuracy and soft-virtual cross
  sections at N$^3$LO}},
  \href{https://doi.org/10.1016/j.nuclphysb.2014.09.012}{\emph{Nucl. Phys. B}
  {\bfseries 888} (2014) 75} [\href{https://arxiv.org/abs/1405.4827}{{\ttfamily
  1405.4827}}].

\bibitem{Catani:1996yz}
S.~Catani, M.~L. Mangano, P.~Nason and L.~Trentadue, \emph{{The Resummation of
  soft gluons in hadronic collisions}},
  \href{https://doi.org/10.1016/0550-3213(96)00399-9}{\emph{Nucl. Phys. B}
  {\bfseries 478} (1996) 273}
  [\href{https://arxiv.org/abs/hep-ph/9604351}{{\ttfamily hep-ph/9604351}}].

\bibitem{Contopanagos:1996nh}
H.~Contopanagos, E.~Laenen and G.~F. Sterman, \emph{{Sudakov factorization and
  resummation}},
  \href{https://doi.org/10.1016/S0550-3213(96)00567-6}{\emph{Nucl. Phys. B}
  {\bfseries 484} (1997) 303}
  [\href{https://arxiv.org/abs/hep-ph/9604313}{{\ttfamily hep-ph/9604313}}].

\bibitem{Magnea:2000ss}
L.~Magnea, \emph{{Analytic resummation for the quark form-factor in QCD}},
  \href{https://doi.org/10.1016/S0550-3213(00)00623-4}{\emph{Nucl. Phys. B}
  {\bfseries 593} (2001) 269}
  [\href{https://arxiv.org/abs/hep-ph/0006255}{{\ttfamily hep-ph/0006255}}].

\bibitem{Catani:2003zt}
S.~Catani, D.~de~Florian, M.~Grazzini and P.~Nason, \emph{{Soft gluon
  resummation for Higgs boson production at hadron colliders}},
  \href{https://doi.org/10.1088/1126-6708/2003/07/028}{\emph{JHEP} {\bfseries
  07} (2003) 028} [\href{https://arxiv.org/abs/hep-ph/0306211}{{\ttfamily
  hep-ph/0306211}}].

\bibitem{Manohar:2003vb}
A.~V. Manohar, \emph{{Deep inelastic scattering as $x \to 1$ using soft
  collinear effective theory}},
  \href{https://doi.org/10.1103/PhysRevD.68.114019}{\emph{Phys. Rev. D}
  {\bfseries 68} (2003) 114019}
  [\href{https://arxiv.org/abs/hep-ph/0309176}{{\ttfamily hep-ph/0309176}}].

\bibitem{Eynck:2003fn}
T.~O. Eynck, E.~Laenen and L.~Magnea, \emph{{Exponentiation of the Drell-Yan
  cross-section near partonic threshold in the DIS and MS-bar schemes}},
  \href{https://doi.org/10.1088/1126-6708/2003/06/057}{\emph{JHEP} {\bfseries
  06} (2003) 057} [\href{https://arxiv.org/abs/hep-ph/0305179}{{\ttfamily
  hep-ph/0305179}}].

\bibitem{Moch:2005ba}
S.~Moch, J.~A.~M. Vermaseren and A.~Vogt, \emph{{Higher-order corrections in
  threshold resummation}},
  \href{https://doi.org/10.1016/j.nuclphysb.2005.08.005}{\emph{Nucl. Phys. B}
  {\bfseries 726} (2005) 317}
  [\href{https://arxiv.org/abs/hep-ph/0506288}{{\ttfamily hep-ph/0506288}}].

\bibitem{Moch:2005ky}
S.~Moch and A.~Vogt, \emph{{Higher-order soft corrections to lepton pair and
  Higgs boson production}},
  \href{https://doi.org/10.1016/j.physletb.2005.09.061}{\emph{Phys. Lett. B}
  {\bfseries 631} (2005) 48}
  [\href{https://arxiv.org/abs/hep-ph/0508265}{{\ttfamily hep-ph/0508265}}].

\bibitem{Laenen:2005uz}
E.~Laenen and L.~Magnea, \emph{{Threshold resummation for electroweak
  annihilation from DIS data}},
  \href{https://doi.org/10.1016/j.physletb.2005.10.038}{\emph{Phys. Lett. B}
  {\bfseries 632} (2006) 270}
  [\href{https://arxiv.org/abs/hep-ph/0508284}{{\ttfamily hep-ph/0508284}}].

\bibitem{Ravindran:2005vv}
V.~Ravindran, \emph{{On Sudakov and soft resummations in QCD}},
  \href{https://doi.org/10.1016/j.nuclphysb.2006.04.008}{\emph{Nucl. Phys. B}
  {\bfseries 746} (2006) 58}
  [\href{https://arxiv.org/abs/hep-ph/0512249}{{\ttfamily hep-ph/0512249}}].

\bibitem{Ravindran:2006cg}
V.~Ravindran, \emph{{Higher-order threshold effects to inclusive processes in
  QCD}}, \href{https://doi.org/10.1016/j.nuclphysb.2006.06.025}{\emph{Nucl.
  Phys. B} {\bfseries 752} (2006) 173}
  [\href{https://arxiv.org/abs/hep-ph/0603041}{{\ttfamily hep-ph/0603041}}].

\bibitem{Idilbi:2006dg}
A.~Idilbi, X.-d. Ji and F.~Yuan, \emph{{Resummation of threshold logarithms in
  effective field theory for DIS, Drell-Yan and Higgs production}},
  \href{https://doi.org/10.1016/j.nuclphysb.2006.07.002}{\emph{Nucl. Phys. B}
  {\bfseries 753} (2006) 42}
  [\href{https://arxiv.org/abs/hep-ph/0605068}{{\ttfamily hep-ph/0605068}}].

\bibitem{Becher:2006mr}
T.~Becher, M.~Neubert and B.~D. Pecjak, \emph{{Factorization and Momentum-Space
  Resummation in Deep-Inelastic Scattering}},
  \href{https://doi.org/10.1088/1126-6708/2007/01/076}{\emph{JHEP} {\bfseries
  01} (2007) 076} [\href{https://arxiv.org/abs/hep-ph/0607228}{{\ttfamily
  hep-ph/0607228}}].

\bibitem{deFlorian:2012za}
D.~de~Florian and J.~Mazzitelli, \emph{{A next-to-next-to-leading order
  calculation of soft-virtual cross sections}},
  \href{https://doi.org/10.1007/JHEP12(2012)08}{\emph{JHEP} {\bfseries 12}
  (2012) 088} [\href{https://arxiv.org/abs/1209.0673}{{\ttfamily 1209.0673}}].

\bibitem{Bonvini:2015ira}
M.~Bonvini, S.~Marzani, J.~Rojo, L.~Rottoli, M.~Ubiali, R.~D. Ball et~al.,
  \emph{{Parton distributions with threshold resummation}},
  \href{https://doi.org/10.1007/JHEP09(2015)191}{\emph{JHEP} {\bfseries 09}
  (2015) 191} [\href{https://arxiv.org/abs/1507.01006}{{\ttfamily
  1507.01006}}].

\bibitem{Ajjath:2020rci}
A.~H. Ajjath, G.~Das, M.~C. Kumar, P.~Mukherjee, V.~Ravindran and K.~Samanta,
  \emph{{Resummed Drell-Yan cross-section at N$^{3}$LL}},
  \href{https://doi.org/10.1007/JHEP10(2020)153}{\emph{JHEP} {\bfseries 10}
  (2020) 153} [\href{https://arxiv.org/abs/2001.11377}{{\ttfamily
  2001.11377}}].

\bibitem{Das:2022zie}
G.~Das, C.~Dey, M.~C. Kumar and K.~Samanta, \emph{{Threshold enhanced cross
  sections for colorless productions}},
  \href{https://doi.org/10.1103/PhysRevD.107.034038}{\emph{Phys. Rev. D}
  {\bfseries 107} (2023) 034038}
  [\href{https://arxiv.org/abs/2210.17534}{{\ttfamily 2210.17534}}].

\bibitem{Das:2019btv}
G.~Das, S.-O. Moch and A.~Vogt, \emph{{Soft corrections to inclusive
  deep-inelastic scattering at four loops and beyond}},
  \href{https://doi.org/10.1007/JHEP03(2020)116}{\emph{JHEP} {\bfseries 03}
  (2020) 116} [\href{https://arxiv.org/abs/1912.12920}{{\ttfamily
  1912.12920}}].

\bibitem{Das:2019uvh}
G.~Das, S.-O. Moch and A.~Vogt, \emph{{Soft corrections to inclusive DIS at
  four loops and beyond}},
  \href{https://doi.org/10.22323/1.352.0010}{\emph{PoS} {\bfseries DIS2019}
  (2019) 010} [\href{https://arxiv.org/abs/1908.03071}{{\ttfamily
  1908.03071}}].

\bibitem{Das:2020adl}
G.~Das, S.~Moch and A.~Vogt, \emph{{Approximate four-loop QCD corrections to
  the Higgs-boson production cross section}},
  \href{https://doi.org/10.1016/j.physletb.2020.135546}{\emph{Phys. Lett. B}
  {\bfseries 807} (2020) 135546}
  [\href{https://arxiv.org/abs/2004.00563}{{\ttfamily 2004.00563}}].

\bibitem{Ajjath:2021lvg}
A.~H. Ajjath, P.~Mukherjee, V.~Ravindran, A.~Sankar and S.~Tiwari,
  \emph{{Next-to SV resummed Drell\textendash{}Yan cross section beyond
  leading-logarithm}},
  \href{https://doi.org/10.1140/epjc/s10052-022-10174-7}{\emph{Eur. Phys. J. C}
  {\bfseries 82} (2022) 234}
  [\href{https://arxiv.org/abs/2107.09717}{{\ttfamily 2107.09717}}].

\bibitem{Dittmaier:2001ay}
S.~Dittmaier and M.~Kr\"amer, \emph{{Electroweak radiative corrections to W
  boson production at hadron colliders}},
  \href{https://doi.org/10.1103/PhysRevD.65.073007}{\emph{Phys. Rev. D}
  {\bfseries 65} (2002) 073007}
  [\href{https://arxiv.org/abs/hep-ph/0109062}{{\ttfamily hep-ph/0109062}}].

\bibitem{Baur:2001ze}
U.~Baur, O.~Brein, W.~Hollik, C.~Schappacher and D.~Wackeroth,
  \emph{{Electroweak radiative corrections to neutral current Drell-Yan
  processes at hadron colliders}},
  \href{https://doi.org/10.1103/PhysRevD.65.033007}{\emph{Phys. Rev. D}
  {\bfseries 65} (2002) 033007}
  [\href{https://arxiv.org/abs/hep-ph/0108274}{{\ttfamily hep-ph/0108274}}].

\bibitem{Baur:2004ig}
U.~Baur and D.~Wackeroth, \emph{{Electroweak radiative corrections to $p
  \bar{p} \to W^\pm \to \ell^\pm \nu$ beyond the pole approximation}},
  \href{https://doi.org/10.1103/PhysRevD.70.073015}{\emph{Phys. Rev. D}
  {\bfseries 70} (2004) 073015}
  [\href{https://arxiv.org/abs/hep-ph/0405191}{{\ttfamily hep-ph/0405191}}].

\bibitem{Arbuzov:2005dd}
A.~Arbuzov, D.~Bardin, S.~Bondarenko, P.~Christova, L.~Kalinovskaya, G.~Nanava
  et~al., \emph{{One-loop corrections to the Drell-Yan process in SANC. I. The
  Charged current case}},
  \href{https://doi.org/10.1140/epjc/s2006-02505-y}{\emph{Eur. Phys. J. C}
  {\bfseries 46} (2006) 407}
  [\href{https://arxiv.org/abs/hep-ph/0506110}{{\ttfamily hep-ph/0506110}}].

\bibitem{CarloniCalame:2006zq}
C.~M. Carloni~Calame, G.~Montagna, O.~Nicrosini and A.~Vicini, \emph{{Precision
  electroweak calculation of the charged current Drell-Yan process}},
  \href{https://doi.org/10.1088/1126-6708/2006/12/016}{\emph{JHEP} {\bfseries
  12} (2006) 016} [\href{https://arxiv.org/abs/hep-ph/0609170}{{\ttfamily
  hep-ph/0609170}}].

\bibitem{Zykunov:2005tc}
V.~A. Zykunov, \emph{{Weak radiative corrections to Drell-Yan process for large
  invariant mass of di-lepton pair}},
  \href{https://doi.org/10.1103/PhysRevD.75.073019}{\emph{Phys. Rev. D}
  {\bfseries 75} (2007) 073019}
  [\href{https://arxiv.org/abs/hep-ph/0509315}{{\ttfamily hep-ph/0509315}}].

\bibitem{CarloniCalame:2007cd}
C.~M. Carloni~Calame, G.~Montagna, O.~Nicrosini and A.~Vicini, \emph{{Precision
  electroweak calculation of the production of a high transverse-momentum
  lepton pair at hadron colliders}},
  \href{https://doi.org/10.1088/1126-6708/2007/10/109}{\emph{JHEP} {\bfseries
  10} (2007) 109} [\href{https://arxiv.org/abs/0710.1722}{{\ttfamily
  0710.1722}}].

\bibitem{Arbuzov:2007db}
A.~Arbuzov, D.~Bardin, S.~Bondarenko, P.~Christova, L.~Kalinovskaya, G.~Nanava
  et~al., \emph{{One-loop corrections to the Drell--Yan process in SANC. (II).
  The Neutral current case}},
  \href{https://doi.org/10.1140/epjc/s10052-008-0531-8}{\emph{Eur. Phys. J. C}
  {\bfseries 54} (2008) 451} [\href{https://arxiv.org/abs/0711.0625}{{\ttfamily
  0711.0625}}].

\bibitem{Dittmaier:2009cr}
S.~Dittmaier and M.~Huber, \emph{{Radiative corrections to the neutral-current
  Drell-Yan process in the Standard Model and its minimal supersymmetric
  extension}}, \href{https://doi.org/10.1007/JHEP01(2010)060}{\emph{JHEP}
  {\bfseries 01} (2010) 060} [\href{https://arxiv.org/abs/0911.2329}{{\ttfamily
  0911.2329}}].

\bibitem{Bonciani:2020tvf}
R.~Bonciani, F.~Buccioni, N.~Rana and A.~Vicini, \emph{{Next-to-Next-to-Leading
  Order Mixed QCD-Electroweak Corrections to on-Shell Z Production}},
  \href{https://doi.org/10.1103/PhysRevLett.125.232004}{\emph{Phys. Rev. Lett.}
  {\bfseries 125} (2020) 232004}
  [\href{https://arxiv.org/abs/2007.06518}{{\ttfamily 2007.06518}}].

\bibitem{Bonciani:2021zzf}
R.~Bonciani, L.~Buonocore, M.~Grazzini, S.~Kallweit, N.~Rana, F.~Tramontano
  et~al., \emph{{Mixed Strong-Electroweak Corrections to the Drell-Yan
  Process}}, \href{https://doi.org/10.1103/PhysRevLett.128.012002}{\emph{Phys.
  Rev. Lett.} {\bfseries 128} (2022) 012002}
  [\href{https://arxiv.org/abs/2106.11953}{{\ttfamily 2106.11953}}].

\bibitem{Bonciani:2021iis}
R.~Bonciani, F.~Buccioni, N.~Rana and A.~Vicini, \emph{{On-shell Z boson
  production at hadron colliders through $\mathcal{O}(\alpha\alpha_{s})$}},
  \href{https://doi.org/10.1007/JHEP02(2022)095}{\emph{JHEP} {\bfseries 02}
  (2022) 095} [\href{https://arxiv.org/abs/2111.12694}{{\ttfamily
  2111.12694}}].

\bibitem{Armadillo:2022bgm}
T.~Armadillo, R.~Bonciani, S.~Devoto, N.~Rana and A.~Vicini, \emph{{Two-loop
  mixed QCD-EW corrections to neutral current Drell-Yan}},
  \href{https://doi.org/10.1007/JHEP05(2022)072}{\emph{JHEP} {\bfseries 05}
  (2022) 072} [\href{https://arxiv.org/abs/2201.01754}{{\ttfamily
  2201.01754}}].

\bibitem{Duhr:2020seh}
C.~Duhr, F.~Dulat and B.~Mistlberger, \emph{{Drell-Yan Cross Section to Third
  Order in the Strong Coupling Constant}},
  \href{https://doi.org/10.1103/PhysRevLett.125.172001}{\emph{Phys. Rev. Lett.}
  {\bfseries 125} (2020) 172001}
  [\href{https://arxiv.org/abs/2001.07717}{{\ttfamily 2001.07717}}].

\bibitem{Duhr:2020sdp}
C.~Duhr, F.~Dulat and B.~Mistlberger, \emph{{Charged current Drell-Yan
  production at N$^{3}$LO}},
  \href{https://doi.org/10.1007/JHEP11(2020)143}{\emph{JHEP} {\bfseries 11}
  (2020) 143} [\href{https://arxiv.org/abs/2007.13313}{{\ttfamily
  2007.13313}}].

\bibitem{Duhr:2021vwj}
C.~Duhr and B.~Mistlberger, \emph{{Lepton-pair production at hadron colliders
  at N$^{3}$LO in QCD}},
  \href{https://doi.org/10.1007/JHEP03(2022)116}{\emph{JHEP} {\bfseries 03}
  (2022) 116} [\href{https://arxiv.org/abs/2111.10379}{{\ttfamily
  2111.10379}}].

\bibitem{Anastasiou:2003yy}
C.~Anastasiou, L.~J. Dixon, K.~Melnikov and F.~Petriello, \emph{{Dilepton
  rapidity distribution in the Drell-Yan process at NNLO in QCD}},
  \href{https://doi.org/10.1103/PhysRevLett.91.182002}{\emph{Phys. Rev. Lett.}
  {\bfseries 91} (2003) 182002}
  [\href{https://arxiv.org/abs/hep-ph/0306192}{{\ttfamily hep-ph/0306192}}].

\bibitem{Anastasiou:2003ds}
C.~Anastasiou, L.~J. Dixon, K.~Melnikov and F.~Petriello, \emph{{High precision
  QCD at hadron colliders: Electroweak gauge boson rapidity distributions at
  NNLO}}, \href{https://doi.org/10.1103/PhysRevD.69.094008}{\emph{Phys. Rev. D}
  {\bfseries 69} (2004) 094008}
  [\href{https://arxiv.org/abs/hep-ph/0312266}{{\ttfamily hep-ph/0312266}}].

\bibitem{Catani:2009sm}
S.~Catani, L.~Cieri, G.~Ferrera, D.~de~Florian and M.~Grazzini, \emph{{Vector
  boson production at hadron colliders: a fully exclusive QCD calculation at
  NNLO}}, \href{https://doi.org/10.1103/PhysRevLett.103.082001}{\emph{Phys.
  Rev. Lett.} {\bfseries 103} (2009) 082001}
  [\href{https://arxiv.org/abs/0903.2120}{{\ttfamily 0903.2120}}].

\bibitem{Melnikov:2006kv}
K.~Melnikov and F.~Petriello, \emph{{Electroweak gauge boson production at
  hadron colliders through $O(\alpha_s^2)$}},
  \href{https://doi.org/10.1103/PhysRevD.74.114017}{\emph{Phys. Rev. D}
  {\bfseries 74} (2006) 114017}
  [\href{https://arxiv.org/abs/hep-ph/0609070}{{\ttfamily hep-ph/0609070}}].

\bibitem{Gavin:2012sy}
R.~Gavin, Y.~Li, F.~Petriello and S.~Quackenbush, \emph{{W Physics at the LHC
  with FEWZ 2.1}},
  \href{https://doi.org/10.1016/j.cpc.2012.09.005}{\emph{Comput. Phys. Commun.}
  {\bfseries 184} (2013) 208}
  [\href{https://arxiv.org/abs/1201.5896}{{\ttfamily 1201.5896}}].

\bibitem{Ravindran:2006bu}
V.~Ravindran, J.~Smith and W.~L. van Neerven, \emph{{QCD threshold corrections
  to di-lepton and Higgs rapidity distributions beyond $N^{2}$ LO}},
  \href{https://doi.org/10.1016/j.nuclphysb.2007.01.005}{\emph{Nucl. Phys. B}
  {\bfseries 767} (2007) 100}
  [\href{https://arxiv.org/abs/hep-ph/0608308}{{\ttfamily hep-ph/0608308}}].

\bibitem{Ravindran:2007sv}
V.~Ravindran and J.~Smith, \emph{{Threshold corrections to rapidity
  distributions of $Z$ and $W^\pm$ bosons beyond $N^{2}$ LO at hadron
  colliders}}, \href{https://doi.org/10.1103/PhysRevD.76.114004}{\emph{Phys.
  Rev. D} {\bfseries 76} (2007) 114004}
  [\href{https://arxiv.org/abs/0708.1689}{{\ttfamily 0708.1689}}].

\bibitem{Ahmed:2014uya}
T.~Ahmed, M.~K. Mandal, N.~Rana and V.~Ravindran, \emph{{Rapidity Distributions
  in Drell-Yan and Higgs Productions at Threshold to Third Order in QCD}},
  \href{https://doi.org/10.1103/PhysRevLett.113.212003}{\emph{Phys. Rev. Lett.}
  {\bfseries 113} (2014) 212003}
  [\href{https://arxiv.org/abs/1404.6504}{{\ttfamily 1404.6504}}].

\bibitem{Anastasiou:2014vaa}
C.~Anastasiou, C.~Duhr, F.~Dulat, E.~Furlan, T.~Gehrmann, F.~Herzog et~al.,
  \emph{{Higgs boson gluon\textendash{}fusion production at threshold in
  N$^3$LO QCD}},
  \href{https://doi.org/10.1016/j.physletb.2014.08.067}{\emph{Phys. Lett. B}
  {\bfseries 737} (2014) 325}
  [\href{https://arxiv.org/abs/1403.4616}{{\ttfamily 1403.4616}}].

\bibitem{Moch:2004pa}
S.~Moch, J.~A.~M. Vermaseren and A.~Vogt, \emph{{The Three loop splitting
  functions in QCD: The Nonsinglet case}},
  \href{https://doi.org/10.1016/j.nuclphysb.2004.03.030}{\emph{Nucl. Phys. B}
  {\bfseries 688} (2004) 101}
  [\href{https://arxiv.org/abs/hep-ph/0403192}{{\ttfamily hep-ph/0403192}}].

\bibitem{Vogt:2004mw}
A.~Vogt, S.~Moch and J.~A.~M. Vermaseren, \emph{{The Three-loop splitting
  functions in QCD: The Singlet case}},
  \href{https://doi.org/10.1016/j.nuclphysb.2004.04.024}{\emph{Nucl. Phys. B}
  {\bfseries 691} (2004) 129}
  [\href{https://arxiv.org/abs/hep-ph/0404111}{{\ttfamily hep-ph/0404111}}].

\bibitem{Gehrmann:2010ue}
T.~Gehrmann, E.~W.~N. Glover, T.~Huber, N.~Ikizlerli and C.~Studerus,
  \emph{{Calculation of the quark and gluon form factors to three loops in
  QCD}}, \href{https://doi.org/10.1007/JHEP06(2010)094}{\emph{JHEP} {\bfseries
  06} (2010) 094} [\href{https://arxiv.org/abs/1004.3653}{{\ttfamily
  1004.3653}}].

\bibitem{Li:2012wna}
Y.~Li and F.~Petriello, \emph{{Combining QCD and electroweak corrections to
  dilepton production in FEWZ}},
  \href{https://doi.org/10.1103/PhysRevD.86.094034}{\emph{Phys. Rev. D}
  {\bfseries 86} (2012) 094034}
  [\href{https://arxiv.org/abs/1208.5967}{{\ttfamily 1208.5967}}].

\bibitem{Catani:2007vq}
S.~Catani and M.~Grazzini, \emph{{An NNLO subtraction formalism in hadron
  collisions and its application to Higgs boson production at the LHC}},
  \href{https://doi.org/10.1103/PhysRevLett.98.222002}{\emph{Phys. Rev. Lett.}
  {\bfseries 98} (2007) 222002}
  [\href{https://arxiv.org/abs/hep-ph/0703012}{{\ttfamily hep-ph/0703012}}].

\bibitem{Catani:2010en}
S.~Catani, G.~Ferrera and M.~Grazzini, \emph{{W Boson Production at Hadron
  Colliders: The Lepton Charge Asymmetry in NNLO QCD}},
  \href{https://doi.org/10.1007/JHEP05(2010)006}{\emph{JHEP} {\bfseries 05}
  (2010) 006} [\href{https://arxiv.org/abs/1002.3115}{{\ttfamily 1002.3115}}].

\bibitem{Catani:2011qz}
S.~Catani, L.~Cieri, D.~de~Florian, G.~Ferrera and M.~Grazzini, \emph{{Diphoton
  production at hadron colliders: a fully-differential QCD calculation at
  NNLO}}, \href{https://doi.org/10.1103/PhysRevLett.108.072001}{\emph{Phys.
  Rev. Lett.} {\bfseries 108} (2012) 072001}
  [\href{https://arxiv.org/abs/1110.2375}{{\ttfamily 1110.2375}}].

\bibitem{Chen:2021vtu}
X.~Chen, T.~Gehrmann, N.~Glover, A.~Huss, T.-Z. Yang and H.~X. Zhu,
  \emph{{Dilepton Rapidity Distribution in Drell-Yan Production to Third Order
  in QCD}}, \href{https://doi.org/10.1103/PhysRevLett.128.052001}{\emph{Phys.
  Rev. Lett.} {\bfseries 128} (2022) 052001}
  [\href{https://arxiv.org/abs/2107.09085}{{\ttfamily 2107.09085}}].

\bibitem{Chen:2022lwc}
X.~Chen, T.~Gehrmann, N.~Glover, A.~Huss, T.-Z. Yang and H.~X. Zhu,
  \emph{{Transverse Mass Distribution and Charge Asymmetry in W Boson
  Production to Third Order in QCD}},
  \href{https://arxiv.org/abs/2205.11426}{{\ttfamily 2205.11426}}.

\bibitem{Laenen:1992ey}
E.~Laenen and G.~F. Sterman, \emph{{Resummation for Drell-Yan differential
  distributions}},  in \emph{{7th Meeting of the APS Division of Particles
  Fields}}, pp.~987--989, 11, 1992.

\bibitem{Mukherjee:2006uu}
A.~Mukherjee and W.~Vogelsang, \emph{{Threshold resummation for W-boson
  production at RHIC}},
  \href{https://doi.org/10.1103/PhysRevD.73.074005}{\emph{Phys. Rev. D}
  {\bfseries 73} (2006) 074005}
  [\href{https://arxiv.org/abs/hep-ph/0601162}{{\ttfamily hep-ph/0601162}}].

\bibitem{Sterman:2000pt}
G.~F. Sterman and W.~Vogelsang, \emph{{Threshold resummation and rapidity
  dependence}},
  \href{https://doi.org/10.1088/1126-6708/2001/02/016}{\emph{JHEP} {\bfseries
  02} (2001) 016} [\href{https://arxiv.org/abs/hep-ph/0011289}{{\ttfamily
  hep-ph/0011289}}].

\bibitem{Bolzoni:2006ky}
P.~Bolzoni, \emph{{Threshold resummation of Drell-Yan rapidity distributions}},
  \href{https://doi.org/10.1016/j.physletb.2006.10.064}{\emph{Phys. Lett. B}
  {\bfseries 643} (2006) 325}
  [\href{https://arxiv.org/abs/hep-ph/0609073}{{\ttfamily hep-ph/0609073}}].

\bibitem{Bonvini:2010tp}
M.~Bonvini, S.~Forte and G.~Ridolfi, \emph{{Soft gluon resummation of Drell-Yan
  rapidity distributions: Theory and phenomenology}},
  \href{https://doi.org/10.1016/j.nuclphysb.2011.01.023}{\emph{Nucl. Phys. B}
  {\bfseries 847} (2011) 93} [\href{https://arxiv.org/abs/1009.5691}{{\ttfamily
  1009.5691}}].

\bibitem{Catani:1989ne}
S.~Catani and L.~Trentadue, \emph{{Resummation of the QCD Perturbative Series
  for Hard Processes}},
  \href{https://doi.org/10.1016/0550-3213(89)90273-3}{\emph{Nucl. Phys. B}
  {\bfseries 327} (1989) 323}.

\bibitem{Westmark:2017uig}
D.~Westmark and J.~F. Owens, \emph{{Enhanced threshold resummation formalism
  for lepton pair production and its effects in the determination of parton
  distribution functions}},
  \href{https://doi.org/10.1103/PhysRevD.95.056024}{\emph{Phys. Rev. D}
  {\bfseries 95} (2017) 056024}
  [\href{https://arxiv.org/abs/1701.06716}{{\ttfamily 1701.06716}}].

\bibitem{Banerjee:2017cfc}
P.~Banerjee, G.~Das, P.~K. Dhani and V.~Ravindran, \emph{{Threshold resummation
  of the rapidity distribution for Higgs production at NNLO+NNLL}},
  \href{https://doi.org/10.1103/PhysRevD.97.054024}{\emph{Phys. Rev. D}
  {\bfseries 97} (2018) 054024}
  [\href{https://arxiv.org/abs/1708.05706}{{\ttfamily 1708.05706}}].

\bibitem{Banerjee:2018vvb}
P.~Banerjee, G.~Das, P.~K. Dhani and V.~Ravindran, \emph{{Threshold resummation
  of the rapidity distribution for Drell-Yan production at NNLO+NNLL}},
  \href{https://doi.org/10.1103/PhysRevD.98.054018}{\emph{Phys. Rev. D}
  {\bfseries 98} (2018) 054018}
  [\href{https://arxiv.org/abs/1805.01186}{{\ttfamily 1805.01186}}].

\bibitem{DASbbH}
G.~Das, \textit{et. al.}, \emph{{Higgs rapidity in bottom annihilation at NNLL
  and beyond}}, {\emph{\normalfont{in preparation}} }.

\bibitem{DASggH}
G.~Das, \textit{et. al.}, \emph{{Higgs rapidity at N3LL and its uncertainty}},
  {\emph{\normalfont{in preparation}} }.

\bibitem{Banerjee:2018mkm}
P.~Banerjee, G.~Das, P.~K. Dhani and V.~Ravindran, \emph{{Threshold resummation
  in the rapidity distribution for a colorless particle production at the
  LHC}}, \href{https://doi.org/10.22323/1.303.0043}{\emph{PoS} {\bfseries
  LL2018} (2018) 043} [\href{https://arxiv.org/abs/1807.04583}{{\ttfamily
  1807.04583}}].

\bibitem{Ahmed:2020amh}
T.~Ahmed, A.~A. H., P.~Mukherjee, V.~Ravindran and A.~Sankar, \emph{{Rapidity
  distribution at soft-virtual and beyond for $n$-colorless particles to
  ${N}^4$LO in QCD}},
  \href{https://doi.org/10.1140/epjc/s10052-021-09658-9}{\emph{Eur. Phys. J. C}
  {\bfseries 81} (2021) 943}
  [\href{https://arxiv.org/abs/2010.02980}{{\ttfamily 2010.02980}}].

\bibitem{Lustermans:2019cau}
G.~Lustermans, J.~K.~L. Michel and F.~J. Tackmann, \emph{{Generalized Threshold
  Factorization with Full Collinear Dynamics}},
  \href{https://arxiv.org/abs/1908.00985}{{\ttfamily 1908.00985}}.

\bibitem{Ebert:2017uel}
M.~A. Ebert, J.~K.~L. Michel and F.~J. Tackmann, \emph{{Resummation Improved
  Rapidity Spectrum for Gluon Fusion Higgs Production}},
  \href{https://doi.org/10.1007/JHEP05(2017)088}{\emph{JHEP} {\bfseries 05}
  (2017) 088} [\href{https://arxiv.org/abs/1702.00794}{{\ttfamily
  1702.00794}}].

\bibitem{Bauer:2000ew}
C.~W. Bauer, S.~Fleming and M.~E. Luke, \emph{{Summing Sudakov logarithms in $B
  \to X_s \gamma $in effective field theory.}},
  \href{https://doi.org/10.1103/PhysRevD.63.014006}{\emph{Phys. Rev. D}
  {\bfseries 63} (2000) 014006}
  [\href{https://arxiv.org/abs/hep-ph/0005275}{{\ttfamily hep-ph/0005275}}].

\bibitem{Bauer:2000yr}
C.~W. Bauer, S.~Fleming, D.~Pirjol and I.~W. Stewart, \emph{{An Effective field
  theory for collinear and soft gluons: Heavy to light decays}},
  \href{https://doi.org/10.1103/PhysRevD.63.114020}{\emph{Phys. Rev. D}
  {\bfseries 63} (2001) 114020}
  [\href{https://arxiv.org/abs/hep-ph/0011336}{{\ttfamily hep-ph/0011336}}].

\bibitem{Bauer:2001ct}
C.~W. Bauer and I.~W. Stewart, \emph{{Invariant operators in collinear
  effective theory}},
  \href{https://doi.org/10.1016/S0370-2693(01)00902-9}{\emph{Phys. Lett. B}
  {\bfseries 516} (2001) 134}
  [\href{https://arxiv.org/abs/hep-ph/0107001}{{\ttfamily hep-ph/0107001}}].

\bibitem{Bauer:2001yt}
C.~W. Bauer, D.~Pirjol and I.~W. Stewart, \emph{{Soft collinear factorization
  in effective field theory}},
  \href{https://doi.org/10.1103/PhysRevD.65.054022}{\emph{Phys. Rev. D}
  {\bfseries 65} (2002) 054022}
  [\href{https://arxiv.org/abs/hep-ph/0109045}{{\ttfamily hep-ph/0109045}}].

\bibitem{Bauer:2002nz}
C.~W. Bauer, S.~Fleming, D.~Pirjol, I.~Z. Rothstein and I.~W. Stewart,
  \emph{{Hard scattering factorization from effective field theory}},
  \href{https://doi.org/10.1103/PhysRevD.66.014017}{\emph{Phys. Rev. D}
  {\bfseries 66} (2002) 014017}
  [\href{https://arxiv.org/abs/hep-ph/0202088}{{\ttfamily hep-ph/0202088}}].

\bibitem{scetIain}
C.~W. Bauer and I.~W. Stewart, \emph{{The Soft-Collinear Effective Theory}},
  {\emph{\href{https://courses.edx.org/c4x/MITx/8.EFTx/asset/notes_scetnotes.pdf}{(Link)}}
  }.

\bibitem{Becher:2014oda}
T.~Becher, A.~Broggio and A.~Ferroglia, \emph{{Introduction to Soft-Collinear
  Effective Theory}}, vol.~896. Springer, 2015,
  \href{https://doi.org/10.1007/978-3-319-14848-9}{10.1007/978-3-319-14848-9},
  [\href{https://arxiv.org/abs/1410.1892}{{\ttfamily 1410.1892}}].

\bibitem{ellis_stirling_webber_1996}
R.~K. Ellis, W.~J. Stirling and B.~R. Webber, \emph{QCD and Collider Physics},
  Cambridge Monographs on Particle Physics, Nuclear Physics and Cosmology.
  Cambridge University Press, 1996,
  \href{https://doi.org/10.1017/CBO9780511628788}{10.1017/CBO9780511628788}.

\bibitem{Vogt:2004ns}
A.~Vogt, \emph{{Efficient evolution of unpolarized and polarized parton
  distributions with QCD-PEGASUS}},
  \href{https://doi.org/10.1016/j.cpc.2005.03.103}{\emph{Comput. Phys. Commun.}
  {\bfseries 170} (2005) 65}
  [\href{https://arxiv.org/abs/hep-ph/0408244}{{\ttfamily hep-ph/0408244}}].

\bibitem{Kulesza:2002rh}
A.~Kulesza, G.~F. Sterman and W.~Vogelsang, \emph{{Joint resummation in
  electroweak boson production}},
  \href{https://doi.org/10.1103/PhysRevD.66.014011}{\emph{Phys. Rev. D}
  {\bfseries 66} (2002) 014011}
  [\href{https://arxiv.org/abs/hep-ph/0202251}{{\ttfamily hep-ph/0202251}}].

\bibitem{Buckley:2014ana}
A.~Buckley, J.~Ferrando, S.~Lloyd, K.~Nordstr\"om, B.~Page, M.~R\"ufenacht
  et~al., \emph{{LHAPDF6: parton density access in the LHC precision era}},
  \href{https://doi.org/10.1140/epjc/s10052-015-3318-8}{\emph{Eur. Phys. J. C}
  {\bfseries 75} (2015) 132} [\href{https://arxiv.org/abs/1412.7420}{{\ttfamily
  1412.7420}}].

\bibitem{Bailey:2020ooq}
S.~Bailey, T.~Cridge, L.~A. Harland-Lang, A.~D. Martin and R.~S. Thorne,
  \emph{{Parton distributions from LHC, HERA, Tevatron and fixed target data:
  MSHT20 PDFs}},
  \href{https://doi.org/10.1140/epjc/s10052-021-09057-0}{\emph{Eur. Phys. J. C}
  {\bfseries 81} (2021) 341}
  [\href{https://arxiv.org/abs/2012.04684}{{\ttfamily 2012.04684}}].

\bibitem{Hahn:2004fe}
T.~Hahn, \emph{{CUBA: A Library for multidimensional numerical integration}},
  \href{https://doi.org/10.1016/j.cpc.2005.01.010}{\emph{Comput. Phys. Commun.}
  {\bfseries 168} (2005) 78}
  [\href{https://arxiv.org/abs/hep-ph/0404043}{{\ttfamily hep-ph/0404043}}].

\bibitem{Hahn:2014fua}
T.~Hahn, \emph{{Concurrent Cuba}},
  \href{https://doi.org/10.1088/1742-6596/608/1/012066}{\emph{J. Phys. Conf.
  Ser.} {\bfseries 608} (2015) 012066}
  [\href{https://arxiv.org/abs/1408.6373}{{\ttfamily 1408.6373}}].

\bibitem{ParticleDataGroup:2020ssz}
{\scshape Particle Data Group} collaboration, P.~A. Zyla et~al., \emph{{Review
  of Particle Physics}},
  \href{https://doi.org/10.1093/ptep/ptaa104}{\emph{PTEP} {\bfseries 2020}
  (2020) 083C01}.

\bibitem{Das:2019bxi}
G.~Das, M.~C. Kumar and K.~Samanta, \emph{{Resummed inclusive cross-section in
  ADD model at N$^{3}$LL}},
  \href{https://doi.org/10.1007/JHEP10(2020)161}{\emph{JHEP} {\bfseries 10}
  (2020) 161} [\href{https://arxiv.org/abs/1912.13039}{{\ttfamily
  1912.13039}}].

\bibitem{Das:2020gie}
G.~Das, M.~C. Kumar and K.~Samanta, \emph{{Resummed inclusive cross-section in
  Randall-Sundrum model at NNLO+NNLL}},
  \href{https://doi.org/10.1007/JHEP07(2020)040}{\emph{JHEP} {\bfseries 07}
  (2020) 040} [\href{https://arxiv.org/abs/2004.03938}{{\ttfamily
  2004.03938}}].

\bibitem{Das:2020pzo}
G.~Das, M.~C. Kumar and K.~Samanta, \emph{{Precision QCD phenomenology of
  exotic spin-2 search at the LHC}},
  \href{https://doi.org/10.1007/JHEP04(2021)111}{\emph{JHEP} {\bfseries 04}
  (2021) 111} [\href{https://arxiv.org/abs/2011.15121}{{\ttfamily
  2011.15121}}].

\bibitem{Cacciari:2011ze}
M.~Cacciari and N.~Houdeau, \emph{{Meaningful characterisation of perturbative
  theoretical uncertainties}},
  \href{https://doi.org/10.1007/JHEP09(2011)039}{\emph{JHEP} {\bfseries 09}
  (2011) 039} [\href{https://arxiv.org/abs/1105.5152}{{\ttfamily 1105.5152}}].

\bibitem{Bonvini:2020xeo}
M.~Bonvini, \emph{{Probabilistic definition of the perturbative theoretical
  uncertainty from missing higher orders}},
  \href{https://doi.org/10.1140/epjc/s10052-020-08545-z}{\emph{Eur. Phys. J. C}
  {\bfseries 80} (2020) 989}
  [\href{https://arxiv.org/abs/2006.16293}{{\ttfamily 2006.16293}}].

\bibitem{McGowan:2022nag}
J.~McGowan, T.~Cridge, L.~A. Harland-Lang and R.~S. Thorne, \emph{{Approximate
  N$^{3}$LO parton distribution functions with theoretical uncertainties:
  MSHT20aN$^3$LO PDFs}},
  \href{https://doi.org/10.1140/epjc/s10052-023-11236-0}{\emph{Eur. Phys. J. C}
  {\bfseries 83} (2023) 185}
  [\href{https://arxiv.org/abs/2207.04739}{{\ttfamily 2207.04739}}].

\bibitem{Accardi:2016ndt}
A.~Accardi et~al., \emph{{A Critical Appraisal and Evaluation of Modern PDFs}},
  \href{https://doi.org/10.1140/epjc/s10052-016-4285-4}{\emph{Eur. Phys. J. C}
  {\bfseries 76} (2016) 471}
  [\href{https://arxiv.org/abs/1603.08906}{{\ttfamily 1603.08906}}].

\bibitem{Altarelli:2008aj}
G.~Altarelli, R.~D. Ball and S.~Forte, \emph{{Small x Resummation with Quarks:
  Deep-Inelastic Scattering}},
  \href{https://doi.org/10.1016/j.nuclphysb.2008.03.003}{\emph{Nucl. Phys. B}
  {\bfseries 799} (2008) 199}
  [\href{https://arxiv.org/abs/0802.0032}{{\ttfamily 0802.0032}}].

\bibitem{Caola:2010kv}
F.~Caola, S.~Forte and S.~Marzani, \emph{{Small x resummation of rapidity
  distributions: The Case of Higgs production}},
  \href{https://doi.org/10.1016/j.nuclphysb.2011.01.001}{\emph{Nucl. Phys. B}
  {\bfseries 846} (2011) 167}
  [\href{https://arxiv.org/abs/1010.2743}{{\ttfamily 1010.2743}}].

\bibitem{Bonvini:2016wki}
M.~Bonvini, S.~Marzani and T.~Peraro, \emph{{Small-$x$ resummation from HELL}},
  \href{https://doi.org/10.1140/epjc/s10052-016-4445-6}{\emph{Eur. Phys. J. C}
  {\bfseries 76} (2016) 597}
  [\href{https://arxiv.org/abs/1607.02153}{{\ttfamily 1607.02153}}].

\bibitem{DelDuca:2017twk}
V.~Del~Duca, E.~Laenen, L.~Magnea, L.~Vernazza and C.~D. White,
  \emph{{Universality of next-to-leading power threshold effects for colourless
  final states in hadronic collisions}},
  \href{https://doi.org/10.1007/JHEP11(2017)057}{\emph{JHEP} {\bfseries 11}
  (2017) 057} [\href{https://arxiv.org/abs/1706.04018}{{\ttfamily
  1706.04018}}].

\bibitem{Ebert:2018lzn}
M.~A. Ebert, I.~Moult, I.~W. Stewart, F.~J. Tackmann, G.~Vita and H.~X. Zhu,
  \emph{{Power Corrections for N-Jettiness Subtractions at ${\cal
  O}(\alpha_s)$}}, \href{https://doi.org/10.1007/JHEP12(2018)084}{\emph{JHEP}
  {\bfseries 12} (2018) 084}
  [\href{https://arxiv.org/abs/1807.10764}{{\ttfamily 1807.10764}}].

\bibitem{Beneke:2018gvs}
M.~Beneke, A.~Broggio, M.~Garny, S.~Jaskiewicz, R.~Szafron, L.~Vernazza et~al.,
  \emph{{Leading-logarithmic threshold resummation of the Drell-Yan process at
  next-to-leading power}},
  \href{https://doi.org/10.1007/JHEP03(2019)043}{\emph{JHEP} {\bfseries 03}
  (2019) 043} [\href{https://arxiv.org/abs/1809.10631}{{\ttfamily
  1809.10631}}].

\bibitem{Bahjat-Abbas:2019fqa}
N.~Bahjat-Abbas, D.~Bonocore, J.~Sinninghe~Damst\'e, E.~Laenen, L.~Magnea,
  L.~Vernazza et~al., \emph{{Diagrammatic resummation of leading-logarithmic
  threshold effects at next-to-leading power}},
  \href{https://doi.org/10.1007/JHEP11(2019)002}{\emph{JHEP} {\bfseries 11}
  (2019) 002} [\href{https://arxiv.org/abs/1905.13710}{{\ttfamily
  1905.13710}}].

\bibitem{Ajjath:2020lwb}
A.~H. Ajjath, P.~Mukherjee, V.~Ravindran, A.~Sankar and S.~Tiwari,
  \emph{{Next-to-soft corrections for Drell-Yan and Higgs boson rapidity
  distributions beyond N$^3$LO}},
  \href{https://doi.org/10.1103/PhysRevD.103.L111502}{\emph{Phys. Rev. D}
  {\bfseries 103} (2021) L111502}
  [\href{https://arxiv.org/abs/2010.00079}{{\ttfamily 2010.00079}}].

\bibitem{Ajjath:2021pre}
A.~H. Ajjath, P.~Mukherjee, V.~Ravindran, A.~Sankar and S.~Tiwari,
  \emph{{Next-to-soft-virtual resummed rapidity distribution for the Drell-Yan
  process to NNLO+NNLL\textasciimacron{}}},
  \href{https://doi.org/10.1103/PhysRevD.106.034005}{\emph{Phys. Rev. D}
  {\bfseries 106} (2022) 034005}
  [\href{https://arxiv.org/abs/2112.14094}{{\ttfamily 2112.14094}}].

\bibitem{Alekhin:2021xcu}
S.~Alekhin, A.~Kardos, S.~Moch and Z.~Tr\'ocs\'anyi, \emph{{Precision studies
  for Drell\textendash{}Yan processes at NNLO}},
  \href{https://doi.org/10.1140/epjc/s10052-021-09361-9}{\emph{Eur. Phys. J. C}
  {\bfseries 81} (2021) 573}
  [\href{https://arxiv.org/abs/2104.02400}{{\ttfamily 2104.02400}}].

\bibitem{Vermaseren:2000nd}
J.~A.~M. Vermaseren, \emph{{New features of FORM}},
  \href{https://arxiv.org/abs/math-ph/0010025}{{\ttfamily math-ph/0010025}}.

\bibitem{Ruijl:2017dtg}
B.~Ruijl, T.~Ueda and J.~Vermaseren, \emph{{FORM version 4.2}},
  \href{https://arxiv.org/abs/1707.06453}{{\ttfamily 1707.06453}}.

\bibitem{Gross:1973id}
D.~J. Gross and F.~Wilczek, \emph{{Ultraviolet Behavior of Nonabelian Gauge
  Theories}}, \href{https://doi.org/10.1103/PhysRevLett.30.1343}{\emph{Phys.
  Rev. Lett.} {\bfseries 30} (1973) 1343}.

\bibitem{Politzer:1973fx}
H.~D. Politzer, \emph{{Reliable Perturbative Results for Strong
  Interactions?}},
  \href{https://doi.org/10.1103/PhysRevLett.30.1346}{\emph{Phys. Rev. Lett.}
  {\bfseries 30} (1973) 1346}.

\bibitem{Caswell:1974gg}
W.~E. Caswell, \emph{{Asymptotic Behavior of Nonabelian Gauge Theories to Two
  Loop Order}}, \href{https://doi.org/10.1103/PhysRevLett.33.244}{\emph{Phys.
  Rev. Lett.} {\bfseries 33} (1974) 244}.

\bibitem{Jones:1974mm}
D.~R.~T. Jones, \emph{{Two Loop Diagrams in Yang-Mills Theory}},
  \href{https://doi.org/10.1016/0550-3213(74)90093-5}{\emph{Nucl. Phys. B}
  {\bfseries 75} (1974) 531}.

\bibitem{Egorian:1978zx}
E.~Egorian and O.~V. Tarasov, \emph{{Two Loop Renormalization of the {QCD} in
  an Arbitrary Gauge}}, {\emph{Teor. Mat. Fiz.} {\bfseries 41} (1979) 26}.

\bibitem{Tarasov:1980au}
O.~V. Tarasov, A.~A. Vladimirov and A.~Y. Zharkov, \emph{{The Gell-Mann-Low
  Function of QCD in the Three Loop Approximation}},
  \href{https://doi.org/10.1016/0370-2693(80)90358-5}{\emph{Phys. Lett. B}
  {\bfseries 93} (1980) 429}.

\bibitem{Larin:1993tp}
S.~A. Larin and J.~A.~M. Vermaseren, \emph{{The Three loop QCD Beta function
  and anomalous dimensions}},
  \href{https://doi.org/10.1016/0370-2693(93)91441-O}{\emph{Phys. Lett. B}
  {\bfseries 303} (1993) 334}
  [\href{https://arxiv.org/abs/hep-ph/9302208}{{\ttfamily hep-ph/9302208}}].

\bibitem{vanRitbergen:1997va}
T.~van Ritbergen, J.~A.~M. Vermaseren and S.~A. Larin, \emph{{The Four loop
  beta function in quantum chromodynamics}},
  \href{https://doi.org/10.1016/S0370-2693(97)00370-5}{\emph{Phys. Lett. B}
  {\bfseries 400} (1997) 379}
  [\href{https://arxiv.org/abs/hep-ph/9701390}{{\ttfamily hep-ph/9701390}}].

\bibitem{Czakon:2004bu}
M.~Czakon, \emph{{The Four-loop QCD beta-function and anomalous dimensions}},
  \href{https://doi.org/10.1016/j.nuclphysb.2005.01.012}{\emph{Nucl. Phys. B}
  {\bfseries 710} (2005) 485}
  [\href{https://arxiv.org/abs/hep-ph/0411261}{{\ttfamily hep-ph/0411261}}].

\bibitem{Baikov:2016tgj}
P.~A. Baikov, K.~G. Chetyrkin and J.~H. K\"uhn, \emph{{Five-Loop Running of the
  QCD coupling constant}},
  \href{https://doi.org/10.1103/PhysRevLett.118.082002}{\emph{Phys. Rev. Lett.}
  {\bfseries 118} (2017) 082002}
  [\href{https://arxiv.org/abs/1606.08659}{{\ttfamily 1606.08659}}].

\bibitem{Herzog:2017ohr}
F.~Herzog, B.~Ruijl, T.~Ueda, J.~A.~M. Vermaseren and A.~Vogt, \emph{{The
  five-loop beta function of Yang-Mills theory with fermions}},
  \href{https://doi.org/10.1007/JHEP02(2017)090}{\emph{JHEP} {\bfseries 02}
  (2017) 090} [\href{https://arxiv.org/abs/1701.01404}{{\ttfamily
  1701.01404}}].

\bibitem{Luthe:2017ttg}
T.~Luthe, A.~Maier, P.~Marquard and Y.~Schroder, \emph{{The five-loop Beta
  function for a general gauge group and anomalous dimensions beyond Feynman
  gauge}}, \href{https://doi.org/10.1007/JHEP10(2017)166}{\emph{JHEP}
  {\bfseries 10} (2017) 166}
  [\href{https://arxiv.org/abs/1709.07718}{{\ttfamily 1709.07718}}].

\bibitem{Henn:2019swt}
J.~M. Henn, G.~P. Korchemsky and B.~Mistlberger, \emph{{The full four-loop cusp
  anomalous dimension in $\mathcal{N}=4$ super Yang-Mills and QCD}},
  \href{https://doi.org/10.1007/JHEP04(2020)018}{\emph{JHEP} {\bfseries 04}
  (2020) 018} [\href{https://arxiv.org/abs/1911.10174}{{\ttfamily
  1911.10174}}].

\bibitem{Huber:2019fxe}
T.~Huber, A.~von Manteuffel, E.~Panzer, R.~M. Schabinger and G.~Yang,
  \emph{{The four-loop cusp anomalous dimension from the $N=4$ Sudakov form
  factor}}, \href{https://doi.org/10.1016/j.physletb.2020.135543}{\emph{Phys.
  Lett. B} {\bfseries 807} (2020) 135543}
  [\href{https://arxiv.org/abs/1912.13459}{{\ttfamily 1912.13459}}].

\bibitem{vonManteuffel:2020vjv}
A.~von Manteuffel, E.~Panzer and R.~M. Schabinger, \emph{{Cusp and collinear
  anomalous dimensions in four-loop QCD from form factors}},
  \href{https://doi.org/10.1103/PhysRevLett.124.162001}{\emph{Phys. Rev. Lett.}
  {\bfseries 124} (2020) 162001}
  [\href{https://arxiv.org/abs/2002.04617}{{\ttfamily 2002.04617}}].

\bibitem{Larin:1996wd}
S.~A. Larin, P.~Nogueira, T.~van Ritbergen and J.~A.~M. Vermaseren, \emph{{The
  Three loop QCD calculation of the moments of deep inelastic structure
  functions}}, \href{https://doi.org/10.1016/S0550-3213(97)80038-7}{\emph{Nucl.
  Phys. B} {\bfseries 492} (1997) 338}
  [\href{https://arxiv.org/abs/hep-ph/9605317}{{\ttfamily hep-ph/9605317}}].

\bibitem{Vermaseren:2005qc}
J.~A.~M. Vermaseren, A.~Vogt and S.~Moch, \emph{{The Third-order QCD
  corrections to deep-inelastic scattering by photon exchange}},
  \href{https://doi.org/10.1016/j.nuclphysb.2005.06.020}{\emph{Nucl. Phys. B}
  {\bfseries 724} (2005) 3}
  [\href{https://arxiv.org/abs/hep-ph/0504242}{{\ttfamily hep-ph/0504242}}].

\end{thebibliography}\endgroup
